\documentclass[a4paper,10pt]{article}
\usepackage{jheppub}
\usepackage[utf8]{inputenc}
\usepackage{amsmath}
\usepackage{amssymb}
\usepackage{tikz-feynman}%[compat=1.1.0]
\title{\boldmath Long distance behavior of $O(N)$-model correlators in de Sitter space and the resummation of secular terms }

\author[a]{Diana L\'opez Nacir}
\author[b]{Francisco D. Mazzitelli}
\author[c]{Leonardo G. Trombetta}

\affiliation[a]{Departamento de F\'\i sica and IFIBA, FCEyN UBA, Facultad de Ciencias Exactas y Naturales, 
 Ciudad Universitaria, Pabell\' on I, 1428 Buenos Aires, Argentina}
\affiliation[b]{Centro At\'omico Bariloche, Instituto Balseiro and CONICET,\\ Comisi\'on Nacional de Energ\'\i a At\'omica, Av. Bustillo 9500, R8402AGP Bariloche, Argentina.}
\affiliation[c]{Scuola Normale Superiore, Piazza dei Cavalieri 7, 56126, Pisa, Italy\\
INFN – Sezione di Pisa, 56200, Pisa, Italy}

\emailAdd{dnacir@df.uba.ar}
\emailAdd{fdmazzi@cab.cnea.gov.ar}
\emailAdd{leonardo.trombetta@sns.it}

\abstract{ We analyze the long distance behavior of the two-point functions for an interacting scalar $O(N)$ model in de Sitter spacetime.  Following our previous work, this behavior is analyzed by analytic continuation of the Euclidean correlators, which are computed by treating the homogeneous zero mode exactly and using a partial resummation of the interactions between the zero and the non-zero modes.  We focus on massless fields and  present an alternative derivation of our method, which involves a double expansion in $1/N$ and the coupling constant of the theory. This derivation is simpler than the previous one  and can be directly extended  for fields with negative squared-mass. 
We extend our previous results by computing the long wavelength limit of the  two-point functions at next-to-leading order in $1/N$ and at leading order in the coupling constant, which involves a further resummation  of Feynman diagrams (needed when the two-point functions are analitically continued). We prove that, after this 
extra resummation,  the two-point functions
vanish in the long distance limit.}

\begin{document}

\maketitle

\section{Introduction}

Since its formulation, quantum field theory (QFT) in de Sitter (dS) spacetime has been a subject that has received a lot of interest and attention. One clear reason is that  in a dS spacetime, being a  maximally symmetric curved space,  it is possible to obtain  exact solutions and  it is relatively simpler to quantize fields.  Physical motivations  increased  when realizing   its application to the early universe, with the emergence of the inflationary paradigm, and  also because of the discovery of the late time accelerated expansion of the Universe. 
This  subject is  clearly worth studying further.

The exponential expansion of the metric in dS spacetime produces an effective growth in the couplings of the theory, that turns out to be crucial in the infrared (IR). For instance,
when considering a  self-interacting scalar field of mass $m$ in a $\lambda\phi^4$ theory, the effective coupling becomes $\lambda H^2/m^2$, and therefore the usual 
perturbative calculations break down for light fields \cite{Burgess}. For massless free fields, minimally coupled scalar fields do not admit a dS invariant vacuum state 
\cite{Allen}.

The understanding of IR effects in QFT  in dS spacetime is still  incomplete. 
For  massless interacting  scalar fields in dS spacetime, this requires the use of nonperturbative techniques. 
There has been a lot of progress in  understanding   the leading IR effects. Indeed,  there is more than one framework in which nonperturbative results have been obtained. 
 For instance,  the  so-called dynamical mass  of  a quantum scalar field $\phi$ with classical potential $\lambda\phi^4/8$ has been computed using  the well known stochastic approach \cite{Starobinsky1,Starobinsky2} and  also by formulating the theory on Euclidean dS space (a $4$-sphere) \cite{Rajaraman1}, yielding 
 \begin{equation}
 m_{\rm dyn}^2=\frac{\sqrt{3\lambda}H^2\Gamma\left(\frac{1}{4}\right)}{8\pi\Gamma\left(\frac{3}{4}\right)},
 \end{equation}
which is clearly nonperturbative in $\lambda$. As was  originally emphasized in \cite{BLHU},  using the analogy with  finite-size effects in condensed matter, the  generation of  a dynamical mass represents an important and characteristic IR effect of the theory.  Indeed, by splitting the field into an IR  part, consisting of the near-homogeneous sector of the field, $\phi_{IR}$, plus the remaining ultraviolet (UV) part, $\phi=\phi_{IR}+\phi_{UV}$, it can be shown that the dynamical mass  is related to the curvature of  the effective potential   in the symmetric phase, and to the leading IR order (leading order in  $\sqrt\lambda$) of the correlator of the IR part of the field,
  \begin{equation}\langle\phi_{IR} (x)\phi_{IR}(x')\rangle\simeq\frac{3 H^3}{8\pi^2 m_{\rm dyn}^{2}}\,.\end{equation}
The above nonperturbative result  is important to cure the divergence appearing in the  free massless correlator. However,  corrections beyond  this constant  IR contribution  are necessary to  understand the long distance  limit  of  the     correlators. 
This last claim is of course irrelevant for the sphere, which is compact, but it is crucial for the theory in dS spacetime.
  
  For a theory with only one scalar field,  a nonperturbative method    that allows a systematic  calculations of such corrections  is   very difficult and, to our knowledge, it is still lacking. 
 One possibility recently explored involves the use of nonperturbative renormalization group techniques \cite{Guilleux}.
There are also nonperturbative calculations based on  stochastic methods  from which it is possible to obtain a   numerical solution that  predicts  how the   correlator of the IR sector    $\langle\phi_{IR} (x)\phi_{IR} (x')\rangle$  decays when  the dS-invariant distance between the two points, $r(x,x')$, approaches infinity.
 To achieve this, the method is based on  an unequal treatment of the 
 IR and UV parts of the field.   It is not clear  however how robust  this result is against the way in which the field is  split  and it is unknown how to compute corrections to the leading order result in a systematic way.  The relation between the different approaches (stochastic, Lorentzian, Euclidean), has been the subject of many works \cite{relation1,relation2,relation3,Tokuda:2017fdh,Tokuda:2018eqs}.

The $O(N)$-symmetric model has been considered before in the large $N$ limit \cite{LargeN1,LargeN2,LargeN3,LargeN4,euclideo-1,euclideo-2,Gautier1,Gautier2}. In particular,  in \cite{euclideo-1,euclideo-2} we have presented a  systematic nonperturbative resummation scheme  in the case the number of fields $N$ is sufficiently large  so that it is possible to  assume a double expansion in $1/N$ and $\sqrt\lambda$.  It is based on the theory on the sphere.  It consists on a reorganization of the perturbative expansion and on an analytic continuation to the dS. The main advantages are: it improves the IR behavior of correlation functions in dS; it can be systematically improved, and  the renormalization process is well understood.
%; it is  technically simpler than other known alternative methods, such as those  based on finding approximate solutions to the  Schwinger-Dyson equations \cite{Gautier}.
  
In this paper we consider  the same $O(N)$-symmetric model as in  \cite{euclideo-1,euclideo-2}.  We  start by summarizing (in Sec.~\ref{EdS}) the main relevant properties of QFT in dS spacetime, and its analytical continuation to the sphere. In Sec.~\ref{SimplerRess} we present a simpler derivation of the reorganized perturbative  (in $1/N$ and $\lambda^{1/2}$) expansion, and we  provide diagramatic rules (which we call  resummed Feynman rules) to systematically  account for corrections.  
In Sec.~\ref{sec-all-N} we analyze  the  long wavelength  limit  of the correlators in the large-$N$ limit up to next-to-leading  order in  $1/N$.
We show that, when analytically continued to dS spacetime, an additional nonperturbative resummation is necessary to take into account all relevant 
diagrams at long distances. After this resummation, the correlators tend to zero as $ r(x,x') \to \infty$. The next to leading IR contribution, which determines
the decay law as $r(x,x')\to \infty$, is evaluated using the resummation of Sec.~\ref{SimplerRess} for all values of $N$. 
 We present our conclusions in Sec.~\ref{Conc}.  
 The Appendices provide additional details of the calculations and the  extension of the resummation scheme to  negative-squared-mass fields.

\section{QFT in  Euclidean de Sitter space}\label{EdS}

The line element of Euclidean de Sitter space can be obtained from the one of dS spacetime in the so-called global coordinates,
\begin{equation}
 ds^2 =-  dt^2 + \frac{1}{H^2} \cosh^2(H t) d\Omega^2 ,
\end{equation}
   by the analytical continuation to imaginary time $t \to i ( \tau  - \pi/2H )$. The periodicity condition $\tau = \tau + 2\pi H^{-1}$ must be imposed in order to avoid a multivalued metric. This leads to the metric of a $d$-sphere of radius $H^{-1}$
\begin{equation}
 ds^2 = H^{-2} \left[ d \theta^2 + \sin(\theta)^2 d\Omega^2 \right], \label{d-sphere}
\end{equation}
with $\theta = H\tau$.

We consider a $O(N)$ scalar model with quartic self-interaction living in this metric with Euclidean action given by
\begin{equation}
 S = \int d^d x \sqrt{g} \left[ \frac{1}{2} \phi_a \left( -\square + m^2 \right) \phi_a + \frac{\lambda}{8N} ( \phi_a \phi_a )^2 \right], \label{Eaction}
\end{equation}
where $\phi_a$ are the components of an element of the adjoint representation of the $O(N)$ group, with $a=1,..,N$. The sum over repeated indices is implied. 
% Also, a possible nonminimal coupling with the curvature $\xi$ is included in the mass parameter $m^2 = \tilde{m}^2 + \xi d(d-1) H^2$. 
% 
% In $d$ dimensions the field coupling constant has units of $H^{4-d}$, thus can be expressed as $\lambda = \mu^{4-d} \lambda_4 $, with $\lambda_4$ a dimensionless constant and $\mu$ a scale with mass dimensions.

% Euclidean de Sitter
%  space is obtained from Lorentzian de Sitter
%  space in global coordinates by performing an analytical continuation $t \to -i ( \tau - \pi/2H )$ and a compactification in imaginary time $\tau = \tau + 2\pi H^{-1}$. The resulting metric is that of a $d$-sphere of radius $H^{-1}$
% \begin{equation}
%  ds^2 = H^{-2} \left[ d \theta^2 + \sin(\theta)^2 d\Omega^2 \right],
% \end{equation}
% where $\theta = H\tau$. Due to the symmetries and compactness of this space, the field can be expanded in $d$-dimensional spherical harmonics
Following the properties of the metric in Eq.~\eqref{d-sphere}, the field can be expanded in a discrete set of modes
\begin{equation}
 \phi_a(x) = \sum_{\vec{L}} \phi_{\vec{L},a} Y_{\vec{L}}(x). 
\end{equation}
It can be easily seen that the free propagator in the symmetric phase is then expressed in the following way
\begin{eqnarray}
 G^{(m)}_{ab}(r) &=& = \delta_{ab} H^d \sum_{\vec{L}} \frac{Y_{\vec{L}}(x) Y^*_{\vec{L}}(x')}{H^2 L(L+d-1) + m^2} = \delta_{ab} \left( \frac{1}{V_d m^2} + \hat{G}^{(m)}(r) \right), \label{free-prop}
\end{eqnarray}
where the superscript $m$ stands for the mass,
\begin{equation}
V_d = \frac{1}{|Y_{\vec{0}}|^2 H^d} = \int d^dx \sqrt{g} = \frac{2\pi^{\frac{d+1}{2}}}{\Gamma\left(\frac{d+1}{2}\right)H^d}
\end{equation} is the surface area of the $d$-sphere, and $r$ is the dS invariant distance,
\begin{equation}
 r(x,x') = 2\left( 1 - \cos(\theta) \cos(\theta') - \sin(\theta) \sin(\theta') \vec{w} \cdot \vec{w}' \right),
\end{equation}
with $\vec{w}$ and $\vec{w}'$ unit vectors on the $d-1$-sphere. In the last equality of Eq.~\eqref{free-prop} we explicitly separated the zero-mode ($\vec{L}=\vec{0}$) part of the propagator, which   has an IR divergence for $m \to 0$, thereby defining $\hat{G}^{(m)}(r)$ as the inhomogeneous (i.e. $L>0$) part, which is instead finite in that limit. 

\subsection{Analytical continuation} \label{sec:analytic-cont}

After analytic continuation back to dS, the dS invariant distance can be immediately computed in the standard cosmological patch. Using conformal coordinates,
\begin{equation}
ds^2=\frac{1}{H^2\eta^2}\left[-d\eta^2+\delta_{ij}dx^idx^j\right],
\end{equation} we obtain
\begin{equation}
r(x,x')= \frac{-(\eta - \eta')^2 + (\vec{x} - \vec{x}')^2}{\eta \eta'}. \end{equation}
There are no ambiguities with this procedure for invariant functions depending only on two points $x$ and $x'$ through $r(x,x')$.

For massive fields, the dS free propagator decays at large distances as
\begin{equation}
 G^{(m)}(r) \simeq \frac{1}{V_d m^2} r^{-\frac{m^2}{d-1}}.
  \label{late-times-massive}
\end{equation}
Moreover, it was also shown \cite{Marolf:2010zp,Marolf:2010nz,Hollands:2010pr} that loop corrections to the two-point function of a massive field also enjoy such decay at large distances. In this way, the standard perturbative expansion is well defined for massive fields in dS, by analytically continuation from the sphere.

In the case of massless fields the story is different. Firstly, we must work with the modified propagator $\hat{G}^{(0)}(r)$, as defined in Eq.~\eqref{free-prop} by substracting the (divergent) zero-mode contribution to the standard propagator. Then, although not divergent for $m \to 0$, it still suffers from IR effects once taken to dS in the form of a secular behavior at long distances. Indeed, using Eq.~\eqref{late-times-massive},
% \begin{equation}
%  \hat{G}^{(m)}(y) \simeq \frac{1}{V_d m^2} \left( y^{-\frac{m^2}{d-1}} - 1 \right), \label{late-times-massive}
% \end{equation}
\begin{equation}
 \hat{G}^{(0)}(r) \simeq \lim_{m \to 0} \frac{1}{V_d m^2} \left( r^{-\frac{m^2}{d-1}} - 1 \right) = - \frac{\log(r)}{V_d (d-1)}. \label{late-times-massless}
\end{equation}
This secular behavior slips into each loop correction, worsening as the number of loops increases \cite{Burgess}. Then, unlike the massive case, the standard perturbative expansion breaks down at large distances/late times in dS. This demands a nonperturbative treatment that not only gives a mass to the zero modes, but also to the inhomogeneous modes.

\section{Reorganizing the perturbative expansion}\label{SimplerRess}

In the context of the QFT on the sphere, let us first consider the approach introduced in Ref.~\cite{BLHU, Rajaraman1}, and later extended in Ref.~\cite{BenekeMoch}, which addresses the divergence for $m \to 0$ in the full propagator $G$ by means of a reformulation of the theory in terms of $\hat{G}$. The main point of the approach is to split the field as $\phi_a(x) =  \phi_{0a} + \hat{\phi}_a(x)$, and treat the constant zero modes $\phi_{0a}$ nonperturbatively, separately from the inhomogeneous parts $\hat{\phi}_a(x)$. 
% This prompts to separate the propagator as well,
% \begin{equation}
%  G^{(m)}(x,x') = G_0^{(m)} + \hat{G}^{(m)}(x,x'),
% \end{equation}
% where now $\hat{G}^{(m)}$ has the property of being finite in the IR ($m \to 0$).  
The interaction part of the action \eqref{Eaction} is separated as follows:
\begin{equation}
 S_{int} = \frac{\lambda V_d}{8N}   |\vec{\phi}_0|^4 + S^{(2)}_{int}[\phi_{0a},\hat{\phi}_a], \label{Sint2}
\end{equation}
where     $S^{(2)}_{int}$ contains at least two powers of $\hat{\phi}_a$ (note that the term linear in $\hat{\phi}_a$ vanishes identically by orthogonality).
% and $V_d$ is the total volume of Euclidean de Sitter
%  space in $d$-dimensions, which thanks to the  compactification is finite and equal to the hypersurface area of a $d$-sphere
% \begin{equation}
%  V_d = \int d^d x \sqrt{g} = \frac{2 \pi^{\frac{d+1}{2}}}{\Gamma\left( \frac{d+1}{2} \right) H^d} = \frac{1}{|Y_{\vec{0}}|^2 H^d},
% \end{equation}
% where  $\Gamma$ is Euler's Gamma function.
% The explicit form of $ \tilde{S}_{int}$ will be written below.
Then, since the zero modes are constant, the path integral over them turns into an ordinary integral, which can be performed exactly (i.e. nonperturbatively in the coupling constant $\lambda$). The generating functional becomes
\begin{eqnarray}
   Z[J_0, \hat{J}] &=& \mathcal{N} \int d^N \hspace{-0.13cm} \phi_0 \int \mathcal{D}\hat{\phi} \, e^{-S - \int_x  (\vec{J}_{0} \cdot \vec{\phi}_{0} + \hat{J}_a \hat{\phi}_a )} \notag \\
   &=& exp\left( -S^{(2)}_{int}\left[\frac{\delta}{\delta J_0},\frac{\delta}{\delta \hat{J}}\right] \right) Z_0[J_0] \hat{Z}_{f}[\hat{J}],
   \label{Z-int}
\end{eqnarray}
where $J_{0a}$ and $\hat{J}_a$ are external sources and we introduced the shorthand notation $\int_x = \int d^d x \sqrt{g}$. The zero part $Z_0[J_0]$ is defined as the exact generating functional of the theory with the zero modes alone,
\begin{equation}
 Z_0[J_0] = \frac{\int d^N\hspace{-0.12cm}\phi_0 \, e^{-V_d \left[\frac{\lambda}{8N}  |\vec{\phi}_0|^4 + \frac{m^2}{2}  |\vec{\phi}_0|^2 + \vec{J}_{0} \cdot \vec{\phi}_{0} \right]}}{\int d^N\hspace{-0.10cm}\phi_0 \, e^{-V_d \left[ \frac{\lambda}{8N}  |\vec{\phi}_0|^4 + \frac{m^2}{2}  |\vec{\phi}_0|^2 \right]}}.
\end{equation}

The simplest example of application is to compute the variance of the zero modes. For a massless field $m=0$, the well known finite result is obtained,
\begin{equation}
 \langle \phi_{0a} \phi_{0b} \rangle_0 = \frac{\int d^N\hspace{-0.12cm}\phi_0 \, \phi_{0a} \phi_{0b} \, e^{- \frac{V_d \lambda}{8N}  |\vec{\phi}_0|^4 }}{\int d^N\hspace{-0.10cm}\phi_0 \, e^{- \frac{V_d \lambda}{8N}  |\vec{\phi}_0|^4 }} = \delta_{ab} \sqrt{\frac{2}{V_d \lambda}} \frac{2}{\sqrt{N}} \frac{\Gamma\left( \frac{N+2}{4} \right)}{\Gamma\left( \frac{N}{4} \right)} \equiv \frac{ \delta_{ab}}{V_d m_{\rm dyn}^2},
\end{equation}
which allows the identification of a dynamical mass $m_{\rm dyn}^2$ by analogy with the free massive field.
% dynamical mass, related to
% \begin{equation}
%  m_{\rm dyn}^2 = \frac{N}{V_d \langle \phi_0^2 \rangle} = \sqrt{\frac{N \lambda}{2V_d}} \frac{1}{2} \frac{\Gamma\left[ \frac{N}{4} \right]}{\Gamma\left[ \frac{N+2}{4} \right]}. \label{mdyn0-allN}
% \end{equation}
This result is valid at LO in $\sqrt{\lambda}$ and for all $N$. Corrections coming from the inhomogeneous modes can then be computed by treating $S^{(2)}_{int}$ in Eq.~\eqref{Z-int} perturbatively. 

Although now the correlator is finite for $m \to 0$, the perturbation theory built with a massless $\hat{G}^{(0)}$ is still ill defined when analytically continued to dS and at long distances/late times, due to the divergent behavior described  in Eq.~\eqref{late-times-massless}. Solving this issue requires further resummations of contributions that also involve the inhomogeneous modes.
 A subclass of such contributions come from the  terms in $S^{(2)}_{int}$ that are quadratic in both $\hat{\phi}_a$ and $\phi_{0a}$, which dress $\hat{G}$ with a nonperturbative mass. 
 This treatment was introduced by us in Ref.~\cite{euclideo-1} for the $O(N)$-model (while in Ref.~\cite{Hollands2012} it was considered for the special case $N=1$).
%  This treatment was originally introduced by us in Ref.~\cite{euclideo-1} 
 Here we will present a simpler formulation, which has the added benefit of being more easily generalizable to other cases, such as for negative squared-mass. 
% in Section~\ref{sec-neg}.

\subsection{Resummation of bi-quadratic terms}
In the spirit of the separation of the interaction part of the action done in Eq.~\eqref{Sint2}, we further isolate the bi-quadratic terms,
\begin{equation}
 S^{(2)}_{int} = \frac{\lambda}{8N} \int d^d x \sqrt{g} \left[ 2 |\vec{\phi}_0|^2 |\hat{\phi}|^2 + 4 (\vec{\phi}_{0} \cdot \vec{\hat{\phi}})^2 \right] + S^{(3)}_{int}, 
\end{equation}
where now $S^{(3)}_{int}$ contains terms with at least three powers of $\hat{\phi}_a$.
The main idea is then to include the bi-quadratic terms in the definition of the propagator $\hat{G}$.  The generating functional becomes
\begin{eqnarray}
Z[\vec{J}_0, \hat{J}] &=& \mathcal{N} \exp\left(-S^{(3)}_{int}\left[\frac{\delta}{\delta J_0},\frac{\delta}{\delta \hat{J}}\right]\right) \int d^N \hspace{-0.13cm} \phi_0 \, e^{-V_d \left[ \frac{m^2}{2} |\vec{\phi}_0|^2 + \frac{\lambda}{8N} |\vec{\phi}_0|^4 + \vec{J}_{0} \cdot \vec{\phi}_{0} \right]} \notag \\
&&\times \int \mathcal{D}\hat{\phi} \, \exp\left( -\frac{1}{2} \iint_{x,y} \hat{\phi}_a  \hat{G}^{-1}_{ab}(\vec{\phi}_0) \hat{\phi}_b  + \int_x \hat{J}_a \hat{\phi}_a \right), \label{Z-resum} 
% \notag \\
% &=& \mathcal{N} e^{-S^{(3)}_{int}\left[\frac{\delta}{\delta J_0},\frac{\delta}{\delta \hat{J}}\right]} \left( \hat{Z}_{f}\left[\hat{J}, m^2 \right] \sqrt{ \det \hat{G}_{rs}^{(m)} } \right)_{m(\delta/\delta J_0)} Z_0[J_0],
\end{eqnarray}
where the $\vec{\phi}_{0}$-dependent inverse propagator of the $\hat{\phi}_a$ fields is given by
\begin{equation}
 \hat{G}^{-1}_{ab}(\vec{\phi}_0) = -\delta_{ab} \square + m_{ab}^2(\vec{\phi}_0)\,,
\end{equation}
with the following mass matrix
\begin{eqnarray}
 m_{ab}^2(\vec{\phi}_{0}) &=& m^2 \delta_{ab} + \frac{\lambda}{2N} \left( \delta_{ab} \delta_{cd} + \delta_{ac} \delta_{bd} + \delta_{ad} \delta_{bc} \right) \phi_{0c} \phi_{0d} \notag \\
 &=& m_1^2 \, P_{ab} + m_2^2 \, \epsilon_a \epsilon_b,
\end{eqnarray}
where in the second line we have split the matrix into the parallel and transverse components with respect to the $\epsilon_a \equiv \phi_{0a}/|\vec{\phi}_0|$ direction, by means of the projector $P_{ab} = \delta_{ab} - \epsilon_a \epsilon_b$, and we defined
\begin{eqnarray}
 m_1^2 &= m^2 + \frac{\lambda}{2N} |\vec{\phi}_0|^2, \\
 m_2^2 &= m^2 + \frac{3\lambda}{2N} |\vec{\phi}_0|^2. 
\end{eqnarray}
This tells us there are $(N-1)$ inhomogeneous fields with mass $m_1^2$ and a single one with mass $m_2^2$. Diagonalizing the mass matrix $m_{ab}^2$, we can factorize the free $\hat{\phi}_a$ part of the path integral (last factor of Eq.~\eqref{Z-resum})
\begin{eqnarray}
\int \mathcal{D}\hat{\phi}^{(1)}_i \, e^{ -\frac{1}{2} \iint_{x,y} \hat{\phi}^{(1)}_i  (\hat{G}_1)^{-1} \hat{\phi}^{(1)}_i  + \int_x \hat{J}^{(1)}_i \hat{\phi}^{(1)}_i } \times \int \mathcal{D}\hat{\phi}^{(2)} \, e^{ -\frac{1}{2} \iint_{x,y} \hat{\phi}^{(2)} (\hat{G}_2)^{-1} \hat{\phi}^{(2)} + \int_x \hat{J}^{(2)} \hat{\phi}^{(2)} } \notag \\
\sim \sqrt{ \det \hat{G}_1}^{N-1} \hat{Z}_{f}\left[\hat{J}^{(1)}_i, m_1^2 \right] \sqrt{ \det \hat{G}_2} \, \hat{Z}_{f}\left[\hat{J}^{(2)}, m_2^2 \right],  
\end{eqnarray}
where $\hat{\phi}_i^{(1)} \equiv P_{ib} \hat{\phi}_b$ and $\hat{\phi}^{(2)} \equiv \epsilon_a \hat{\phi}_a$ are the transverse and parallel components with respect to the $\vec{\phi}_{0}$-direction respectively, and now the indexes $i,j$ run from $1$ to $N-1$. We also use the shorthand $\hat{G}_\alpha \equiv \hat{G}^{(m_\alpha)}$, with $\alpha = 1,2$. 
% This factorization means there is no mixing terms between the $\hat{\phi}^{(1)}_i$'s and $\hat{\phi}^{(2)}$ at this level.
% , and therefore any mixed two-point correlator $\langle \hat{\phi}^{(1)}_i(x) \hat{\phi}^{(2)}(x') \rangle$ vanishes. This stands true when including the other interactions. 
Putting this back into Eq.~\eqref{Z-resum} we obtain
\begin{equation}
 Z[\vec{J}_0, \hat{J}_i^{(1)}, \hat{J}^{(2)}] = \exp\left(-S^{(3)}_{int}\left[\frac{\delta}{\delta J_0},\frac{\delta}{\delta \hat{J}}\right]\right) \mathcal{Z}[\vec{J}_0, \hat{J}_i^{(1)}, \hat{J}^{(2)}],
\end{equation}
% and ignoring for now the perturbative corrections coming from $S^{(3)}_{int}$,
%$\sim \lambda \phi_0 \hat{\varphi}^3$ and $\sim \lambda \hat{\varphi}^4$, since they are higher order in $\lambda$),
where we are introducing a new ``free'' generating functional, 
\begin{eqnarray}
\mathcal{Z}[\vec{J}_0, \hat{J}^{(1)}_i, \hat{J}^{(2)}] &=& \mathcal{N} \Biggl\langle \sqrt{ \det \hat{G}_1}^{N-1} \sqrt{ \det \hat{G}_2} \, \hat{Z}_1[\hat{J}^{(1)}_i] \,  \hat{Z}_2[\hat{J}^{(2)}] \Biggr\rangle_{0}^{\vec{J}_0}, \notag \\
&\equiv& \Bigl\langle \hat{Z}_1[\hat{J}^{(1)}_i] \,  \hat{Z}_2[\hat{J}^{(2)}] \Bigr\rangle_{\bar{0}}^{\vec{J}_0}
\label{Z-LO}
\end{eqnarray}
where $\hat{Z}_{\alpha} \equiv \hat{Z}_{f}\left[\hat{J}^{(\alpha)}, m_\alpha^2 \right]$ is the free generating functional of a single inhomogeneous field of mass $m_\alpha$, normalized to $\hat{Z}_\alpha[0]=1$, and we have defined the notation 
\begin{equation}
 \langle \dots \rangle_{\bar{0}}^{\vec{J}_0} \equiv \frac{\Biggl\langle \sqrt{ \det \hat{G_1}}^{N-1} \sqrt{ \det \hat{G}_2} \dots \Biggr\rangle_{0}^{\vec{J}_0}}{\Biggl\langle \sqrt{ \det \hat{G}_1}^{N-1} \sqrt{ \det \hat{G}_2} \Biggr\rangle_{0}}, \label{zero-mode-exp-value}
\end{equation}
choosing the normalization $\mathcal{N}$ such that $\mathcal Z[0] = 1$. The superindex $\vec{J}_0$ indicates that the $\bar{0}$-expectation value is taken over the zero modes in the presence of an external source $\vec{J}_0$. 

Notice that the functional derivatives with respect to the full $\hat{J}_a$ can be split in terms of derivatives with respect to $\hat{J}^{(1)}_i$ and $\hat{J}^{(2)}$,
\begin{eqnarray}
 \frac{\delta^2}{\delta \hat{J}_a(x) \delta \hat{J}_b(x')} = \frac{P_{ab}}{(N-1)} \delta_{ij} \frac{\delta^2}{\delta \hat{J}^{(1)}_i(x) \delta \hat{J}^{(1)}_j(x')} + \epsilon_a \epsilon_b \frac{\delta^2}{\delta \hat{J}^{(2)}(x) \delta \hat{J}^{(2)}(x')}, \label{derivs-J-splitted}
\end{eqnarray}
where we are taking advantage of the already manifest $O(N-1)$-symmetry of the fields of mass $m_1$, in the plane orthogonal to $\vec{\phi}_0$. Afterwards, when applying these derivatives either to $\hat{Z}_1$ or $\hat{Z}_2$, we will drop the superindices $(1)$ and $(2)$ as there will be no ambiguity on which $\hat{J}$ is which.

Later on, when setting $\vec{J}_0 = 0$, the expectation value taken over the zero modes has the effect of averaging over the direction of $\vec{\phi}_0$, giving rise to manifestly $O(N)$-symmetric expressions for the correlators.

\subsection{Resummed Feynman rules}

After having performed the above resummation  of the bi-quadratic interactions,  we now have a new ``free'' generating functional Eq.~\eqref{Z-LO} where the propagators $\hat{G}_\alpha$ are massive, with  masses that depend on $|\vec{\phi_0}|$.  If the remaining interaction terms in $S^{(3)}_{int}$  are  treated perturbatively,   the new type of perturbative corrections  can be computed using new Feynman rules, which we call Resummed Feynman rules, or R-Feynman rules for short. The R-Feynman  diagrams are built with  the dressed propagators $\hat{G}_1$ and $\hat{G}_2$ as internal lines. The last step is to evaluate the weighted average over the zero modes. The  derivation of the new rules is straightforward, but with an important new ingredient that we will discuss below. First, let us write the interactions explicitly (according to the decomposition in the fields $\hat{\phi}_1$ and $\hat{\phi}_2$),
\begin{eqnarray}
 S^{(3)}_{int} &=& \int d^d x \sqrt{g} \left[ \frac{\lambda}{2N} (\vec{\phi}_0 \cdot\vec{ \hat{\phi}}) |\vec{\hat{\phi}}|^2 + \frac{\lambda}{8N} |\vec{\hat{\phi}}|^4 \right]
 \notag \\
 &=& \int d^d x \sqrt{g} \left[ \frac{\lambda |\vec{\phi}_0|}{2N} \hat{\phi}_2 \left(|\vec{\hat{\phi}}_1|^2 + \hat{\phi}_2^2 \right) + \frac{\lambda}{8N} \left( |\vec{\hat{\phi}}_1|^4 + 2 |\vec{\hat{\phi}}_1|^2 \hat{\phi}_2^2 + \hat{\phi}_2^4 \right) \right].
 \label{Sint3}
\end{eqnarray}
In what concerns the Feynman rules, the $|\vec{\phi_0}|$ factor in front of the first two terms shall not be regarded as a ``leg'' of those vertices, but rather just as a coefficient\footnote{ In the terminology of the standard perturbative expansion,  the integration over the zero modes in $\langle \dots \rangle_{\bar{0}}$ takes care of summing over all possible ways of connecting the legs associated to $\phi_{0a}$,  both the ones that here appear explicitly in the vertices in Fig.~\ref{fig:vertices}  as well as those that are implicit in the masses of the resummed propagators $m_1^2(\vec{\phi_0})$ and $m_2^2(\vec{\phi_0})$.}. The rules are summarized in Figs. \ref{fig:props-hat} and \ref{fig:vertices}.
\begin{figure}[h!]
 \centering
\begin{tikzpicture}
  \begin{feynman}
    \vertex (a) {\(x,i\)};
    \vertex [right=1.2cm of a](b);
    \vertex [right=1.2cm of b] (c) {\(x',j\)};
    \vertex [below=0.5cm of b](d) {\(\quad\,\, \hat{G}_1(x,x') \delta_{ij}\)};
        
    \diagram* {
      (a) -- [very thick,solid] (b) -- [very thick,solid] (c),
    };
  \end{feynman}
\end{tikzpicture}
\hspace{1cm}
\begin{tikzpicture}
  \begin{feynman}
    \vertex (a) {\(x\)};
    \vertex [right=1.2cm of a](b);
    \vertex [right=1.2cm of b] (c) {\(x'\)};
    \vertex [below=0.5cm of b](d) {\(\quad \hat{G}_2(x,x')\)};
        
    \diagram* {
      (a) -- [very thick,dash dot] (c),
    };
  \end{feynman}
\end{tikzpicture}
 \caption{Massive inhomogeneous propagators $\hat{G}_1$ and $\hat{G}_2$ are used in the internal lines.}
 \label{fig:props-hat}
\end{figure}
% \begin{figure}[h!]
%  \centering
%  \includegraphics{Images/props-hat.png}
%  % props-hat.png: 552x85 px, 300dpi, 4.67x0.72 cm, bb=0 0 132 20
%  \caption{Propagators $\hat{G}_1$ and $\hat{G}_2$.}
%  \label{fig:props-hat}
% \end{figure}
\begin{figure}[h!]
 \centering
\begin{tikzpicture}
  \begin{feynman}
    \vertex (c);
    \vertex at ($(c) + (-0.848cm, 0.848cm)$) (v1);
    \vertex at ($(c) + (0.848cm, 0.848cm)$) (v2);
    \vertex at ($(c) + (-0.848cm, -0.848cm)$) (v3);
    \vertex at ($(c) + (0.848cm, -0.848cm)$) (v4);

    \vertex at ($(v1) + (-0.2cm, 0)$) (l1) {\(i\)};
    \vertex at ($(v2) + (0.2cm, 0)$) (l2) {\(j\)};
    \vertex at ($(v3) + (-0.2cm, 0)$) (l3) {\(k\)};
    \vertex at ($(v4) + (0.2cm, 0)$) (l4) {\(l\)};
      
    \vertex at ($(c) + (0, -1.5cm)$) (l) {\(-\frac{\lambda}{8N} \frac{A_{ijkl}}{3}\)};  
      
    \diagram* {
      (v1) -- [very thick,solid] (v4),
      (v2) -- [very thick,solid] (v3),
    };
  \end{feynman}
\end{tikzpicture}
\hspace{0.4cm}
\begin{tikzpicture}
  \begin{feynman}
    \vertex (c);
    \vertex at ($(c) + (-0.848cm, 0.848cm)$) (v1);
    \vertex at ($(c) + (0.848cm, 0.848cm)$) (v2);
    \vertex at ($(c) + (-0.848cm, -0.848cm)$) (v3);
    \vertex at ($(c) + (0.848cm, -0.848cm)$) (v4);

    \vertex at ($(v1) + (-0.2cm, 0)$) (l1) {\(i\)};
    \vertex at ($(v2) + (0.2cm, 0)$) (l2) {\(j\)};
          
    \vertex at ($(c) + (0, -1.5cm)$) (l) {\(-\frac{\lambda}{4N} \delta_{ij}\)};  
      
    \diagram* {
      (c) -- [very thick,solid] (v1),
      (c) -- [very thick,solid] (v2),
      (c) -- [very thick,dash dot] (v3),
      (c) -- [very thick,dash dot] (v4),
    };
  \end{feynman}
\end{tikzpicture}
\hspace{0.4cm}
\begin{tikzpicture}
  \begin{feynman}
    \vertex (c);
    \vertex at ($(c) + (-0.848cm, 0.848cm)$) (v1);
    \vertex at ($(c) + (0.848cm, 0.848cm)$) (v2);
    \vertex at ($(c) + (-0.848cm, -0.848cm)$) (v3);
    \vertex at ($(c) + (0.848cm, -0.848cm)$) (v4);

    \vertex at ($(c) + (0, -1.5cm)$) (l) {\(-\frac{\lambda}{8N}\)};  
      
    \diagram* {
      (v1) -- [very thick,dash dot] (v4),
      (v2) -- [very thick,dash dot] (v3),
    };
  \end{feynman}
\end{tikzpicture}
\vspace{0.5cm}
\\
\begin{tikzpicture}
  \begin{feynman}
    \vertex (c);
    \vertex at ($(c) + (0, 1.2cm)$) (v1);
    \vertex at ($(c) + (-1.039cm, -0.6cm)$) (v2);
    \vertex at ($(c) + (1.039cm, -0.6cm)$) (v3);
        
    \vertex at ($(v2) + (-0.2cm, 0)$) (l1) {\(i\)};
    \vertex at ($(v3) + (0.2cm, 0)$) (l2) {\(j\)};
      
    \vertex at ($(c) + (0, -1.5cm)$) (l) {\(-\frac{\lambda |\vec{\phi}_0|}{2N} \delta_{ij}\)};  
      
    \diagram* {
      (c) -- [very thick,dash dot] (v1),
      (c) -- [very thick,solid] (v2),
      (c) -- [very thick,solid] (v3),
    };
  \end{feynman}
\end{tikzpicture}
\hspace{0.4cm}
\begin{tikzpicture}
  \begin{feynman}
    \vertex (c);
    \vertex at ($(c) + (0, 1.2cm)$) (v1);
    \vertex at ($(c) + (-1.039cm, -0.6cm)$) (v2);
    \vertex at ($(c) + (1.039cm, -0.6cm)$) (v3);
      
    \vertex at ($(c) + (0, -1.5cm)$) (l) {\(-\frac{\lambda |\vec{\phi}_0|}{2N}\)};  
      
    \diagram* {
      (c) -- [very thick,dash dot] (v1),
      (c) -- [very thick,dash dot] (v2),
      (c) -- [very thick,dash dot] (v3),
    };
  \end{feynman}
\end{tikzpicture}
 \caption{Vertices. The factor $A_{ijkl}/3=\left( \delta_{ij} \delta_{kl} + \delta_{ik} \delta_{jl} + \delta_{il} \delta_{jk} \right)/3$ in the first vertex accounts for the possible permutations of the indices.} 
 \label{fig:vertices}
\end{figure}
% \begin{figure}[h!]
%  \centering
%  \includegraphics{Images/vertices.png}
%  % vertices.png: 831x639 px, 300dpi, 7.04x5.41 cm, bb=0 0 199 153
%  \caption{Vertices. The factor $A_{ijkl}/3$ in the first vertex is a totally symmetric tensor that accounts for the possible permutations of the indices.} 
%  \label{fig:vertices}
% \end{figure} 

There is an important difference with traditional Feynman rules to consider here. As a consequence of the definition of the new ``free'' generating functional in Eq.~\eqref{Z-LO}
as a weighted average over the zero modes, there is no direct cancellation of disconnected graphs when computing perturbative corrections. Indeed, consider a correction $\Delta \hat{Z}$ to the generating functional of the inhomogeneous modes, which at LO is just $\hat{Z}_1 \hat{Z}_2$, prior to the zero mode average. The corrected complete\footnote{It is complete in the sense that  both zero and inhomogeneous modes are taken into account.} generating functional $\mathcal{Z}'$ now reads
\begin{equation}
 \mathcal{Z}'[\vec{J}_0, \hat{J}^{(1)}_i, \hat{J}^{(2)}] = \Bigl\langle \hat{Z}_1[\hat{J}^{(1)}_i] \,  \hat{Z}_2[\hat{J}^{(2)}] + \Delta \hat{Z}[\hat{J}^{(1)}_i,\hat{J}^{(2)}] \Bigr\rangle_{\bar{0}}^{\vec{J}_0}, 
\label{Z-NLO}
\end{equation}
which corrects Eq.~\eqref{Z-LO}. Now consider a corrected n-point function of inhomogeneous fields computed from the previous  expression, 
\begin{eqnarray}
\frac{1}{\mathcal{Z}'} \frac{\delta^n \mathcal{Z}'}{\delta \hat{J}_{a_1}(x_1) \dots \delta \hat{J}_{a_n}(x_n)} \Bigg|_{J=0} &=& \Biggl\langle \frac{\delta^n (\hat{Z}_1 \hat{Z}_2)}{\delta \hat{J}_{a_1}(x_1) \dots \delta \hat{J}_{a_n}(x_n)} \Bigg|_{\hat{J}=0} \Biggr\rangle_{\bar{0}} \notag \\
&&+ \Biggl\langle \frac{\delta^n \Delta \hat{Z}}{\delta \hat{J}_{a_1}(x_1) \dots \delta \hat{J}_{a_n}(x_n)} \Bigg|_{\hat{J}=0} \Biggr\rangle_{\bar{0}} \notag \\
&&- \Bigl\langle \Delta \hat{Z} \Bigr\rangle_{\bar{0}} \Biggl\langle \frac{\delta^n (\hat{Z}_1 \hat{Z}_2)}{\delta \hat{J}_{a_1}(x_1) \dots \delta \hat{J}_{a_n}(x_n)} \Bigg|_{J=0} \Biggr\rangle_{\bar{0}}\,,  \label{n-pt-NLO}
\end{eqnarray}
where we used the normalization $\hat{Z}_\alpha[0]=1$, with $\alpha=1,2$, and we also treated the correction $\Delta \hat{Z}$ perturbatively. The first term in the right-hand side is the leading contribution obtained from Eq.~\eqref{Z-LO}, while the second and third terms are the corrections. In the usual case, the second term contains both connected and disconnected contributions, the latter of which are cancelled by the third term. However, in the current situation this does not occur due to the weighting over the zero modes. Indeed, 
\begin{equation}
 \Biggl\langle \Delta \hat{Z} \frac{\delta^n (\hat{Z}_1 \hat{Z}_2)}{\delta \hat{J}_{a_1}(x_1) \dots \delta \hat{J}_{a_n}(x_n)} \Bigg|_{J=0} \Biggr\rangle_{\bar{0}} \neq \Bigl\langle \Delta \hat{Z} \Bigr\rangle_{\bar{0}} \Biggl\langle \frac{\delta^n (\hat{Z}_1 \hat{Z}_2)}{\delta \hat{J}_{a_1}(x_1) \dots \delta \hat{J}_{a_n}(x_n)} \Bigg|_{J=0} \Biggr\rangle_{\bar{0}}. 
\end{equation}
By adding and subtracting the left-hand side of the above equation to Eq.~\eqref{n-pt-NLO}, we can identify two contributions to the correction of the n-point function as follows: a connected part
\begin{equation}
 \Delta \langle \hat{\phi}_{a_1}(x_1) \dots \hat{\phi}_{a_n}(x_n) \rangle_{C} = \Biggl\langle \frac{\delta^n \Delta \hat{Z}}{\delta \hat{J}_{a_1}(x_1) \dots \delta \hat{J}_{a_n}(x_n)} \Bigg|_{\hat{J}=0} - \Delta \hat{Z} \frac{\delta^n (\hat{Z}_1 \hat{Z}_2)}{\delta \hat{J}_{a_1}(x_1) \dots \delta \hat{J}_{a_n}(x_n)} \Bigg|_{J=0} \Biggr\rangle_{\bar{0}}\,, \label{n-pt-C}
\end{equation}
which is built in the standard way with the connected R-Feynman diagrams using the above rules; and a 0-connected part
\begin{eqnarray}
\Delta \langle \hat{\phi}_{a_1}(x_1) \dots \hat{\phi}_{a_n}(x_n) \rangle_{^{0} C} &=& \Biggl\langle \Delta \hat{Z} \frac{\delta^n (\hat{Z}_1 \hat{Z}_2)}{\delta \hat{J}_{a_1}(x_1) \dots \delta \hat{J}_{a_n}(x_n)} \Bigg|_{J=0} \Biggr\rangle_{\bar{0}}\nonumber\\
& &-\Bigl\langle \Delta \hat{Z} \Bigr\rangle_{\bar{0}} \Biggl\langle \frac{\delta^n (\hat{Z}_1 \hat{Z}_2)}{\delta \hat{J}_{a_1}(x_1) \dots \delta \hat{J}_{a_n}(x_n)} \Bigg|_{J=0} \Biggr\rangle_{\bar{0}}, \notag \\ \label{n-pt-0C}
\end{eqnarray}
which accounts for graphs that are disconnected according to the R-Feynman rules (that is, graphs that are not connected by lines associated to the propagators  $\hat{G}_\alpha$), but  when written  in terms of the original perturbation theory  they are actually connected by the lines that  would correspond  to the zero-modes.
% propagators $G_0$.  
% We refer to these as $0$-connected diagrams. 
This means that in general, one has to include both contributions \eqref{n-pt-C} and \eqref{n-pt-0C} when computing corrections to the correlators using this formalism. However, as we show in Appendix \ref{app-laplace},  the 0-connected parts \eqref{n-pt-0C} are suppressed by extra powers of $\lambda$ with respect to the connected parts \eqref{n-pt-C}.   This will become clearer next, after considering an example in which we compute the 0-connected parts and verify  they are of higher order in $\lambda$.
 For this reason,  we will not need to compute the 0-connected parts  for the computations we will consider here.

\subsection{Resummed two-point functions}\label{R2PF}

Let us apply the formalism just described to compute the two-point functions. We have to consider  two distinct contributions,
\begin{equation}
 \langle \phi_{a}(x) \phi_{b}(x') \rangle = \langle \phi_{0a} \phi_{0b} \rangle + \langle \hat{\phi}_{a}(x) \hat{\phi}_{b}(x') \rangle. \label{2-pt-f}
\end{equation}
First ignoring corrections from $S^{(3)}_{int}$, we only need  to take derivatives of $\mathcal{Z}[\vec{J}_0, \hat{J}^{(1)}_i, \hat{J}^{(2)}] $ in Eq.~\eqref{Z-LO}. 
For the constant part, i.e. the first term of Eq.~\eqref{2-pt-f}, we take two derivatives with respect to $\vec{J}_{0a}$ and then set all external sources to zero. Exploiting also the $O(N)$-symmetry, it can be expressed as
\begin{equation}
\langle \phi_{0a} \phi_{0b} \rangle = \frac{\delta_{ab}}{N} \langle  |\vec{\phi}_0|^2 \rangle_{\bar{0}}.\label{2pt-phi0-LO}
\end{equation} 
For the inhomogeneous part $\langle \hat{\phi}_a(x) \hat{\phi}_b(x') \rangle$, we use Eq.~\eqref{derivs-J-splitted}  applied  to Eq.~\eqref{Z-LO}, obtaining
% In order to see this, consider the free path integral over the non-zero modes of a pair of $\hat{\phi}$ fields
% \begin{eqnarray}
% \int \mathcal{D}\hat{\phi} \, \hat{\phi}_a(x) \hat{\phi}_b(x') e^{-\frac{1}{2} \iint_{x,y} \hat{\phi}_c  \hat{G}^{-1}_{cd}(\vec{\phi}_0) \hat{\phi}_d } \notag \\
% \sim \sqrt{\det \hat{G}^{(m_1)}}^{N-1} \sqrt{ \det \hat{G}^{(m_2)}} \left[ \hat{G}^{(m_1)}(y) P_{ab} + \hat{G}^{(m_2)}(y) \frac{\phi_{0a} \phi_{0b}}{|\vec{\phi}_0|^2} \right],
% \end{eqnarray}
% \begin{eqnarray}
% \int \mathcal{D}\hat{\phi} \, \hat{\phi}_a(x) \hat{\phi}_b(x') e^{-\frac{1}{2} \iint_{x,y} \hat{\phi}_c  \hat{G}^{-1}_{cd}(\vec{\phi}_0) \hat{\phi}_d } \sim \hat{G}^{(m_1)}(y) P_{ab} + \hat{G}^{(m_2)}(y) \frac{\phi_{0a} \phi_{0b}}{|\vec{\phi}_0|^2},
% \end{eqnarray}
% with $y=y(x,x')$ denoting the invariant dS distance. 
% With this under consideration, the inhomogeneous part of the two-point function reads
\begin{equation}
\langle \hat{\phi}_a(x) \hat{\phi}_b(x') \rangle = \frac{1}{\mathcal{Z}[0]} \frac{\delta^2 \mathcal{Z}}{\delta \hat{J}_a(x) \delta \hat{J}_b(x')}\Bigg|_{J=0} = \Biggl\langle \hat{G}_1(r) P_{ab} + \hat{G}_2 (r) \epsilon_a \epsilon_b \Biggr\rangle_{\bar{0}}, \label{2pt-hatphi-LO}
% \langle \hat{\phi}_a(x) \hat{\phi}_b(x') \rangle = \frac{\Biggl\langle \sqrt{\det \hat{G}^{(m_1)}}^{N-1} \sqrt{ \det \hat{G}^{(m_2)}} \left[ \hat{G}^{(m_1)}(y) P_{ab} + \hat{G}^{(m_2)}(y) \frac{\phi_{0a} \phi_{0b}}{|\vec{\phi}_0|^2} \right] \Biggr\rangle_0}{\Biggl\langle \sqrt{\det \hat{G}^{(m_1)}}^{N-1} \sqrt{ \det \hat{G}^{(m_2)}} \Biggr\rangle_0},
\end{equation}
% Even though there is a splitting in two types of inhomogeneous fields with different masses $m_1^2$ and $m_2^2$ according to the direction of $\vec{\phi_0}$, the subsequent integration over $\vec{\phi}_0$ will average over directions in the field space and lead to a $O(N)$-symmetric result. 
% Indeed, 
where now it is straightforward to perform the integration over the directions of $\vec{\phi}_0$. Indeed, using
\begin{subequations}\label{symmetry-prop}
 \begin{eqnarray}
 \int d\Omega_N \, P_{ab} \, f(|\vec{\phi}_0|) &=& \Omega_N \, \delta_{ab} \left( 1 - \frac{1}{N} \right) \, f(|\vec{\phi}_0|), \\
 \int d\Omega_N \, \epsilon_a \epsilon_b \, f(|\vec{\phi}_0|) &=& \Omega_N \, \frac{\delta_{ab}}{N} \, f(|\vec{\phi}_0|),
\end{eqnarray}
\end{subequations}
we arrive at
\begin{eqnarray}
\langle \hat{\phi}_a(x) \hat{\phi}_b(x') \rangle &=& \frac{\delta_{ab}}{N} \Biggl\langle (N-1) \hat{G}_1(r) + \hat{G}_2(r) \Biggr\rangle_{\bar{0}}, \label{2pt-resummed-implicit} 
\end{eqnarray}
which is now explicitly $O(N)$-symmetric. The remaining step is to compute the zero-mode expectation value. Prior to elaborating on how to do this, let us briefly discuss some perturbative corrections to these expressions. 

% Putting it all together, the full two-point function without any corrections from $S^{(3)}_{int}$ is
% \begin{eqnarray}
% \langle \phi_a(x) \phi_b(x') \rangle &=& \frac{\delta_{ab}}{N} \Biggl\langle \phi_0^2 + (N-1) \hat{G}_1(y) + \hat{G}_2(y) \Biggr\rangle_{\bar{0}}. \label{full-resummed-2pt-f}
% \end{eqnarray}

% , which for $m=0$ evaluate Eq.~\eqref{resummed-2-pt-func} for $\bar{u}(m=0)$ of Eq.~\eqref{bar-u-massless}. The effective masses are then
\subsubsection{Corrections} \label{sec-pert}

As a usage example of the R-Feynman rules, we  compute the corrections  coming from some of the interactions in $S^{(3)}_{int}$. 
The simplest ones are local diagrams with at most one of the quartic vertices in Eq.~\eqref{Sint3}.

\begin{figure}[h!]
 \centering
\begin{tikzpicture}
  \begin{feynman}
    \vertex (n1) [blob,very thick,fill=white,minimum size=1cm] {};
    \vertex [right=1cm of n1] (n2) [blob,very thick,fill=white,minimum size=1cm] {};
  \end{feynman}
\end{tikzpicture}
\hspace{0.8cm}
\begin{tikzpicture}
  \begin{feynman}
    \vertex (n1) [blob,very thick,fill=white,minimum size=1cm] {};
    \vertex [right=1cm of n1] (n2) [blob,very thick,fill=white,minimum size=1cm,dash dot] {};
  \end{feynman}
\end{tikzpicture}
\hspace{0.8cm}
\begin{tikzpicture}
  \begin{feynman}
    \vertex (n1) [blob,very thick,fill=white,minimum size=1cm,dash dot] {};
    \vertex [right=1cm of n1] (n2) [blob,very thick,fill=white,minimum size=1cm,dash dot] {};
  \end{feynman}
\end{tikzpicture}
 \caption{ Bubble diagrams contributing to $ \Delta \hat{Z}$ at lowest order.}
  \label{fig:deltaZ-bubbles}
\end{figure}
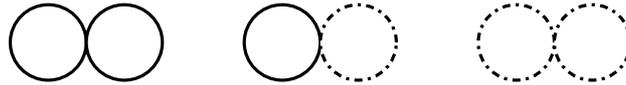

We start by noticing that  according to the R-Feynman rules we cannot build connected diagrams that correct $\langle \phi_{0a} \phi_{0b} \rangle$,
\begin{equation}
  \Delta \langle \phi_{0a} \phi_{0b} \rangle_{C} \equiv 0. \label{2pt-phi0-NLO-conn}
\end{equation}
There are, however, corrections in the form of 0-connected diagrams from the fact that there are bubble diagrams (i.e. corrections to $\hat{Z}_1 \hat{Z}_2$) that do not cancel in general. Indeed, 
 \begin{eqnarray}
 \Delta \langle \phi_{0a} \phi_{0b} \rangle_{^0 C} &=& \langle \Delta \hat{Z} \, \phi_{0a} \phi_{0b} \rangle_{\bar{0}} - \langle \Delta \hat{Z} \rangle_{\bar{0}} \langle \phi_{0a} \phi_{0b} \rangle_{\bar{0}}, \label{2pt-phi0-NLO-disc}
  \end{eqnarray}
where $\Delta \hat{Z}$ is given by the bubble diagrams shown in Fig. \ref{fig:deltaZ-bubbles},
\begin{equation}
 \Delta \hat{Z} = -\frac{\lambda}{8} \Biggl( (N+2) [\hat{G}_1]^2 + 2 [\hat{G}_1] [\hat{G}_2] + \frac{3}{N} [\hat{G}_2]^2 \Biggr). \label{deltaZ-bubbles}
\end{equation}
% \begin{figure}[h!]
%  \centering
%  %\includegraphics{Images/...}
%  % daisy-wide.png: 1189x130 px, 300dpi, 10.07x1.10 cm, bb=0 0 285 31
%  \caption{ Bubble diagrams contributing to $ \Delta \hat{Z}$.}
%   \label{fig:deltaZ-bubbles}
% \end{figure}
 
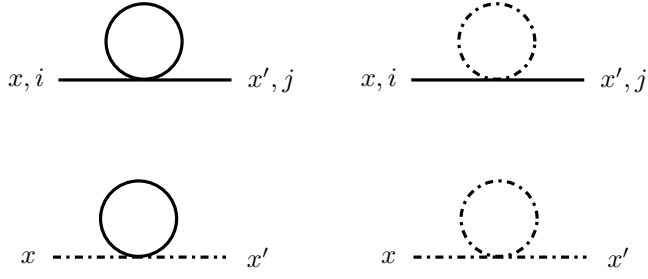
\begin{figure}[h!]
 \centering
\begin{tikzpicture}
  \begin{feynman}
    \vertex (a) {};
    \vertex [right=1.2cm of a](b);
    \vertex [right=1.2cm of b] (c) {};
    
    \vertex at ($(b) + (0.05cm, 0.51cm)$) (n) [blob,very thick,fill=white,minimum size=1cm] {};
    \vertex at ($(a) + (-0.3cm, 0)$) (l1) {\(x,i\)};
    \vertex at ($(c) + (0.4cm, 0)$) (l2) {\(x',j\)};
        
    \diagram* {
      (a) -- [very thick,solid] (b) -- [very thick,solid] (c),
    };
  \end{feynman}
\end{tikzpicture}
\hspace{0.4cm}
\begin{tikzpicture}
  \begin{feynman}
    \vertex (a) {};
    \vertex [right=1.2cm of a](b);
    \vertex [right=1.2cm of b] (c) {};
    
    \vertex at ($(b) + (0.05cm, 0.51cm)$) (n) [blob,very thick,fill=white,minimum size=1cm,dash dot] {};
    \vertex at ($(a) + (-0.3cm, 0)$) (l1) {\(x,i\)};
    \vertex at ($(c) + (0.4cm, 0)$) (l2) {\(x',j\)};
        
    \diagram* {
      (a) -- [very thick,solid] (b) -- [very thick,solid] (c),
    };
  \end{feynman}
\end{tikzpicture}
\vspace{1cm}
\\
\hspace{-0.3cm}
\begin{tikzpicture}
  \begin{feynman}
    \vertex (a) {};
    \vertex [right=1.2cm of a](b);
    \vertex [right=1.2cm of b] (c) {};
    
    \vertex at ($(b) + (0.05cm, 0.51cm)$) (n) [blob,very thick,fill=white,minimum size=1cm] {};
    \vertex at ($(a) + (-0.2cm, 0)$) (l1) {\(x\)};
    \vertex at ($(c) + (0.3cm, 0)$) (l2) {\(x'\)};
        
    \diagram* {
      (a) -- [very thick,dash dot] (b) -- [very thick,dash dot] (c),
    };
  \end{feynman}
\end{tikzpicture}
\hspace{1cm}
\begin{tikzpicture}
  \begin{feynman}
    \vertex (a) {};
    \vertex [right=1.2cm of a](b);
    \vertex [right=1.2cm of b] (c) {};
    
    \vertex at ($(b) + (0.05cm, 0.51cm)$) (n) [blob,very thick,fill=white,minimum size=1cm,dash dot] {};
    \vertex at ($(a) + (-0.2cm, 0)$) (l1) {\(x\)};
    \vertex at ($(c) + (0.3cm, 0)$) (l2) {\(x'\)};
        
    \diagram* {
      (a) -- [very thick,dash dot] (b) -- [very thick,dash dot] (c),
    };
  \end{feynman}
\end{tikzpicture}
 \caption{Connected diagrams correcting the two point function of the daisy type. The first two are proportional to $\delta_{ij}$ (and  using Eq.~(\ref{derivs-J-splitted}) contribute proportional to $P_{ab}$), while the last two enter proportionally to $\epsilon_a \epsilon_b$.}
 \label{fig:connected-daisy}
\end{figure}

Let us now consider corrections to the inhomogeneous part of the two-point function, Eq.~\eqref{2pt-hatphi-LO}. The needed connected R-Feynman diagrams are shown in Fig. \ref{fig:connected-daisy}.
% , both the connected (a) and 0-connected (b) ones
% \begin{figure}[h!]
%  \centering
%  \includegraphics{Images/daisy-wide.png}
%  % daisy-wide.png: 1189x130 px, 300dpi, 10.07x1.10 cm, bb=0 0 285 31
%  \caption{Connected diagrams correcting the two point function of the daisy type. The first two are proportional to $\delta_{ij}$ (and  using Eq.~\ref{derivs-J-splitted} contribute proportional to $P_{ab}$), while the last two enter proportionally to $\epsilon_a \epsilon_b$.}
%  \label{fig:connected-daisy}
% \end{figure}
Notice that from Eq.~\eqref{derivs-J-splitted}, the diagrams with two  external lines associated to $\hat{\phi}_1$ contribute proportional to $P_{ab}$, while those where the external lines correspond to $\hat{\phi}_2$ go with a factor of $\epsilon_a \epsilon_b$. Therefore, using the R-Feynman rules the connected part reads
\begin{eqnarray}
 \Delta \langle \hat{\phi}_a(x) \hat{\phi}_b(x') \rangle_C &=& \Biggl\langle P_{ab} \left( \frac{\lambda}{2N}(N+1) [\hat{G}_1] \frac{\partial \hat{G}_1(r)}{\partial m^2} + \frac{\lambda}{2N}[\hat{G}_2] \frac{\partial \hat{G}_1(r)}{\partial m^2} \right) \notag \\
 &&\quad+ \epsilon_a \epsilon_b \left( \frac{\lambda}{2N} (N-1) [\hat{G}_1] \frac{\partial \hat{G}_2(r)}{\partial m^2} + \frac{\lambda}{2N} 3 [\hat{G}_2] \frac{\partial \hat{G}_2(r)}{\partial m^2} \right) \Biggr\rangle_{\bar{0}},
\end{eqnarray}
where we are using the   property
\begin{equation}
 \frac{\partial \hat{G}(x,x')}{\partial m^2} = - \int_z \, \hat{G}(x,z) \hat{G}(z,x').
\end{equation}
Further simplification using properties \eqref{symmetry-prop} and dropping terms of order $\mathcal{O}(1/N^2)$ leads to
\begin{eqnarray}
 \Delta \langle \hat{\phi}_a(x) \hat{\phi}_b(x') \rangle_C \simeq \delta_{ab} \frac{\lambda}{2} \Biggl\langle [\hat{G}_1] \frac{\partial \hat{G}_1(r)}{\partial m^2} + \frac{1}{N} \Biggl( [\hat{G}_2] \frac{\partial \hat{G}_1(r)}{\partial m^2} + [\hat{G}_1] \frac{\partial \hat{G}_2(r)}{\partial m^2} \Biggr) \Biggr\rangle_{\bar{0}}. \label{2pt-hatphi-NLO-conn}
\end{eqnarray}
Finally, for the 0-connected part, according to the definition \eqref{n-pt-0C} for $n=2$, we have
\begin{equation}
 \Delta \langle \hat{\phi}_a(x) \hat{\phi}_b(x') \rangle_{^0 C} = \Biggl\langle \Delta \hat{Z} \frac{\delta^2 (\hat{Z}_1 \hat{Z}_2)}{\delta \hat{J}_{a}(x) \delta \hat{J}_{b}(x')} \Bigg|_{J=0} \Biggr\rangle_{\bar{0}} -  \Bigl\langle \Delta \hat{Z} \Bigr\rangle_{\bar{0}} \Biggl\langle \frac{\delta^2 (\hat{Z}_1 \hat{Z}_2)}{\delta \hat{J}_{a}(x) \delta \hat{J}_{b}(x')} \Bigg|_{J=0} \Biggr\rangle_{\bar{0}},
%  \lambda \, \delta_{ab} \Bigl( \langle f \, g \rangle_{\bar{0}} - \langle f \rangle_{\bar{0}} \langle g \rangle_{\bar{0}} \Bigr).
\label{2pt-hatphi-NLO-disc} 
\end{equation}
where  
\begin{equation}
 \frac{\delta^2 (\hat{Z}_1 \hat{Z}_2)}{\delta \hat{J}_{a}(x) \delta \hat{J}_{b}(x')} \Bigg|_{J=0} = \hat{G}_1(r) P_{ab} + \hat{G}_2(r) \epsilon_a \epsilon_b. \label{2pt-hatphi-LO-2} 
\end{equation}

In order to    evaluate  the corrected  two-point functions, we need to compute explicitly  the integrals over the zero-modes  that define the   expectation values on the right hand sides of Eqs.~(\ref{2pt-phi0-LO}), (\ref{2pt-resummed-implicit}), (\ref{2pt-phi0-NLO-disc}), (\ref{2pt-hatphi-NLO-conn}), and (\ref{2pt-hatphi-NLO-disc}).   This can be systematically performed order by order assuming a double expansion in $\lambda$ and $1/N$, as we show next. 
 
\subsection{Computation of zero-mode expectation values in a double $\lambda$ and $1/N$-expansion}

In order to perform a  systematic evaluation of the expectation values over the zero modes $\langle \dots \rangle_{\bar{0}}$,  order by order in $1/N$,    
we can use  the saddle-point approximation (Laplace method).  Indeed, by a simple change of variables $u = \lambda  |\vec{\phi}_0|^2/2N$, the entire $N$ dependence can be collected in an overall factor in the exponential, so that for a generic function of $u$, $g(u)$, the  zero-mode expectation values can be written as
\begin{eqnarray}
\langle g(u) \rangle_{\bar{0}} &=& \frac{\int_{0}^{\infty} du \, D(u) \, g(u) \, e^{N h(u)} }{ \int_{0}^{\infty} du \, D(u) \, e^{N h(u)}},
\end{eqnarray}
where we have introduced the following functions
\begin{eqnarray}
 h(u) &=& \frac{1}{2} \log(u) + \frac{1}{2} \log(\det \hat{G}_1) - \frac{m^2 V_d}{\lambda} u - \frac{V_d}{2\lambda} u^2, \label{h-func} \\
 D(u) &=& \frac{1}{u} \sqrt{ \frac{\det \hat{G}_2}{\det \hat{G}_1}}. \label{d-func}
\end{eqnarray}
We explicitly emphasize the dependence of the propagators $\hat{G}_1$ and $\hat{G}_2$ with $u$ through their masses $m_1^2 = m^2 + u$ and $m_2^2 = m^2 + 3u$. 

Assuming an expansion for large values of $N$, as described in the Appendix~\ref{app-laplace}, we can  approximate
\begin{equation}
\langle \, g(u) \, \rangle_{\bar{0}} \simeq g(\bar{u}) + \frac{1}{N} \left[ B^{(1)}_1 g' + B^{(1)}_2 g'' \right]_{\bar{u}} + \mathcal{O}\left( \frac{1}{N^2} \right), \label{laplace-exp}
\end{equation}where a prime means  a derivative with respect to $u$.
The coefficients $B^{(1)}_i(u)$, as well as the higher order ones, are built from the functions $h(u)$ and $D(u)$ and their derivatives, while $\bar{u}$ is the solution of $h'(\bar{u}) = 0$, provided that $h''(\bar{u}) < 0$. For our function $h(u)$ defined in Eq.~\eqref{h-func}, $\bar{u}$ is the solution of
\begin{equation}
 \left[ 1 + \frac{\lambda}{2} \frac{\partial [\hat{G}^{(m)}]}{\partial m^2} \right] \bar{u}^2 +  \left[ m^2 + \frac{\lambda}{2} [\hat{G}^{(m)}] \right] \bar{u} - \frac{\lambda}{2V_d} = 0, 
\end{equation}
where we have approximated $[\hat{G}_1]$ for small $\bar{u}$ at linear order, assuming that $\bar{u} \ll H^2$. The coincidence limit of the UV propagator is divergent and shall be renormalized in the standard way, however we defer this discussion to the Appendix~\ref{app-ren}. After the renormalization procedure, the expressions   of  these quantities turn out to be the same, but with  the  coincidence limit of the propagators and their derivatives replaced by the corresponding  finite counterpart, and the constants $m$ and $\lambda$ by the renormalized quantities.  From now on, we assume the replacements have been done, but for the sake of simplicity, we  use the same notation  for the finite quantities.
%\footnote{From now on these coincidence limits are assumed to be already renormalized and finite, although we avoid to denote this explicitly with any label.}. 
The relevant (positive) solution is then
\begin{equation}
 \bar{u} = \frac{- \left( m^2 + \frac{\lambda}{2} [\hat{G}^{(m)}] \right) + \sqrt{\left( m^2 + \frac{\lambda}{2} [\hat{G}^{(m)}] \right)^2 + \frac{2\lambda}{V_d} \left( 1 + \frac{\lambda}{2} \frac{\partial [\hat{G}^{(m)}]}{\partial m^2} \right)}}{2 \left( 1 + \frac{\lambda}{2} \frac{\partial [\hat{G}^{(m)}]}{\partial m^2} \right)}. \label{ubar}
\end{equation}

Another  property,  shown in Appendix~\ref{app-laplace}, that will  be useful for the computation of 0-connected parts is
\begin{eqnarray}
&&\langle \, g(u) k(u) \, \rangle_{\bar{0}} - \langle \, g(u) \, \rangle_{\bar{0}} \langle \, k(u) \, \rangle_{\bar{0}} \simeq \frac{C^{(1)}_{2} }{N} g' k' \label{1/N-expanded-generic-product-text} \\
&&\quad \quad+ \frac{1}{N^2} \Biggl[ C^{(2)}_{2}  \, g' k' + C^{(2)}_{3}  \left(g' k''+ g'' k'\right) + C^{(2)}_{4}  \left(g' k''' + g'' k'' + g''' k' \right) \Biggr] + \mathcal{O}\left( \frac{1}{N^3} \right),  \notag
\end{eqnarray}
where $g(u)$ and $k(u)$ are arbitrary (although sufficiently smooth) functions of $u$
% ,  a superscript ``${}^{(j)}$'' on a function  stands for $j$-derivatives with respect to $u$
, and  all quantities  on the right-hand side are evaluated at $\bar{u}$. The coefficients $C^{(1)}_2(u)$ and $C^{(2)}_i(u)$ depend of $h(u)$, $D(u)$ and their derivatives and are given in the Appendix~\ref{app-laplace}. Notice that in this case it is important to keep the $N^{-2}$ terms due to the typical extra overall factor of $N$ in the bubble diagrams (see for example Eq.~\eqref{deltaZ-bubbles}).

\subsubsection{Massless case}

We now focus  on the  case of a massless, minimally coupled field we are interested in. It can be readily seen that setting $m=0$ it leads to an expansion in powers of $\sqrt{\lambda}$. Indeed, expanding Eq.~\eqref{ubar} up to order $\lambda$ we have
% As discussed in Section \ref{sec-pert}, ignoring perturbative corrections is only accurate up to next-to-leading order, 
\begin{equation}
 \bar{u}(m=0) = \sqrt{\frac{\lambda}{2V_d}} - \frac{\lambda}{4} [\hat{G}^{(0)}] + \mathcal{O}(\lambda^{3/2}). \label{bar-u-massless} 
\end{equation}
Then, evaluating the coefficients of the expansions Eqs. \eqref{laplace-exp} and \eqref{1/N-expanded-generic-product-text} at this $\bar{u}$, allows them to be expanded in a similar fashion
% If we focus on the massless case ($m=0$), actually the expansion is in powers of $\sqrt{\lambda}$. Also, in this case there will be perturbative corrections coming from the $\hat{\phi}^4$ interaction starting at order $(\sqrt{\lambda})^2$, so one should not trust these LO results beyond that. 
% according to Eqs.~\eqref{double-exp-massless},
% \begin{subequations}
% \begin{eqnarray}
%  B^{(1)}_{1}  &\simeq& -\frac{1}{2} \sqrt{\frac{\lambda}{2V_d}} - \frac{3\lambda [\hat{G}^{(0)}]}{8} + \mathcal{O}(\lambda^{3/2}),\\ 
%  B^{(1)}_{2}  &\simeq& \frac{\lambda}{4V_d} + \mathcal{O}(\lambda^{3/2}).
% \end{eqnarray} 
% \end{subequations} 
\begin{subequations}\label{double-exp-massless}
\begin{eqnarray}
 B^{(1)}_{1}  &\simeq& -\frac{1}{2} \sqrt{\frac{\lambda}{2V_d}} - \frac{3\lambda [\hat{G}^{(0)}]}{8} + \mathcal{O}(\lambda^{3/2}),\\ 
 B^{(1)}_{2}  &\simeq& \frac{\lambda}{4V_d} + \mathcal{O}(\lambda^{3/2}),\\
 C^{(1)}_{2}  &\simeq& \frac{\lambda}{2V_d} + \mathcal{O}(\lambda^{3/2}),\\
 C^{(2)}_{2}  &\simeq& -\frac{\lambda}{4V_d} + \mathcal{O}(\lambda^{3/2}),\\
 C^{(2)}_{3}  &\simeq& -\frac{\lambda^2 [\hat{G}^{(0)}]}{8 V_d} + \mathcal{O}(\lambda^{5/2}),\\
 C^{(2)}_{4}  &\simeq& \frac{\lambda^2}{8V_d^2} + \mathcal{O}(\lambda^{5/2}).
\end{eqnarray} 
\end{subequations}
% Finally, provided $f'  g'$ are not enhanced as $\lambda\to 0$, for the massless case the following property holds,
% \begin{equation}
%  \langle \, f \, g \, \rangle_{\bar{0}} = \langle \, f \, \rangle_{\bar{0}} \langle \, g \, \rangle_{\bar{0}} + \mathcal{O}(\lambda, N^{-1}), \label{factorization}
% \end{equation}
%  as proven in Appendix \ref{app-laplace}. 

With all these ingredients we can now compute the two-point functions up to next-to-next-to LO (NNLO) in $\sqrt{\lambda}$ and next-to LO (NLO) in $1/N$. At such order, it is enough to evaluate  the sum of the  contributions in Eqs. ~(\ref{2pt-phi0-LO}) and  (\ref{2pt-resummed-implicit}),  and the connected correction to the inhomogeneous part Eq.~\eqref{2pt-hatphi-NLO-conn} (recall there is no such correction for the constant part). For completion, in what follows, we also show separately the 0-connected corrections, both the constant
Eq.~\eqref{2pt-phi0-NLO-disc} and the inhomogeneous Eq.~\eqref{2pt-hatphi-NLO-disc} corrections, and we verify that they are of higher order in $\sqrt{\lambda}$ with respect to the parts we keep. 
% to However, at this order we can safely disregard the contributions coming from the 0-connected parts, Eqs.~\eqref{2pt-hatphi-NLO-disc} and \eqref{2pt-phi0-NLO-disc}, due to the property \eqref{factorization}. 

The results are the following. For the constant part
\begin{equation}
 \langle \phi_{0a} \phi_{0b} \rangle = \delta_{ab} \left[ \sqrt{\frac{2}{\lambda V_d}} - \frac{[\hat{G}^{(0)}]}{2} - \frac{1}{2N} \left( \sqrt{\frac{2}{\lambda V_d}} + \frac{3[\hat{G}^{(0)}]}{2} \right) \right]_{\bar{u}}; \label{resummed-2-pt-func-phi0}
\end{equation} while for the inhomogeneous part
\begin{eqnarray}
\langle \hat{\phi}_a(x) \hat{\phi}_b(x') \rangle+\Delta\langle \hat{\phi}_a(x) \hat{\phi}_b(x') \rangle_C &=& \delta_{ab} \Biggl\{ \hat{G}_1(r) + \frac{\lambda}{2} [\hat{G}_1] \frac{\partial \hat{G}_1(r)}{\partial m^2} + \frac{1}{N} \Biggl[ \hat{G}_2(r) - \hat{G}_1(r) \notag \\
&&+ \left( -\frac{1}{2} \sqrt{\frac{\lambda}{2V_d}} - \frac{3\lambda [\hat{G}^{(0)}]}{8} + \frac{\lambda [\hat{G}_2]}{2} \right) \frac{\partial \hat{G}_1(r)}{\partial m^2} + \frac{\lambda}{4V_d} \frac{\partial^2 \hat{G}_1(r)}{\partial (m^2)^2} \notag \\
&&+ \frac{\lambda}{2} [\hat{G}_1] \frac{\partial \hat{G}_2(r)}{\partial m^2}\Biggl] \Biggr\}_{\bar{u}}. \label{resummed-2-pt-func}
\end{eqnarray}
where we have dropped terms of order $\lambda^{3/2}$. At this point we have arrived to a result that is the equivalent of the main result of our previous work \cite{euclideo-1}\footnote{See Eq.~(5.26) of this reference.}, once the new expression is properly expressed in terms of propagators with masses $\sqrt{\lambda/2V_d}$ and $3 \sqrt{\lambda/2V_d}$ by expanding $\bar{u}$ as in Eq.~\eqref{bar-u-massless}\footnote{Notice that in Ref.~\cite{euclideo-1} a further approximation was assumed to be valid, namely $\lambda \, \partial \hat{G}^{(\sqrt{3}m_{\rm dyn})}(r)/\partial m^2 \simeq \lambda \, \partial \hat{G}^{(m_{\rm dyn})}(r) /\partial m^2 + \mathcal{O}(\lambda^{3/2})$, where $m_{\rm dyn}^2 = \sqrt{\lambda/(2V_d)}$. This is correct on the sphere, but it actually fails in dS at sufficiently long distances. }
%In order to get full compatibility between both results, it is necessary to apply this approximation also here.}. 
Also, the coincidence propagators need to be expressed in terms of the massless one, $\lambda [\hat{G}_1], \lambda [\hat{G}_2] \simeq \lambda [\hat{G}^{(0)}] + \mathcal{O}(\lambda^{3/2})$.

For the 0-connected parts we take Eqs.~\eqref{2pt-phi0-NLO-disc} and \eqref{2pt-hatphi-NLO-disc}, and use  \eqref{deltaZ-bubbles} and  Eqs.~\eqref{2pt-hatphi-LO-2}  together with properties \eqref{symmetry-prop} and the expansion \eqref{1/N-expanded-generic-product-text} with \eqref{double-exp-massless}. Finally dropping higher order terms in $1/N$ we obtain, for the constant part
\begin{eqnarray}
  \Delta \langle \phi_{0a} \phi_{0b} \rangle_{^0 C} &=& -\frac{\lambda \delta_{ab}}{4 V_d} \left[ [\hat{G}_1] \frac{\partial [\hat{G}_1]}{\partial m^2} + \frac{1}{N} \left( \frac{3}{2} [\hat{G}_1] \frac{\partial [\hat{G}_1]}{\partial m^2} +  [\hat{G}_1] \frac{\partial [\hat{G}_2]}{\partial m^2} +  [\hat{G}_2] \frac{\partial [\hat{G}_1]}{\partial m^2} \right) \right]_{\bar{u}} \notag \\
  &\simeq& -\frac{\lambda \delta_{ab}}{4 V_d} \left( 1 + \frac{7}{2N} \right) [\hat{G}^{(0)}] \frac{\partial [\hat{G}^{(m)}]}{\partial m^2}\Bigg|_{0}, 
\end{eqnarray}
while for the inhomogeneous part
\begin{eqnarray}
 \Delta \langle \hat{\phi}_a(x) \hat{\phi}_b(x') \rangle_{^0 C} &\simeq& - \frac{\lambda^2 \delta_{ab}}{8 V_d} \Bigg\{ \left( 1 + \frac{1}{2N} \right) [\hat{G}_1] \frac{\partial [\hat{G}_1]}{\partial m^2} \frac{\partial \hat{G}_1(r)}{\partial m^2} \notag \\
 &&+ \frac{1}{N} \left( \frac{\partial [\hat{G}_1]}{\partial m^2} [\hat{G}_2] + [\hat{G}_1] \frac{\partial [\hat{G}_2]}{\partial m^2} \right) \frac{\partial \hat{G}_1(r)}{\partial m^2} + \frac{1}{N} [\hat{G}_1] \frac{\partial [\hat{G}_1]}{\partial m^2} \frac{\partial \hat{G}_2(r)}{\partial m^2} \notag\\
 &&+ \frac{\lambda}{4 V_d N} [\hat{G}_1] \frac{\partial [\hat{G}_1]}{\partial m^2} \left( \frac{\partial^3 \hat{G}_1(r)}{\partial (m^2)^3} - V_d [\hat{G}^{(0)}] \frac{\partial^2 \hat{G}_1(r)}{\partial (m^2)^2} \right) \notag \\
 &&+ \frac{\lambda}{4 V_d N}  \left( \left.\frac{\partial [\hat{G}_1]}{\partial m^2} \right.^2 + [\hat{G}_1] \frac{\partial^2 [\hat{G}_1]}{\partial (m^2)^2} \right) \left[ \frac{\partial^2 \hat{G}_1(r)}{\partial (m^2)^2} - V_d [\hat{G}^{(0)}] \frac{\partial \hat{G}_1(r)}{\partial m^2} \right] \notag\\
 &&+ \frac{\lambda}{4 V_d N}  \left( 3 \frac{\partial [\hat{G}_1]}{\partial m^2} \frac{\partial^2 [\hat{G}_1]}{\partial (m^2)^2} + [\hat{G}_1] \frac{\partial^3 [\hat{G}_1]}{\partial (m^2)^3} \right) \frac{\partial \hat{G}_1(r)}{\partial m^2} \Biggr\}_{\bar{u}}, \label{2pt-hatphi-NLO-disc-explicit}
\end{eqnarray}
% \begin{eqnarray}
%  \Delta \langle \hat{\phi}_a(x) \hat{\phi}_b(x') \rangle_{^0 C} &\simeq& - \frac{\lambda^2 \delta_{ab}}{8 V_d} \Bigg\{ \left( 1 + \frac{1}{2N} \right) [\hat{G}_1] \frac{\partial [\hat{G}_1]}{\partial m^2} \frac{\partial \hat{G}_1(r)}{\partial m^2} \notag \\
%  &&+ \frac{1}{N} \left( \frac{\partial [\hat{G}_1]}{\partial m^2} [\hat{G}_2] + [\hat{G}_1] \frac{\partial [\hat{G}_2]}{\partial m^2} \right) \frac{\partial \hat{G}_1(r)}{\partial m^2} \notag\\
%  &&+ \frac{1}{N} [\hat{G}_1] \frac{\partial [\hat{G}_1]}{\partial m^2} \frac{\partial \hat{G}_2(r)}{\partial m^2} + \frac{\lambda}{4 V_d N} [\hat{G}_1] \frac{\partial [\hat{G}_1]}{\partial m^2} \frac{\partial^3 \hat{G}_1(r)}{\partial (m^2)^3} + \dots \Biggr\}, \label{2pt-hatphi-NLO-disc-explicit}
% \end{eqnarray}
% where we also discarded terms with less derivatives with respect to $u$ acting on the $r$ dependent functions. {\textcolor{red}{ No entendi la frase anterior. Habria que decir que significan los puntos suspensivos.}}The reason for this will become clear in a moment. 
In both cases, with these explicit expressions we can now confirm that the 0-connected contributions are suppressed by extra factors of $\lambda$ with respect to 
Eqs.~\eqref{resummed-2-pt-func-phi0} and \eqref{resummed-2-pt-func}.

It is interesting to analyze the large distance behavior of resummed two-point functions obtained after our procedure. First, notice that in the final expression for the inhomogeneous part, Eq.~\eqref{resummed-2-pt-func}, the free inhomogeneous propagators $\hat{G}_1(r)$ and $\hat{G}_2(r)$ are now evaluated for positive masses $M_1^2 \equiv m_1^2(\bar{u})$ and $M_2^2 \equiv m_2^2(\bar{u})$,
\begin{subequations}\label{eff-masses-massless}
\begin{eqnarray}
 M_1^2 &=& \sqrt{\frac{\lambda}{2V_d}} - \frac{\lambda}{4} [\hat{G}^{(0)}] + \mathcal{O}(\lambda^{3/2}), \\
 M_2^2 &=& 3 M_1^2.
\end{eqnarray}
\end{subequations}%\ref{sec:analytic-cont}
Therefore, upon analytic continuation to dS, both approach a constant exponentially at large distances, as discussed in subsection~\ref{sec:analytic-cont} for massive fields (Eq.~\eqref{late-times-massive} minus a constant part), rather than the logarithmic divergence of massless fields (as in Eq.~(\ref{late-times-massless})), that plagued the perturbative expansion prior to our resummation.

However, notice that the result Eq.~\eqref{resummed-2-pt-func} also contains derivatives of $\hat{G}_1(r)$ and $\hat{G}_2(r)$ with respect to their masses, and moreover, the number of derivatives increases with the orders in the double $\sqrt{\lambda}$ and $1/N$-expansion. This is due to the treatment given to the zero-mode weighted integrals (see Eq.~\eqref{laplace-exp}). The behavior of these derivatives in dS and at large distances is easily obtained from Eq.~\eqref{late-times-massive} and found to be
\begin{equation}
 \frac{\partial^p \hat{G}^{(m)}(r)}{\partial (m^2)^p} \simeq \frac{1}{V_d m^2} \left(- \frac{\log(r)}{d-1} \right)^p r^{-\frac{m^2}{d-1}} + \dots,
\end{equation}
meaning that, although there is an overall exponential decay, for large enough distances higher order terms become as relevant as the lower order ones, thus breaking the expansion.   

The origin of this problem is in the assumption used when computing \eqref{resummed-2-pt-func} in the double expansion in $\sqrt{\lambda}$ and $1/N$, which requires  $\sqrt{\lambda} \log(r) \ll N$. A proper analysis of the long distance behavior must then consider the opposite hierarchy $\sqrt{\lambda} \log(r) \gg N$, and will be discussed in the next section. Nevertheless, this does not invalidate the result \eqref{resummed-2-pt-func}, as long as one is only interested in points not too far apart (or in the Euclidean correlators).

Let us close this section by pointing out a related aspect to be  considered when using the results on the sphere to study the large distance (IR) behavior by analytical continuation to dS, which  is the use of power counting rules to select the relevant diagrams. Notice that the power counting in the coupling constant $\lambda$ we use here\footnote{Following previous works on the sphere \cite{Rajaraman1,BenekeMoch}, $ |\vec{\phi}_0| \sim \lambda^{-1/4}$ and $\hat{G}^{(0)} \sim \lambda^0$.} assumes there is no additional nonperturbative IR enhancement. This is certainly the case for the sphere, but is not when the correlators are analytically continued to dS. Indeed,  due to the non compactness of the dS spacetime, it turns out that the counting used here applies only up to a certain maximum distance, but not beyond. This demands going beyond perturbative R-Feynman diagrams to understand the far IR limit by collecting all contributions at a given order in $\lambda$ that are relevant in this limit. In order to make this feasible, an organizing principle such as powers of $1/N$ is necessary. 
%This has been the approach of many other previous works where the computations are performed exclusively in dS \cite{Gautier}. %
We will discuss more on this in the following section.

\section{Long wavelength behavior of the two-point functions for massless fields} \label{sec-all-N}

\subsection{Leading IR contribution: a consistent calculation beyond   $N\to+\infty$}\label{LIRC}

Let us first discuss the limiting value of the two-point functions, once analytically continued back to dS, for large distances/late times, $r\to +\infty$. Contributions to this limit may come either from the zero-mode part, or from the limiting value of the inhomogeneous part. Indeed, according to Eqs.~\eqref{free-prop} and \eqref{late-times-massive}, in this limit the massive UV propagators $\hat{G}_\alpha$ go to
\begin{equation}
 \hat{G}_\alpha(r) \to -\frac{1}{V_d  m_\alpha^2}, \,(\alpha=1,2), 
\end{equation}
with $m_2^2=3m_1^2=3u$.
 However, once all the constant contributions are put together, one expects the full two-point function to approach a vanishing value.

After the resummation of the bi-quadratic terms, the full two-point function without any corrections from $S^{(3)}_{int}$ is
  \begin{eqnarray}
  \langle \phi_a(x) \phi_b(x') \rangle &=& \frac{\delta_{ab}}{N} \Biggl\langle  |\vec{\phi}_0|^2 + (N-1) \hat{G}_{1}(r) + \hat{G}_{2}(r) \Biggr\rangle_{\bar{0}}. \label{full-resummed-2pt-f}
  \end{eqnarray}
Combining the constant contributions, we have
\begin{eqnarray}\label{constant-resummed}
 \langle \phi_a(x) \phi_b(x') \rangle^{LO} &\to& \frac{\delta_{ab}}{N} \Biggl\langle \frac{2N}{\lambda} u - \left(N-1+\frac{1}{3} \right) \frac{1}{V_d u} \Biggr\rangle_{\bar{0}} \simeq -\sqrt{\frac{2}{V_d \lambda}} \frac{4}{3N} + \mathcal{O}(N^{-2}, \lambda^0),
\end{eqnarray}
where we used Laplace's method at NLO in $1/N$ to compute the integrals \footnote{Although in this case the integrals can be computed exactly for any $N$.}.

First observe that this result  vanishes at LO in the large N limit,  $N\to +\infty$. This is because 
the family of diagrams that contribute at large-$N$ is that of the daisy and superdaisy type, which add a local part to the self-energy, and whose leading contribution in $\sqrt{\lambda}$ is already completely taken into account by the exact treatment of the zero modes.

Starting at NLO in $1/N$ we encounter a nonvanishing value. The reason is that at NLO in $1/N$ there are diagrams that contribute at LO in $\sqrt{\lambda}$ that are not of the type included in the resummation above.  Indeed,  the relevant diagrams that must be added to the self-energy are non local and have the form of a bubble-chain.  These are only partially accounted for in our treatment so far, even at the LO in $\sqrt{\lambda}$.  This stems from the fact that, although on the sphere there is a hierarchy between  interactions with $\vec{\phi}_0$ and $\vec{\hat{\phi}}$ in terms of powers of $\sqrt{\lambda}$, upon analytical continuation to dS, when $r \to \infty$ the correlators of inhomogeneous modes receive an enhancement that makes them as relevant as the zero mode correlators,
\begin{equation}\label{newcounting}
 \langle |\vec{\phi_{0}}|^2 \rangle_{\bar{0}} \sim \frac{1}{\sqrt{\lambda}}, \quad\quad \langle \hat{\phi}_a(x) \hat{\phi}_b(x') \rangle_{\bar{0}} \to\sim \frac{\delta_{ab}}{\sqrt{\lambda}}.
\end{equation}
Therefore,  a consistent calculation of the two-point function at large distances at  in dS,    at  LO in $\sqrt{\lambda}$   and   NLO  in $1/N$,  requires the addition of diagrams   correcting   Eq.~\eqref{2pt-hatphi-LO}.  To show this explicitly,   let us   now    evaluate   the  limiting  value of the diagrams when the distance between the two points goes to infinity.  Doing this, we will get a non-vanishing result that should be added to Eq.~\eqref{constant-resummed}.  

In order to  identify the relevant diagrams that potentially contribute to the limiting value at LO in $\sqrt{\lambda}$ we use the scaling in Eq.~\eqref{newcounting} as a  new way of counting the corresponding order in $\sqrt{\lambda}$.  
An strategy to perform the calculation of the  diagrams consists in  the following two steps: I) use the  R-Feynman rules described in Sec. \ref{R2PF} to compute  the  corrections to the propagators  $\hat{G_1}$ and $\hat{G_2}$ (with $|\vec{\phi}_0|$-dependent masses)  and replace the corrected propagators  into Eq.~\eqref{2pt-hatphi-LO}; II) evaluate the two point functions using  \eqref{symmetry-prop},  as for Eq.~\eqref{2pt-resummed-implicit}, and  taking  the expectation value over $|\vec{\phi}_0|$  assuming the  double expansion  in $\sqrt{\lambda}$ and $1/N$. Recall that the 0-connected diagrams are suppressed by extra factors of $\sqrt{\lambda}$.

\begin{figure}[h!]
 \centering
  \begin{tikzpicture}[anchor=base, baseline=-0.1cm]
  \begin{feynman}
    \vertex (a);
    \vertex [right=1.2cm of a] (b);
    \vertex [right=1.2 of b] (c); 
        
    \diagram* {
      (a) -- [very thick,dash dot] (b) -- [very thick,dash dot] (c),
    };

    \vertex at ($(b) + (0cm, 0cm)$) (n2) [blob,very thick,fill=white,minimum size=0.5cm] {};
    \vertex at ($(b) + (0cm, -1cm)$) (l) {\( (a) \)};    
  \end{feynman}
\end{tikzpicture}
\hspace{1cm}
 \begin{tikzpicture}[anchor=base, baseline=-0.1cm]
  \begin{feynman}
    \vertex (a);
    \vertex [right=1.2cm of a] (b);
    \vertex [right=1.2 of b] (c); 
        
    \diagram* {
      (a) -- [very thick,dash dot] (b) -- [very thick,dash dot] (c),
    };

    \vertex at ($(b) + (0cm, 0cm)$) (n2) [blob,very thick,fill=black,minimum size=0.5cm] {};
    \vertex at ($(b) + (0cm, -1cm)$) (l) {\( (a') \)};    
  \end{feynman}
\end{tikzpicture}
\caption{Diagrams contributing to the corrections of $\hat G_2$}
\label{fig:Diana1}
\end{figure}
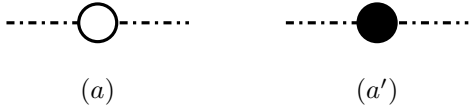

 From 
Eq.~\eqref{2pt-resummed-implicit} it is immediate to conclude  that for $\hat{G}_1$  it is necessary  to compute the NLO correction in $1/N$, while for $\hat{G}_2$ the  LO is enough.   
We illustrate the calculation with the diagram in Fig.~\ref{fig:Diana1}$a$, that corrects the propagator $\hat{G}_2$.
 To begin with, it is easy to see that the diagram does not vanish at $N\to \infty$.  
 
Our method to obtain the  contribution  of a   given diagram in the large-distance limit  is based on  general
 theorems  proved in Refs.~\cite{Marolf:2010zp,Marolf:2010nz}. The theorems  imply that 
 a calculation of a two-point Feynman diagram   using   full massive propagators $G_\alpha$  (i.e. the standard ones) instead of  $\hat G_\alpha$  (both inside the loops and in the external legs)   would give a vanishing result at large distances. The R-Feynman rules involve the modified propagators 
 \begin{equation}
 \hat G_\alpha=G_\alpha-G_\alpha^{(0)},\, (\alpha=1,2),\label{sustit}
 \end{equation}
that differ from the standard massive propagators by the constant contributions of the Euclidean zero modes. Therefore we can make the substitution above into the R-Feynman diagrams and keep, in the large distance limit,  only the contributions that are not connected by $G_\alpha$. This is so because after making the substitution \eqref{sustit}, any two-point  diagram connected by full massive propagators $G_\alpha$, up to constant factors (possibly involving  factors of $G_\alpha^{(0)}$),  can be thought as a particular two-point Feynman diagram which are known  to vanish at large distances   \cite{Marolf:2010zp,Marolf:2010nz}. 

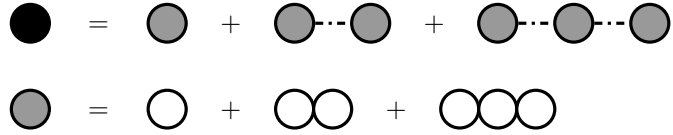
\begin{figure}[h!]
 \centering
 \begin{eqnarray}\nonumber
 \begin{tikzpicture}[anchor=base, baseline=-0.1cm]
  \begin{feynman}
    \vertex (a);        
    \vertex at ($(a) + (0cm, 0cm)$) (n) [blob,very thick,fill=black,minimum size=0.5cm] {};
  \end{feynman}
\end{tikzpicture}
\quad&=&\quad
\begin{tikzpicture}[anchor=base, baseline=-0.1cm]
  \begin{feynman}
    \vertex (a);        
    \vertex at ($(a) + (0cm, 0cm)$) (n) [blob,very thick,fill=black!40!white,minimum size=0.5cm] {};
  \end{feynman}
\end{tikzpicture}
\quad+\quad
\begin{tikzpicture}[anchor=base, baseline=-0.1cm]
  \begin{feynman}
    \vertex (a);
    \vertex at ($(a) + (0cm, 0cm)$) (n1) [blob,very thick,fill=black!40!white,minimum size=0.5cm] {};
    \vertex [right=1cm of n1] (n2) [blob,very thick,fill=black!40!white,minimum size=0.5cm] {};
    
    \diagram* {
      (n1) -- [very thick,dash dot] (n2),
    };    
  \end{feynman}
\end{tikzpicture} 
\quad+\quad
\begin{tikzpicture}[anchor=base, baseline=-0.1cm]
  \begin{feynman}
    \vertex (a);
    \vertex at ($(a) + (0cm, 0cm)$) (n1) [blob,very thick,fill=black!40!white,minimum size=0.5cm] {};
    \vertex [right=1cm of n1] (n2) [blob,very thick,fill=black!40!white,minimum size=0.5cm] {};
    \vertex [right=1cm of n2] (n3) [blob,very thick,fill=black!40!white,minimum size=0.5cm] {};
    
    \diagram* {
      (n1) -- [very thick,dash dot] (n2) -- [very thick,dash dot] (n3),
    };    
  \end{feynman}
\end{tikzpicture}\,
\\\nonumber\\
 \begin{tikzpicture}[anchor=base, baseline=-0.1cm]
\begin{feynman}
    \vertex (a);        
    \vertex at ($(a) + (0cm, 0cm)$) (n) [blob,very thick,fill=black!40!white,minimum size=0.5cm] {};
  \end{feynman}
\end{tikzpicture}
\quad&=&\quad
\begin{tikzpicture}[anchor=base, baseline=-0.1cm]
  \begin{feynman}
    \vertex (a);        
    \vertex at ($(a) + (0cm, 0cm)$) (n) [blob,very thick,fill=white,minimum size=0.5cm] {};
  \end{feynman}
\end{tikzpicture}
\quad+\quad
\begin{tikzpicture}[anchor=base, baseline=-0.1cm]
  \begin{feynman}
    \vertex (a);
    \vertex at ($(a) + (0cm, 0cm)$) (n1) [blob,very thick,fill=white,minimum size=0.5cm] {};
    \vertex [right=0.5cm of n1] (n2) [blob,very thick,fill=white,minimum size=0.5cm] {};
  \end{feynman}
\end{tikzpicture}
\quad+\quad
\begin{tikzpicture}[anchor=base, baseline=-0.1cm]
  \begin{feynman}
    \vertex (a);
    \vertex at ($(a) + (0cm, 0cm)$) (n1) [blob,very thick,fill=white,minimum size=0.5cm] {};
    \vertex [right=0.5cm of n1] (n2) [blob,very thick,fill=white,minimum size=0.5cm] {};
    \vertex [right=0.5cm of n2] (n3) [blob,very thick,fill=white,minimum size=0.5cm] {};
  \end{feynman}
\end{tikzpicture}\,\nonumber
\end{eqnarray}\caption{Diagrams defining the  solid black bubble as a sum of  infinite diagrams involving  partially-dressed  bubbles which are drawn as gray bubbles. The  gray bubble  is in turn defined as a chain of bubbles as shown  in the second line of the figure. }\label{bubblesres}
 \end{figure}

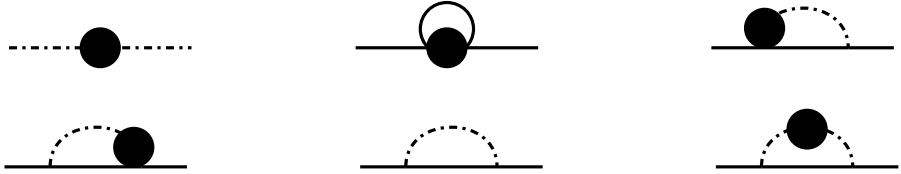
\begin{figure}[h!] \centering
\begin{tikzpicture}[anchor=base, baseline=-0.1cm]
  \begin{feynman}
    \vertex (a);
    \vertex [right=1.2cm of a] (b);
    \vertex [right=1.2 of b] (c); 
        
    \diagram* {
      (a) -- [very thick,dash dot] (b) -- [very thick,dash dot] (c),
    };

    \vertex at ($(b) + (0cm, 0cm)$) (n2) [blob,very thick,fill=black,minimum size=0.5cm] {};
  \end{feynman}
\end{tikzpicture}\hspace{2cm}
  \begin{tikzpicture}[anchor=base, baseline=-0.1cm]
  \begin{feynman}
    \vertex (a);
    \vertex [right=1.2cm of a] (b);
    \vertex [right=1.2 of b] (c); 
    
    \diagram* {
      (a) -- [very thick,solid] (b) -- [very thick,solid] (c),
    };
    
    \vertex at ($(b) + (0cm, 0.25cm)$) (n1) [blob,very thick,fill=white,minimum size=0.7cm] {};
    \vertex at ($(b) + (0cm, 0cm)$) (n2) [blob,very thick,fill=black,minimum size=0.5cm] {};
  \end{feynman}
\end{tikzpicture} \hspace{2cm}
\begin{tikzpicture}[anchor=base, baseline=-0.1cm]
  \begin{feynman}
    \vertex (a);
    \vertex [right=0.6cm of a] (n1);
    \vertex [right=1.2cm of n1] (n2);
    \vertex [right=0.6cm of n2] (b);

    \diagram* {
      (a) -- [very thick,solid] (n1) -- [very thick,solid] (n2) -- [very thick,solid] (b),
       (n1) -- [very thick,dash dot,half left] (n2),
    };
    
    \vertex at ($(n1) + (0.1cm, 0.26cm)$) (c) [blob,very thick,fill=black,minimum size=0.5cm] {};
  \end{feynman}
\end{tikzpicture}\\
\vspace{0.5cm}
 
\begin{tikzpicture}[anchor=base, baseline=-0.1cm]
  \begin{feynman}
    \vertex (a);
    \vertex [right=0.6cm of a] (n1);
    \vertex [right=1.2cm of n1] (n2);
    \vertex [right=0.6cm of n2] (b);

    \diagram* {
      (a) -- [very thick,solid] (n1) -- [very thick,solid] (n2) -- [very thick,solid] (b),
       (n1) -- [very thick,dash dot,half left] (n2),
    };
    
    \vertex at ($(n2) + (-0.1cm, 0.26cm)$) (c) [blob,very thick,fill=black,minimum size=0.5cm] {};
  \end{feynman}
\end{tikzpicture}
  \hspace{2cm}
\begin{tikzpicture}[anchor=base, baseline=-0.1cm]
\begin{feynman}
    \vertex (a);
    \vertex [right=0.6cm of a] (n1);
    \vertex [right=1.2cm of n1] (n2);
    \vertex [right=0.6cm of n2] (b);
        
    \diagram* {
      (a) -- [very thick,solid] (n1) -- [very thick,solid] (n2) -- [very thick,solid] (b),
       (n1) -- [very thick,dash dot,half left] (n2),
    };
  \end{feynman}
\end{tikzpicture}
 \hspace{2cm}
\begin{tikzpicture}[anchor=base, baseline=-0.1cm]
\begin{feynman}spacetime
    \vertex (a);
    \vertex [right=0.6cm of a] (n1);
    \vertex [right=1.2cm of n1] (n2);
    \vertex [right=0.6cm of n2] (b);
        
    \diagram* {
      (a) -- [very thick,solid] (n1) -- [very thick,solid] (n2) -- [very thick,solid] (b),
       (n1) -- [very thick,dash dot,half left] (n2),
    };
    
    \vertex at ($(n1) + (0.6cm, 0.5cm)$) (c) [blob,very thick,fill=black,minimum size=0.5cm] {};
  \end{feynman}
\end{tikzpicture}\caption{Diagrams contributing to the large distance limit of the  two-point functions at  LO in $\sqrt{\lambda}$   and at  NLO in $1/N$. The   solid black bubbles are the dressed bubbles   defined  in Fig.~\ref{bubblesres}.} \label{Diana10Fig}
\end{figure}

There is an infinite number of diagrams that contribute to this order,  that can be obtained by dressing the bubble 
in Fig.~\ref{fig:Diana1}$a$. The dressed diagram in Fig.~\ref{fig:Diana1}$a'$ contains the  chains of bubbles given in Fig.~\ref{bubblesres}.
 Fortunately, 
 this set of diagrams can be computed and resummed, following the same method applied for the one with a single bubble, that is, writing the propagator $\hat G_\alpha$ in terms of the usual massive propagator and the zero-mode contribution. 
 The details are presented in Appendix \ref{bubbles}, as well as the analysis of the corrections to the propagator $\hat{G}_1$. It is shown there that all diagrams given in Fig.~\ref{Diana10Fig} should be included at $\mathcal{O}\left(  \sqrt{\lambda},1/N \right)$, and that when the constant part of those  diagrams   are included, the full two-point function indeed decays to zero in the IR also at NLO in $1/N$.

%Fortunately, this set of diagrams can be computed and resummed, following the same strategy applied for a single bubble, that is, writing the propagator $\hat G_a$ in terms of the usual massive propagator and the zero-mode contribution. The details are presented in Appendix \ref{bubbles}, as well as the analysis of the corrections to the propagator $\hat{G}_1$. It is shown there that all diagrams given in Eqs.~\eqref{Diana10} should be included at $\mathcal{O}\left(  \sqrt{\lambda},1/N \right)$, and that when the constant part of those  diagrams   are included, the full two-point function indeed decays to zero in the IR also at NLO in $1/N$.

\begin{figure}[h!]
 \centering
\begin{tikzpicture}[anchor=base, baseline=-0.1cm]
\begin{feynman}
    \vertex (a);
    \vertex [right=1.2cm of a] (n1);
    \vertex [right=1.2cm of n1] (n2);
    \vertex [right=0.6cm of n2] (n3);
    \vertex [right=0.6cm of n3] (c);
    \vertex [right=1.2cm of n3] (n4); 
    \vertex [right=0.6cm of n4] (n5);
    \vertex [right=1.2cm of n5] (n6);
    \vertex [right=1.2cm of n6] (b);        
        
    \vertex [above=1.38cm of c] (m1);
    \vertex [above=0.5cm of m1] (m2);
        
    \diagram* {
      (a) -- [very thick,solid] (b),
      (n1) -- [very thick,solid,half left] (n6),
      (n2) -- [very thick,solid,half left] (n5),
      (n3) -- [very thick,solid,half left] (n4),
      (m1) -- [very thick,loosely dotted] (m2),
    };    
  \end{feynman}
\end{tikzpicture}
  \caption{Example of a class of R-Feynman diagrams contributing to the two-point functions (correcting $\hat{G}_1$ in this case) at NNLO or higher orders in $1/N$, but at LO in $\sqrt{\lambda}$ at large distances in dS.}
  \label{fig:NNLO}
\end{figure}
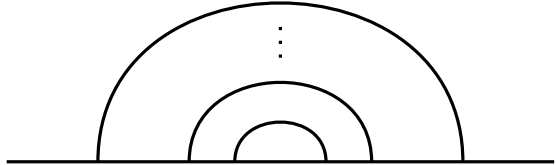

At higher orders, there are even more additional diagrams contributing to the constant limiting value. This can be seen from the fact that the  exact treatment of the zero modes contains contributions at all orders in $1/N$, which  however are not compensated by the limiting value of the diagrams included so far for the inhomogeneous  part, even at the LO in $\sqrt{\lambda}$.  Indeed, while the diagrams  in Fig.~\ref{bubblesres} are enough at NLO in $1/N$, using Eq.(\ref{newcounting}) one can see that  a diagram of the type   shown in Fig.~\ref{fig:NNLO}, which  is NNLO in $1/N$,   should also be taken into account at LO in $\sqrt{\lambda}$.
%  \begin{figure}[h!]\label{fNNLO}
%  \includegraphics[width=5cm]{DiagNNLO}
%   \caption{\textcolor{red}{Podemos poner un diagrama como estos -rotado apropiadamente- o varios (la serie de diagramas de este tipo), como prefieran, como ejemplo(s) de diagrama(s) que contribuye(n) al NNLO en $1/N$ y , en el limite de distancias grandes,    at leading order in $\sqrt{\lambda}$.}}
% \end{figure}

Despite the added complexity to perform a consistent calculation at large distances beyond  $N\to+\infty$, the situation now is nevertheless much better than before resumming the bi-quadratic terms, where the correlators of the inhomogeneous modes were massless and outright divergent in the IR (see Eq.~\eqref{late-times-massless}).

\subsection{Asymptotic IR behaviour of the two-point functions}

The next step is to study the decay of the full two-point function at large distances. To do this properly at a given order in $1/N$, again it would be necessary to include the most relevant $r$-dependent parts of all the diagrams that contribute at that order. We intend to pursue this in a systematic way in future work. Our aim here is to just analyze what kind of decay is expected after resumming the bi-quadratic terms and the exact treatment of the zero modes.

% We have previously derived a general expression for the resummed two-point functions in the $O(N)$-model with any mass $m$ in Eq.~\eqref{2pt-resummed-implicit}, obtaining explicit expressions order by order in a $1/N$ expansion in terms of the free propagator $\hat{G}(y)$ and its derivatives. However, as discussed previously, this expansion eventually breaks down for distances such that $\sqrt{\lambda} \log(y) \gg N$. 
Here, being interested in the asymptotic behavior at large distances and/or late times, we work out a different approach to compute the $r$-dependent part of the two-point function in the $r \to +\infty$ limit for all $N$. For this, we will assume that this limit can be taken before integrating over the zero modes.
% Moreover, in this Section we will focus on the massless field ($m=0$).

For   $r \to +\infty$, we can approximate the analytically-continued (to dS) propagators 
\begin{equation} \label{large-y-approx}
 G_\alpha(r) \simeq  \frac{1}{V_d m_\alpha^2} e^{-m_\alpha^2 y},\, (\alpha=1,2),
\end{equation}
\\provided $m_2^2=3m_1^2=3u \ll H^2$. Here we have defined $y= \log(r)/(d-1)$. Looking only at the $r$-dependent part, having already discussed the constant part, the two-point functions at long distances are given by the following expression
\begin{eqnarray}
\langle \phi_a(x) \phi_b(x') \rangle_{IR} \simeq \frac{\delta_{ab}}{N V_d}  \frac{\int_{0}^{\infty} du \, u^{N/2-2} \left[ (N-1) e^{-u y} + \frac{1}{3} e^{-3 u y}\right] \, e^{\tilde{h}(u)} }{ \int_{0}^{\infty} du \, u^{N/2-1} \, e^{\tilde{h}(u)}}, \label{2pt-all-N}
\end{eqnarray}
where
\begin{eqnarray}
 \tilde{h}(u) &=& \frac{(N-1)}{2} \log(\det \hat{G}_1) + \frac{1}{2} \log(\det \hat{G}_2) - N \frac{V_d}{2\lambda} u^2 \label{htilde-func} \\
 &\simeq& \frac{N}{2} \log(\det \hat{G}^{(0)}) - V_d \frac{(N+2)}{2} [\hat{G}^{(0)}] u - V_d \left( \frac{(N+8)}{4} \frac{\partial[\hat{G}^{(m)}]}{\partial m^2}\Big|_0 + \frac{N}{2\lambda} \right) u^2 \notag \\
 &\equiv& \tilde{h}(0) -A u - \frac{B u^2}{4}.\label{coefhu}
\end{eqnarray}
In the second line we have expanded the propagators for small $u$ up to quadratic order. Here we can see that the integrands of Eq.~\eqref{2pt-all-N} are exponentially suppressed for $u \gg \sqrt{\lambda/2V_d}$, and therefore, for small enough $\lambda$ the approximation of Eqs.~\eqref{large-y-approx} is justified. Notice that for the cases $N \leq 2$ there is a IR divergence in the integral of Eq.~\eqref{2pt-all-N}. This divergence is spurious and is actually absent when one considers both the $r$-dependent and constant parts together.
% (however, an alternative derivation of Eq.~\eqref{2pt-all-N-integrated} is given in Appendix~\ref{app-alternative}). 

The needed integrals are of the form
\begin{equation}
 \int_0^{\infty} du \, u^\alpha \, e^{-A u - \frac{B u^2}{4}} = \frac{A}{B^{1+\alpha/2}} \Gamma(1+\alpha) \, U\left(1+\frac{\alpha}{2},\frac{3}{2},\frac{A^2}{B}\right), \label{hyper-U-integral}
\end{equation}
where $U(a,b,z)$ is the Tricomi confluent hypergeometric function, and $A$ and $B$ are the linear and quadratic in $u$ coefficients of $\tilde{h}(u)$ respectively, defined in Eq.\eqref{coefhu}. With these expressions, we can evaluate Eq.~\eqref{2pt-all-N},
\begin{eqnarray}
\langle \phi_a(x) \phi_b(x') \rangle_{IR} &\simeq& \frac{\delta_{ab}}{N V_d} \frac{\Gamma\left( \frac{N}{2} - 1 \right) \sqrt{B}}{\Gamma\left( \frac{N}{2} \right) U_2(0)} \Biggl[ (N-1) \left( 1 + \frac{y}{A} \right) U_0(y) + \left( \frac{1}{3} + \frac{y}{A} \right) U_0(3 y) \Biggr], \label{2pt-all-N-integrated}
\end{eqnarray}
where we defined $U_t(x)=U\left(\frac{N+t}{4},\frac{3}{2},\frac{(A+x)^2}{B}\right)$ as a shorthand. %, and
% \begin{subequations}
% \begin{eqnarray}
%  A &=& V_d \frac{(N+2)}{2} [\hat{G}^{(0)}],\\
%  B &=& V_d \left[ (N+8) \frac{\partial[\hat{G}^{(m)}]}{\partial m^2}\Big|_0 + \frac{2N}{\lambda} \right].
% \end{eqnarray}
% \end{subequations}
Although we have assumed a large value of $y$ when using Eqs.~\eqref{large-y-approx}, nothing has been said about $N$. For $N \to +\infty$, the limit must be taken carefully 
and before $y \to +\infty$, since the parameter $N$ enters in several arguments of the hypergeometric functions. Doing  this, as crosscheck, we  recover the known result for the asymptotic behavior at large distances of the two-point function, 
\begin{equation}
     \langle \phi_a(x) \phi_b(x') \rangle_{IR}  \simeq  \frac{1}{V_d m_{\rm dyn}^2} e^{-m_{\rm dyn}^2 y},\, (N\to +\infty),
\end{equation}  with $m_{\rm dyn}^2=\sqrt{\lambda/(2V_d)}.$ % The details are worked out in Appendix~\ref{app-largeN}.

We are now in position to study the $y \to +\infty$ asymptotic behavior for fixed $N$. For this we can use that $U_0(x) \simeq B^{N/4} x^{-N/2}$ at large $x$, and therefore the resummed UV two-point function takes the following form,
\begin{eqnarray}
 \langle \phi_a(x) \phi_b(x') \rangle_{IR} &\simeq& \delta_{ab} \, C(N,\lambda) \, y^{1-N/2}, \label{2pt-all-N-asymptotic} 
\end{eqnarray}
%  &\simeq& \delta_{ab} \Biggl\{ \Gamma \left(\frac{N+2}{4}\right) \frac{2 (2 V_d N)^{N/4} \left[ N-1 + 3^{-N/2}\right]}{\sqrt{\pi} V_d N (N-2)} \frac{r^{1-N/2}}{\lambda^{N/4}} - \frac{2 \left(N-\frac{2}{3}\right)}{V_d N (N-2)} \frac{\Gamma \left(\frac{N+2}{4}\right)}{\Gamma \left(\frac{N}{4}\right)} \Biggr\},
where we kept the most relevant
%\footnote{Ignoring possible constant terms of order $N^{-2}$, which according to the previous discussion for the $N^{-1}$ terms, are also expected to vanish if the appropriate diagrams are included at the necessary order.} 
term for $y \to +\infty$. The coefficient is defined as
\begin{equation}
 C(N,\lambda) =\frac{2 B^{\frac{(N+2)}{4}} \left[ N-1 + 3^{-N/2}\right]}{V_d N (N-2) A \, U_2(0)} \sim \frac{1}{\lambda^{N/4}} ,%\\
%  L(N,\lambda) &=& - \frac{2 B^{1/2} \left(N-\frac{2}{3}\right) U_0(0)}{V_d N (N-2)U_2(0)} \sim \frac{1}{\sqrt{\lambda}}, \label{limit-value}
\end{equation} 
with their LO dependence in $\lambda$ explicitly shown. The behavior of the resummed two-point functions at large distances changes drastically with $N$, as seen from Eq.~\eqref{2pt-all-N-asymptotic}. On the one hand, for $N>2$   they  decay  as a power law for $y \to +\infty$. 
On the other hand, for the cases $N \leq 2$, as already mentioned, both the $r$-dependent and constant parts must be kept during the computation. This allows to show that for $N=1,2$ the two-point function diverges as $\sqrt{y}$ and $\log(y)$ respectively\footnote{Although Ref.~\cite{Hollands2012} reports $\log(y)$ for $N=1$.}.
% On the other hand, for the case $N=1$ it can be readily seen that the two-point function diverges as $\sqrt{y}$. For the case of $N=2$ a limit must be taken, leading to a logarithmic divergence $\log(y)$.
These distinct asymptotic behaviors for different values of $N$ are summarized in Table~\ref{comparisson-decays}, where also the standard perturbative results (prior to the resummation of bi-quadratic interactions) are shown for comparison.

{\renewcommand{\arraystretch}{1.5}
\begin{table}
\begin{center}
\begin{tabular}{| c | c | c | c |}\hline
   & Free & $L$-loop & Resummed \\\hline
  $N=1$ & & & $\sqrt{\log(r)}$ \\ \cline{1-1} \cline{4-4}  
  $N=2$ & & & $\log(\log(r))$ \\ \cline{1-1} \cline{4-4}
  $N>2$ & $\log(r)$ &  $\log(r)^L$ & $\log(r)^{-(N-2)/2}$ \\ \cline{1-1} \cline{4-4} 
  $N \to \infty$ & & &  $r^{-m_{\rm dyn}^2/(d-1)}$  \\ \hline
\end{tabular} 
\end{center}
\caption{Asymptotic behavior of the resummed two-point function in the  massless case ($m=0$) for different values of the number of fields $N$ as the de Sitter invariant distance $r \to +\infty$. After resummation of the bi-quadratic interaction terms, the usual logarithmic divergences present in the perturbative calculation (at all $N$) are softened when $N=1,2$ and cured for $N>2$. Here $m_{\rm dyn}^2=\sqrt{\lambda/(2V_d)}.$ }
\label{comparisson-decays}
\end{table}

We now have a more complete picture of the asymptotic behavior at large distances/late times of the resummed two-point functions. Indeed, for large but not strictly infinite $N$, the behavior can initialy be well described by Eq.~\eqref{resummed-2-pt-func} as quasi-exponential. But as $y$ approaches and exceeds $\sim N/\sqrt{\lambda}$, the behavior changes to a power-law of $\log(r)$, as shown in this section, with an exponent  strongly dependent on the value of $N$.

% Let us briefly comment on the limiting value $L(N,\lambda)$. When it exists, i.e. for $N>2$, it is related to the dynamical mass. Indeed, here we are not studying the full two-point functions, but rather their UV parts. It is the full correlator that should decay to zero at large distances, while the UV part, on the other hand, should go to minus $\langle \phi_{0a} \phi_{0b} \rangle_0 \equiv 1/V_d m_{\rm dyn}^2(N)$. Reading $m_{\rm dyn}^2(N)$ from the explicit expression for $L(N,\lambda)$, Eq.~\eqref{limit-value}, we obtain,
% \begin{equation}
%  m_{\rm dyn}^2(N) = - \frac{1}{V_d L(N,\lambda)} \simeq \frac{3 (N-2)}{2 (3 N-2)} \frac{\sqrt{N} \, \Gamma\left(\frac{N}{4}\right)}{\Gamma \left(\frac{N+2}{4}\right)} \sqrt{\frac{\lambda}{2 V_d}} + \mathcal{O}(\lambda),
% \end{equation}
% which, unsurprisingly, only coincides with the actual dynamical mass for $N \to +\infty$, Eq.\eqref{}.
%\frac{N (N-2)U_2(0)}{2 B^{1/2} \left(N-\frac{2}{3}\right) U_0(0)} 

In summary, in this Section we have analyzed the effects of the resummation on the IR behavior of the two-point functions. Given that the bi-quadratic interactions between the zero and UV modes are treated nonperturbatively, the UV modes acquire a $|\vec{\phi_0}|$-dependent mass, and therefore the IR behavior is affected. It is convergent or less divergent, depending on the value of $N$. To understand these results, we go back to Eq.~\eqref{2pt-resummed-implicit}, that shows that the interacting two-point function is a weighted average of free and massive  two-point functions. As the free correlators decay exponentially, one could naively expect the interacting correlator to have the same behavior. However, for massless fields $m=0$ the bi-quadratic interaction induces a mass for the UV modes that is proportional to $u=\lambda |\vec{\phi}_0|^2/2N$ and vanishes at $u=0$. Therefore,  the IR behavior is dominated by the contributions coming from the region $u\ll H^2$ in the integral of Eq.~\eqref{2pt-all-N}. Due to the volume factor
$u^{N/2-1}$ in that equation, these contributions are suppressed for large values of $N$, and the correlators decay more strongly as $N$ increases.

It is unclear whether additional corrections coming from the remaining (infinitely many) R-Feynman diagrams that contribute at each order in $1/N$, when treated nonperturbatively, might change this behavior. This can be particularly important for small $N$, for which the decay given by the current resummation is   milder, or absent altogether. For instance, if the UV modes were to have an additional dynamical mass coming from the interactions in $S_{int}^{(3)}$, this mechanism could dominate the IR behavior.
% even for small values of $N$. %eventually giving an exponential decay for all values of $N$.

% The generating functional must then be corrected as
% \begin{equation}
%  Z = Z^{LO} + \Delta Z^{(1,3)^2} + \Delta Z^{(0,4)} + \dots. \label{pert-Z-negmass}
% \end{equation}

\section{Conclusions}\label{Conc}

In this paper we have considered the same  $O(N)$-symmetric theory and the same reorganization of the perturbation theory on the sphere (in $\lambda$ and $1/N$),  as  in \cite{euclideo-1}.  We have presented  a procedure that allows   the  systematic  perturbative expansion we developed  in \cite{euclideo-1} for massless fields  to be performed in a much simpler way. Another advantage of the present  procedure is that it can be immediately extended to fields with negative mass squared, as we show in Appendix \ref{sec-neg}.  

The results  presented here also extend the previous ones, providing an explicit evaluation of the long wavelength limit of the two-point functions for arbitrary values of the number of fields $N$. The resummed two-point function is given in Eq.~\eqref{2pt-resummed-implicit}, that shows that it can be written as a weighted  average of free, massive two-point functions.
This result  can be interpreted as a spectral representation of the two-point correlation function
of an interacting field,  analogous to the K{\"a}llen-Lehmann representation in Minkowski spacetime \cite{Hollands2012}. Being an average of functions that decay exponentially at large distances, the IR behaviour of the resummed two-point function is improved: it tends to a constant as $r\to \infty$ for all $N>2$. The aymptotic $r$-dependence summarized in Table \ref{comparisson-decays} shows how this constant value is approached. These results should be contrasted with the usual divergent results
obtained in perturbation theory for interacting massless fields.  In the IR, the weighted average  is  dominated by the contributions of free two-point functions of small masses, that become more relevant for low values of $N$. Because of this reason, for $N=1,2$ the resummed two-point functions do not tend to a constant as $r\to\infty$,
but  diverge with a milder divergence than in perturbation theory.

The reorganization of the perturbation theory, implemented by treating exactly the bi-quadratic interaction terms between the constant zero modes and the inhomogeneous 
fields, is not enough to describe the far IR behaviour of the correlation functions. The corrections computed within a combined double expansion in $1/N$ and 
$\sqrt\lambda$  show that higher order terms become as relevant as lower order ones when the correlation functions are analytically continued to dS spacetime, and
evaluated at very long distances $\sqrt\lambda\log(r) \gtrsim N$ (note that this is not a problem when computing the correlator functions on the sphere, which is compact).
We have shown that in order to analyze the behaviour of the two-point functions for such large distances, an additional resummation is mandatory, since an infinite
number of diagrams
that correct the inhomogeneous propagators contribute  at order $\mathcal{O}\left(  \sqrt{\lambda},1/N \right)$ in the double expansion.  
We have identified these diagrams, broadly speaking, as those containing chains of an arbitrary number of bubbles.
% of zero modes \cite{Gautier}.
 The fact that the R-Feynman rules needed in the  
additional resummation  involve the
free massive propagators that come from the first resummation, allowed us to use general results about the IR behaviour of interacting massive propagators
in dS spacetime, and to perform explicitly the resummation of the bubble chains in the large-$r$ limit. Within the double expansion in  $1/N$ and 
$\sqrt\lambda$ we have shown that the contribution of the bubble chains changes the behaviour of the two-point functions, that now tend to zero (instead of a constant)
as $r\to\infty$ up to  $\mathcal{O}\left(  \sqrt{\lambda},1/N \right)$. We expect the resummation of bubble chains to modify the decays in Table \ref{comparisson-decays}
for large values of $N$ as well, when the subleading IR parts are included.  When the  number  of fields is small, the situation is less clear, since even more diagrams would be needed to understand 
the large-$r$ limit.  We consider this is out of the scope of the present paper, but we hope to address this issue in the future.    

Finally, let us comment about the applicability of our method to physical situations that involve $n$-point functions with $n>2$. This is in principle non trivial,  due to  the fact that the method directly applies to the sphere rather than to dS. Although it is immediate  to obtain the different  two-point  functions in dS spacetime given the unique two-point function on the sphere (which is well-known \cite{birrell} and is completely analogous to the same problem in Minkowski space), it is unclear how to proceed in general for n-point functions. Notice however that this problem arises due to causality, there is no such a problem as far as one is interested in computing n-point functions for spatially-separated points.  Therefore, the calculation of n-point functions using our resummation could have several cosmological applications.  We plan to consider this problem, starting with the calculation analogous to that in Ref.\cite{Serreau4poit} for comparison of the methods.

\acknowledgments
% \section{Acknowledgements}
This work has been supported by CONICET ,  ANPCyT and UNCuyo. 
We would like to thank Bei-Lok Hu and Stefan Hollands for useful discussions on related matters. The diagrams were done with the tikz-feynman package \cite{Ellis:2017-tikzfeynman}. 
 
\appendix
\section{Double Expansions in $1/N$ and $\lambda$ of $\langle \dots \rangle_{\bar{0}}$} \label{app-laplace}

The expectation values over the zero modes can be computed in a $1/N$ expansion by a saddle-point approximation (Laplace's method). Consider an integral of the form
\begin{equation}
\int_0^{+\infty} du \, f(u) \, e^{N h(u)}, 
\end{equation}
which is approximated up to NLO in $1/N$ by the following formula
% \begin{eqnarray}
% \sqrt{\frac{2\pi}{|h''(\bar{u})| N}} e^{N h(\bar{u})} \Biggl[ f + \frac{1}{N} \left( -\frac{f^{(2)}}{2h''} + \frac{f \, h^{(4)}}{8(h'')^2} + \frac{f^{(1)} h^{(3)}}{2(h'')^2} - \frac{5f (h^{(3)})^2}{24 (h'')^3} \right) \Biggr], \label{laplace}
% \end{eqnarray}
\begin{eqnarray}
\sqrt{\frac{2\pi}{|h''(\bar{u})| N}} e^{N h(\bar{u})} \Biggl[ f + \frac{1}{N} \left( -\frac{f''}{2h''} + \frac{f \, h''''}{8(h'')^2} + \frac{f' h'''}{2(h'')^2} - \frac{5f (h''')^2}{24 (h'')^3} \right) \Biggr], \label{laplace}
\end{eqnarray}
where it is understood that all the functions are evaluated in the saddle-point $\bar{u}$, i.e. $h'(\bar{u}) = 0$, $h''(\bar{u}) < 0$. If there were more than one point that satisfies these conditions, the solution would involve a sum over them of the previous expression. We can now apply this result to expressions of the kind
\begin{equation}
\langle \, g(u) \, \rangle_{\bar{0}} = \frac{\int_{0}^{\infty} du \, D(u) \, g(u) \, e^{N h(u)} }{ \int_{0}^{\infty} du \, D(u) \, e^{N h(u)}}, 
\end{equation}
by choosing $f(u) = D(u) g(u)$ and expanding the denominator in $1/N$ as well. This last step will ensure the cancelation of any NLO terms without derivatives of $g(u)$. The resulting expression is
\begin{equation}
\langle \, g(u) \, \rangle_{\bar{0}} \simeq g(\bar{u}) + \frac{1}{N} \left[ B^{(1)}_1 g' + B^{(1)}_2 g'' \right]_{\bar{u}}, \label{1/N-expanded-generic}
\end{equation}
where
\begin{subequations}
\begin{eqnarray}
B^{(1)}_1 &=& \left(  \gamma E + \frac{\eta}{2} \right), \\
B^{(1)}_2 &=& \frac{\gamma}{2},
\end{eqnarray} 
\end{subequations}
and $\gamma = \frac{1}{|h''|}$, $\eta = \frac{h'''}{(h'')^2}$, $\sigma = h''''$ and $E = D'/D$. 

Another useful combination that appears when computing perturbative corrections is the following,
% \begin{eqnarray}
% &&\langle \, g(u) k(u) \, \rangle_{\bar{0}} - \langle \, g(u) \, \rangle_{\bar{0}} \langle \, k(u) \, \rangle_{\bar{0}} \simeq  \label{1/N-expanded-generic-product} \\
% &&\Biggl\{ \frac{C^{(1)}_{2} }{N} g' k' + \frac{1}{N^2} \Biggl[ C^{(2)}_{2}  \, g' k' + C^{(2)}_{3}  \left(g' k''+ g'' k'\right) + C^{(2)}_{4}  \left(g' k^{(3)} + g'' k'' + g^{(3)} k' \right) \Biggr] + \dots\Biggr\}_{\bar{u}},  \notag
% \end{eqnarray}
\begin{eqnarray}
&&\langle \, g(u) k(u) \, \rangle_{\bar{0}} - \langle \, g(u) \, \rangle_{\bar{0}} \langle \, k(u) \, \rangle_{\bar{0}} \simeq  \label{1/N-expanded-generic-product} \\
&&\Biggl\{ \frac{C^{(1)}_{2} }{N} g' k' + \frac{1}{N^2} \Biggl[ C^{(2)}_{2}  \, g' k' + C^{(2)}_{3}  \left(g' k''+ g'' k'\right) + C^{(2)}_{4}  \left(g' k''' + g'' k'' + g''' k' \right) \Biggr] + \dots\Biggr\}_{\bar{u}},  \notag
\end{eqnarray}
where
\begin{subequations}
\begin{eqnarray}
C^{(1)}_{2}  &=& \gamma, \\
C^{(2)}_{2}  &=& \eta^2+ \gamma \eta E +\frac{\gamma ^3 \sigma}{2}+\gamma ^2 E',\\
C^{(2)}_{3}  &=& \gamma \eta + \gamma^2 E, \\ 
C^{(2)}_{4}  &=& \frac{1}{2} \gamma ^2.
\end{eqnarray} 
\end{subequations}
% \begin{eqnarray}
% &&\langle \, g(u) k(u) \, \rangle_{\bar{0}} - \langle \, g(u) \, \rangle_{\bar{0}} \langle \, k(u) \, \rangle_{\bar{0}} \simeq \frac{\gamma}{N} g' k' \Bigg|_{\bar{u}} \notag \\
% &&\,\,\,\,\,\,\,\,+ \frac{1}{N^2} \Biggl\{ \left[\gamma ^3 h^{(3)}+\frac{\gamma ^2 D'}{D}\right] \left[g' k''+ g'' k'\right] \notag \\
% &&\,\,\,\,\,\,\,\,\,\,\,\,\,\,\,\,\,\,\,\,\,\,\,\,\,\,\,\,+ g' k' \left[\gamma ^4 (h^{(3)})^2+\gamma ^3 \left(\frac{h^{(3)}
%    D'}{D}+\frac{h^{(4)}}{2}\right)+\gamma ^2
%    \left(\frac{D''}{D}-\frac{D'^2}{D^2}\right)\right] \notag \\
% &&\,\,\,\,\,\,\,\,\,\,\,\,\,\,\,\,\,\,\,\,\,\,\,\,\,\,\,\,+\frac{1}{2} \gamma ^2
%    \left[g' k^{(3)} + g'' k'' + g^{(3)} k' \right] \Biggr\}_{\bar{u}}, \label{1/N-expanded-generic-product}
% \end{eqnarray}
For this last expression it was necessary to extend Eq.~\eqref{laplace} up to NNLO in $1/N$. 

%Let us now consider the expansion of these expressions in powers of $\lambda$. Assuming no enhancement is provided by the functions $g$ and $k$ (no inverse powers of $u$), then this amounts to expand the coefficients after being evaluated at $\bar{u}$.

%\subsection{Massless case} \label{app-factorization}

In the massless case,  $\bar{u} \sim \sqrt{\lambda}$, and the coefficients up to order $\lambda$ are expanded as
\begin{subequations}\label{double-exp-massless-app}
\begin{eqnarray}
 B^{(1)}_{1}  &\simeq& -\frac{1}{2} \sqrt{\frac{\lambda}{2V_d}} - \frac{3\lambda [\hat{G}^{(0)}]}{8} + \mathcal{O}(\lambda^{3/2}),\\ 
 B^{(1)}_{2}  &\simeq& \frac{\lambda}{4V_d} + \mathcal{O}(\lambda^{3/2}),\\
 C^{(1)}_{2}  &\simeq& \frac{\lambda}{2V_d} + \mathcal{O}(\lambda^{3/2}),\\
 C^{(2)}_{2}  &\simeq& -\frac{\lambda}{4V_d} + \mathcal{O}(\lambda^{3/2}),\\
 C^{(2)}_{3}  &\simeq& -\frac{\lambda^2 [\hat{G}^{(0)}]}{8 V_d} + \mathcal{O}(\lambda^{5/2}),\\
 C^{(2)}_{4}  &\simeq& \frac{\lambda^2}{8V_d^2} + \mathcal{O}(\lambda^{5/2}).
\end{eqnarray} 
\end{subequations}
As it turns out, all of the $C$ coefficients start at least at order $\lambda$. When the combination \eqref{1/N-expanded-generic-product} comes from a perturbative calculation, there will be another $\lambda$ in front, meaning that up to higher order terms, it is valid to factorize a mean value over the zero modes of a product, in terms of products of mean values, i.e. $\langle \, g \, k \, \rangle_{\bar{0}} \simeq \langle \, g \, \rangle_{\bar{0}} \langle \, k \, \rangle_{\bar{0}} $, even beyond the large-$N$ limit. This greatly simplifies calculations of perturbative corrections. Indeed, the application of this property to the 0-connected parts  defined in Eq.~\eqref{n-pt-0C}, implies that they are suppressed by extra powers of $\sqrt{\lambda}$ with respect to the connected parts \eqref{n-pt-C}.

\section{Renormalization} \label{app-ren}

The functions $h(u)$ and $\tilde{h}(u)$, defined in Eqs.~\eqref{h-func} and \eqref{htilde-func} respectively, contain divergent contributions from the coincidence limit of the UV propagator $[\hat{G}^{(m)}]$ and its first derivative (second and higher order derivatives are finite in this limit). These can be absorbed in the usual way by the introduction of counterterms associated with each of the parameters $m^2$ and $\lambda$. Consider for instance the function $\tilde{h}(u)$ for an arbitrary mass $m^2$,
\begin{eqnarray}
 \tilde{h}(u) &\simeq& \frac{N}{2} \log(\det \hat{G}^{(m)}) - V_d \left( \frac{N m_B^2}{\lambda_B} + \frac{(N+2)}{2} [\hat{G}^{(m)}] \right) u \notag \\
 &&- V_d \left( \frac{(N+8)}{4} \frac{\partial[\hat{G}^{(m)}]}{\partial m^2} + \frac{N}{2\lambda_B} \right) u^2,
\end{eqnarray}
where the labels $B$ have been added to denote the bare quantities. Splitting these into a finite part   and a conterterm, $\lambda_B=\lambda+\delta\lambda$ and $m_B^2=m^2+\delta m^2$, it is straightforward to see that the following choice of counterterms  allows for perturbative renormalization, 
\begin{subequations}
\begin{eqnarray}
 \delta m^2 = - \frac{\lambda}{2N} (N+2) [\hat{G}^{(m)}]_{div}, \\
 \delta \lambda = \frac{\lambda^2}{2N} (N+8) \left( \frac{\partial[\hat{G}^{(m)}]}{\partial m^2} \right)_{div},
\end{eqnarray} 
\end{subequations}
 leading to
\begin{eqnarray}
 \tilde{h}(u) &\simeq& \frac{N}{2} \log(\det \hat{G}^{(m)}) - V_d \left( \frac{N m^2}{\lambda} + \frac{(N+2)}{2} [\hat{G}^{(m)}]_{ren} \right) u \notag \\
 &&- V_d \left[ \frac{(N+8)}{4} \left(\frac{\partial[\hat{G}^{(m)}]}{\partial m^2} \right)_{fin} + \frac{N}{2\lambda} \right] u^2.
\end{eqnarray}
Although the first term is still divergent, it actually drops from any physical quantity due to the normalization of the generating functional. 
For the sake of clarity, we avoid the $ren$ and $fin$ labels in the main text.

\section{Negative squared-mass case} \label{sec-neg}

 In this Appendix we  consider  an  extension of  the resummation  to cases  with  negative-squared-mass  fields, and  we show that our method can be  applied  without any substantial modification.   Indeed, we can directly evaluate Eq.~\eqref{ubar} for a negative squared-mass $m^2=-\mu^2<0$, and then expand in $\lambda$
\begin{equation}
  \bar{u}(-\mu^2) = \mu^2 + \frac{\lambda}{2V_d \mu^2} - \frac{\lambda}{2} [\hat{G}^{(0)}] - \frac{\lambda \mu^2}{2} \frac{\partial [\hat{G}^{(0)}]}{\partial m^2} + \mathcal{O}\left( \lambda^2 \right). \label{bar-u-negmass}
 \end{equation}  Then, the constant part of the two-point function \eqref{2pt-phi0-LO} now reads, at the leading order in $\lambda$,
\begin{equation}
 \langle \phi_{0a} \phi_{0b} \rangle \simeq \delta_{ab} \frac{2\mu^2}{\lambda},\label{negsqmass}
 %  + \frac{1}{V_d \mu^2} \right) -  \frac{\delta_{ab}}{N} B_1^{(1)},
\end{equation}
% From this expression, together with Eq.~\eqref{M0-variance} and Eq.~\eqref{variance-NLO} evaluated for this $\bar{u}$, we can obtain the dynamical mass $M_0^2$ in a double expansion in $1/N$ and $\lambda$. At leading order in $\lambda$ we have, 
%  \begin{eqnarray}
%   M_0^2(-\mu^2) &=& \frac{\lambda}{2V_d \mu^2} + \mathcal{O}(\lambda^{2}, N^{-2}),
%  \end{eqnarray}
%  \begin{eqnarray}
%   M_0^2(-\mu^2) &=& \frac{\lambda}{2V_d \mu^2} - \frac{\lambda^2}{4 V_d^2 \mu^6}  +\frac{\lambda^2}{4 V_d \mu^4} [\hat{G}^{(0)}] \\
%   &&+ \frac{\lambda^2}{N} \left[ \frac{1}{2V_d^2 \mu^6} + \frac{1}{2V_d \mu^4} \left( [\hat{G}^{(0)}] + 4\mu^2 \frac{\partial [\hat{G}^{(0)}]}{\partial m^2} \right) \right]
%   + \mathcal{O}(\lambda^{3}, N^{-2}). \notag
%  \end{eqnarray}
 which is a well known result. The positive value of the dynamical mass shows that the symmetry is restored in the effective potential due to the IR effects.
 For the two-point function at separate points, we find that the effective masses of the propagators  are
 \begin{eqnarray}
  M_1^2(-\mu^2) &=& \frac{\lambda}{2V_d \mu^2} - \frac{\lambda}{2} [\hat{G}^{(0)}] - \frac{\lambda \mu^2}{2} \frac{\partial [\hat{G}^{(0)}]}{\partial m^2} + \mathcal{O}(\lambda^2), \\
  M_2^2(-\mu^2) &=& 2\mu^2 + 3 M_1^2(-\mu^2).
 \end{eqnarray}
 It is worth remarking that both   masses are strictly positive, showing that this resummation is enough to overcome the divergences appearing for  the tree-level propagators. When analytically continuing the results to dS, again the same discussion as for massless fields applies, that is, the results can only be trusted for points not too far apart. 

Let us  briefly comment on the perturbative corrections. The main aspect to take into account is that now the effective coupling is different than in the massless case. From Eq.~(\ref{negsqmass}) we have  the scaling $\sqrt{\langle |\vec{\phi_0}|^2 \rangle_0} \sim \lambda^{-1/2}$, and therefore the IR enhancement is stronger than for the massless case ($\sqrt{\langle |\vec{\phi_0}|^2 \rangle_0} \sim \lambda^{-1/4}$). Hence, all the interaction terms in $S^{(3)}_{int}$ contribute about the same. Indeed, taking into account that the terms of the form $\lambda (\vec{\phi_0}\cdot\vec{\hat{\phi}}) |\vec{\hat\phi}|^2$ must always be considered in pairs (by symmetry considerations, correlators with an odd number of  factors of $\phi_{0a}$  will vanish), their contribution will go as $\lambda^2 \langle |\vec{\phi_0}|^2 \rangle_0 \sim \lambda$. Given that the effective coupling in this case is also $\lambda$,   the perturbative corrections  contribute  already at the NLO level. Of course,  a   calculation of all contributions at that order will be much more involved than for the massless case, but,   in principle,  it can be  performed  using the R-Feynman rules.

\section{Long distance limit at   $\mathcal{O}\left(  \sqrt{\lambda},1/N \right)$ }\label{bubbles}

In this Appendix we provide the details of the calculation of the long distance limit, $r(x,x')\to \infty$, of the two-point functions  at leading order in $\sqrt{\lambda}$ and  next-to-leading (NLO) order  in $1/N$. As mentioned in Sec. \ref{LIRC}, there is an infinite set of Feynman diagrams contributing  to the two-point functions in this limit.

We need to compute   all    Feynman diagrams which  contribute at  NLO in $1/N$ and are in turn  enhanced at long distances, in the sense  that  they   give a correction at LO in   $\sqrt{\lambda}$ in the limit   $r(x,x')\to \infty$.  The first diagram  given in Fig.~\ref{fig:connected-daisy} is  an example   that contribute at NLO in $1/N$ (and also at LO), but  that  is  clearly NLO in $\sqrt{\lambda}$,  since the propagator in the loop   is evaluated at coincident points and therefore cannot give an  enhancement for   $r(x,x')\to \infty$. Hence, note the diagrams we need contain at least two vertexes.  
To proceed,  we use the  R-Feynman rules described in Sec. \ref{R2PF} to compute corrections  to both propagators $\hat{G_1}$ and $\hat{G_2}$ whose masses depend on $|\vec{\phi_0}|$. Then, the two point functions can be evaluated using Eq.~\eqref{symmetry-prop}. Hence, for $\hat{G}_1$  it is necessary  to compute the NLO correction in $1/N$, while for $\hat{G}_2$ the  LO is enough.   
To count the power of $1/N$ at leading order in the limit $N\to \infty$, it is useful to note that $|\vec{\phi}_0|$ must be considered as counting as $\sqrt{N}$.  In other words $|\vec{\phi}_0|=\sqrt{2Nu/\lambda}$ and $u$, as it turns into an  integration variable,  is considered to be independent of $N$. As usual, each trace with solid lines gives a factor of $N$.     

Let  us first consider the diagrams correcting $\hat{G}_2$.  Taking the above considerations into account,  it is simple to see the diagram in Fig.~\ref{fig:Diana1}$a$ does not vanish at $N\to \infty$. 
% \begin{figure}[h!]
%  \centering
%   \begin{tikzpicture}[anchor=base, baseline=-0.1cm]
%   \begin{feynman}
%     \vertex (a);
%     \vertex [right=1.2cm of a] (b);
%     \vertex [right=1.2 of b] (c); 
%         
%     \diagram* {
%       (a) -- [very thick,dash dot] (b) -- [very thick,dash dot] (c),
%     };
% 
%     \vertex at ($(b) + (0cm, 0cm)$) (n2) [blob,very thick,fill=white,minimum size=0.5cm] {};
%     \vertex at ($(b) + (0cm, -1cm)$) (l) {\( (a) \)};    
%   \end{feynman}
% \end{tikzpicture}
% \hspace{1cm}
%  \begin{tikzpicture}[anchor=base, baseline=-0.1cm]
%   \begin{feynman}
%     \vertex (a);
%     \vertex [right=1.2cm of a] (b);
%     \vertex [right=1.2 of b] (c); 
%         
%     \diagram* {
%       (a) -- [very thick,dash dot] (b) -- [very thick,dash dot] (c),
%     };
% 
%     \vertex at ($(b) + (0cm, 0cm)$) (n2) [blob,very thick,fill=black,minimum size=0.5cm] {};
%     \vertex at ($(b) + (0cm, -1cm)$) (l) {\( (a') \)};    
%   \end{feynman}
% \end{tikzpicture}
% \caption{...}
% \label{fig:App1}
% \end{figure} 
There are two ways of adding corrections to this diagrams that can be thought as dressing the bubble in the middle in two steps:
First, the bubble can be corrected by adding another bubble with a vertex with four solid lines (this gives a trace and therefore a factor $N$ that compensate the one suppressing the vertex). A sum of the chain of bubbles corrected in this way  gives a partially-dressed bubble, which we  draw  as a solid gray circle, 
 \begin{eqnarray}
% a)\quad\quad 
\begin{tikzpicture}[anchor=base, baseline=-0.1cm]
\begin{feynman}
    \vertex (a);        
    \vertex at ($(a) + (0cm, 0cm)$) (n) [blob,very thick,fill=black!40!white,minimum size=0.5cm] {};
  \end{feynman}
\end{tikzpicture}
\quad&=&\quad
\begin{tikzpicture}[anchor=base, baseline=-0.1cm]
  \begin{feynman}
    \vertex (a);        
    \vertex at ($(a) + (0cm, 0cm)$) (n) [blob,very thick,fill=white,minimum size=0.5cm] {};
  \end{feynman}
\end{tikzpicture}
\quad+\quad
\begin{tikzpicture}[anchor=base, baseline=-0.1cm]
  \begin{feynman}
    \vertex (a);
    \vertex at ($(a) + (0cm, 0cm)$) (n1) [blob,very thick,fill=white,minimum size=0.5cm] {};
    \vertex [right=0.5cm of n1] (n2) [blob,very thick,fill=white,minimum size=0.5cm] {};
  \end{feynman}
\end{tikzpicture}
\quad+\quad
\begin{tikzpicture}[anchor=base, baseline=-0.1cm]
  \begin{feynman}
    \vertex (a);
    \vertex at ($(a) + (0cm, 0cm)$) (n1) [blob,very thick,fill=white,minimum size=0.5cm] {};
    \vertex [right=0.5cm of n1] (n2) [blob,very thick,fill=white,minimum size=0.5cm] {};
    \vertex [right=0.5cm of n2] (n3) [blob,very thick,fill=white,minimum size=0.5cm] {};
  \end{feynman}
\end{tikzpicture}
\quad+\quad\dots\quad \equiv \quad \hat{B}\,, \label{App2-a} 
 \end{eqnarray}
 and we will denote as $\hat{B}$.  Second, each partially-dressed bubble can be further dressed by adding two vertexes with two solid lines and a discontinuous one,  as follows:
 \begin{eqnarray}
% b)\quad\quad
 \begin{tikzpicture}[anchor=base, baseline=-0.1cm]
  \begin{feynman}
    \vertex (a);        
    \vertex at ($(a) + (0cm, 0cm)$) (n) [blob,very thick,fill=black,minimum size=0.5cm] {};
  \end{feynman}
\end{tikzpicture}
\quad&=&\quad
\begin{tikzpicture}[anchor=base, baseline=-0.1cm]
  \begin{feynman}
    \vertex (a);        
    \vertex at ($(a) + (0cm, 0cm)$) (n) [blob,very thick,fill=black!40!white,minimum size=0.5cm] {};
  \end{feynman}
\end{tikzpicture}
\quad+\quad
\begin{tikzpicture}[anchor=base, baseline=-0.1cm]
  \begin{feynman}
    \vertex (a);
    \vertex at ($(a) + (0cm, 0cm)$) (n1) [blob,very thick,fill=black!40!white,minimum size=0.5cm] {};
    \vertex [right=1cm of n1] (n2) [blob,very thick,fill=black!40!white,minimum size=0.5cm] {};
    
    \diagram* {
      (n1) -- [very thick,dash dot] (n2),
    };    
  \end{feynman}
\end{tikzpicture} 
\quad+\quad
\begin{tikzpicture}[anchor=base, baseline=-0.1cm]
  \begin{feynman}
    \vertex (a);
    \vertex at ($(a) + (0cm, 0cm)$) (n1) [blob,very thick,fill=black!40!white,minimum size=0.5cm] {};
    \vertex [right=1cm of n1] (n2) [blob,very thick,fill=black!40!white,minimum size=0.5cm] {};
    \vertex [right=1cm of n2] (n3) [blob,very thick,fill=black!40!white,minimum size=0.5cm] {};
    
    \diagram* {
      (n1) -- [very thick,dash dot] (n2) -- [very thick,dash dot] (n3),
    };    
  \end{feynman}
\end{tikzpicture} 
\quad+\quad\dots\quad \equiv \quad \hat{\mathcal{B}}\, .\label{App2-b}
 \end{eqnarray}
  The dressed bubble  (which  is plotted as a black solid circle and we will denote as $\hat{\mathcal{B}}$) is obtained by summing this  chain of diagrams. Therefore, the diagram we need to compute can be represented as in Fig.~\ref{fig:Diana1}$a'$.

 We now  consider the diagrams correcting $\hat{G}_1$. We start by  writing  all the different diagrams   that contribute at order $1/N$ and have the minimum amount of vertexes. These are given in the first line of  Fig.~\ref{fig:App3}.  Then,  as for $\hat{G}_2$, there is actually a set of diagrams contributing at the same order, and the sum of them can be expressed in terms of the same resummed bubbles as in the second line of Fig.~\ref{fig:App3}.
 \begin{figure}[h!]
 \centering
 \begin{tikzpicture}[anchor=base, baseline=-0.1cm]
  \begin{feynman}
    \vertex (a);
    \vertex [right=1.2cm of a] (b);
    \vertex [right=1.2 of b] (c); 
    
    \diagram* {
      (a) -- [very thick,solid] (b) -- [very thick,solid] (c),
    };
    
    \vertex at ($(b) + (0cm, 0.25cm)$) (n1) [blob,very thick,fill=white,minimum size=0.7cm] {};
    \vertex at ($(b) + (0cm, 0cm)$) (n2) [blob,very thick,fill=white,minimum size=0.5cm] {};
    \vertex at ($(b) + (0cm, -1cm)$) (l) {\( (a) \)};    
  \end{feynman}
\end{tikzpicture}
\hspace{2cm}
\begin{tikzpicture}[anchor=base, baseline=-0.1cm]
\begin{feynman}
    \vertex (a);
    \vertex [right=0.6cm of a] (n1);
    \vertex [right=1.2cm of n1] (n2);
    \vertex [right=0.6cm of n2] (b);
        
    \diagram* {
      (a) -- [very thick,solid] (n1) -- [very thick,solid] (n2) -- [very thick,solid] (b),
       (n1) -- [very thick,dash dot,half left] (n2),
    };
    
    \vertex at ($(n1) + (0.6cm, -1cm)$) (l) {\( (b) \)};
  \end{feynman}
\end{tikzpicture}
\hspace{2cm}
\begin{tikzpicture}[anchor=base, baseline=-0.1cm]
  \begin{feynman}
    \vertex (a);
    \vertex [right=0.6cm of a] (n1);
    \vertex [right=1.2cm of n1] (n2);
    \vertex [right=0.6cm of n2] (b);
       
    \diagram* {
      (a) -- [very thick,solid] (n1) -- [very thick,solid] (n2) -- [very thick,solid] (b),
       (n1) -- [very thick,dash dot,half left] (n2),
    };
    
    \vertex at ($(n1) + (0.1cm, 0.26cm)$) (c) [blob,very thick,fill=white,minimum size=0.5cm] {};
    \vertex at ($(n1) + (0.6cm, -1cm)$) (l) {\( (c) \)};
  \end{feynman}
\end{tikzpicture}
\vspace{0.5cm}
\\
 \begin{tikzpicture}[anchor=base, baseline=-0.1cm]
  \begin{feynman}
    \vertex (a);
    \vertex [right=1.2cm of a] (b);
    \vertex [right=1.2 of b] (c); 
    
    \diagram* {
      (a) -- [very thick,solid] (b) -- [very thick,solid] (c),
    };
    
    \vertex at ($(b) + (0cm, 0.25cm)$) (n1) [blob,very thick,fill=white,minimum size=0.7cm] {};
    \vertex at ($(b) + (0cm, 0cm)$) (n2) [blob,very thick,fill=black,minimum size=0.5cm] {};
    \vertex at ($(b) + (0, -1cm)$) (l) {\( (a') \)};
  \end{feynman}
\end{tikzpicture}
\hspace{0.5cm}
\begin{tikzpicture}[anchor=base, baseline=-0.1cm]
\begin{feynman}
    \vertex (a);
    \vertex [right=0.6cm of a] (n1);
    \vertex [right=1.2cm of n1] (n2);
    \vertex [right=0.6cm of n2] (b);
        
    \vertex [right=0.6cm of b] (c);
    \vertex [right=0.6cm of c] (m1);
    \vertex [right=1.2cm of m1] (m2);
    \vertex [right=0.6cm of m2] (d);    
        
    \diagram* {
      (a) -- [very thick,solid] (b),
       (n1) -- [very thick,dash dot,half left] (n2),
       (c) -- [very thick,solid] (d),
       (m1) -- [very thick,dash dot,half left] (m2),
    };
    
    \vertex at ($(m1) + (0.6cm, 0.5cm)$) (e) [blob,very thick,fill=black,minimum size=0.5cm] {};
    \vertex at ($(b) + (0.3cm, -0.1cm)$) (l1) {\( + \)};
    \vertex at ($(l1) + (0, -1cm)$) (l2) {\( (b') \)};
  \end{feynman}
\end{tikzpicture}
\hspace{0.5cm}
\begin{tikzpicture}[anchor=base, baseline=-0.1cm]
  \begin{feynman}
    \vertex (a);
    \vertex [right=0.6cm of a] (n1);
    \vertex [right=1.2cm of n1] (n2);
    \vertex [right=0.6cm of n2] (b);
        
    \diagram* {
      (a) -- [very thick,solid] (n1) -- [very thick,solid] (n2) -- [very thick,solid] (b),
       (n1) -- [very thick,dash dot,half left] (n2),
    };
    
    \vertex at ($(n1) + (0.1cm, 0.26cm)$) (c) [blob,very thick,fill=black,minimum size=0.5cm] {};
    \vertex at ($(n1) + (0.6cm, -1cm)$) (l) {\( (c') \)};
  \end{feynman}
\end{tikzpicture}
\caption{Diagrams that correct the two-point function $\hat G_1$. The first line shows diagrams with up to 4 vertexes. The second line includes bubble resummations. }
\label{fig:App3}
\end{figure}
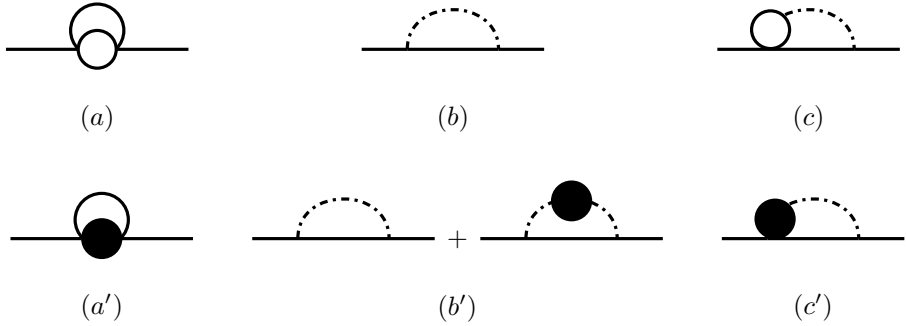
  
Of course, a brute-force  evaluation of the diagrams we have drawn is too difficult,  however, our aim here is to compute the leading order in $\sqrt{\lambda}$  in the long distance limit.  To achieve this, we will explote the knowledge of perturbation theory for massive fields \cite{Marolf:2010zp,Marolf:2010nz}. The main observation is that  the R-Feynman rules are built from modified ``massive'' propagators  $\hat{G}_1$ and $\hat{G}_2$, which differ  from standard massive propagators only by an homogeneous  (constant) contribution (see Fig.~\ref{fig:App4} for the corresponding diagrammatic representation):
 \begin{equation}\label{splitprop}
\hat{G}_\alpha=G_{\alpha}-G^{(0)}_\alpha,\,\,\mbox{(with $\alpha=1,2$)\,,}\end{equation}
where $G_{\alpha}$ is the standard propagator with mass $m_a$ and 
 \begin{equation}G^{(0)}_{\alpha}=\frac{1}{V_d m_\alpha^2}.\end{equation}
 With the use of Eq.~\eqref{splitprop}, the two-point diagrams we need to compute can be split into diagrams where the  two points are connected by at least one massive propagator $G_{\alpha}$ and disconnected diagrams.  It has been shown that any connected  two-point  diagram built with  massive propagators  and standard cubic or quadratic vertexes vanishes at long distances  \cite{Marolf:2010zp,Marolf:2010nz}.  Therefore, we need to keep and compute only the two-point diagrams that once split according to  Eq.~(\ref{splitprop}) have the two points disconnected.
All of these facts will become clearer in a moment,  after considering a few  examples.
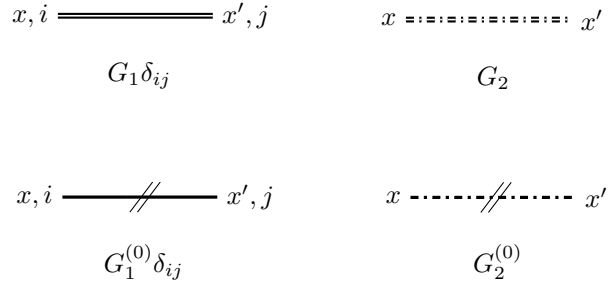
\begin{figure}[h!]
 \centering
\begin{tikzpicture}
  \begin{feynman}
    \vertex (a) {\(x,i\)};
    \vertex [right=1.2cm of a](b);
    \vertex [right=1.2cm of b] (c) {\(x',j\)};
    \vertex [below=0.5cm of b](d) {\(\quad\,\, G_1 \delta_{ij}\)};
        
    \diagram* {
      (a) -- [thick,solid,double] (c), %line width=0.6mm
    };
  \end{feynman}
\end{tikzpicture}
\hspace{1cm}
\begin{tikzpicture}
  \begin{feynman}
    \vertex (a) {\(x\)};
    \vertex [right=1.2cm of a](b);
    \vertex [right=1.2cm of b] (c) {\(x'\)};
    \vertex [below=0.5cm of b](d) {\(\quad G_2\)};
        
    \diagram* {
      (a) -- [thick,dash dot,double] (c), %line width=0.6mm
    };
  \end{feynman}
\end{tikzpicture}
\vspace{1cm}
\\
\begin{tikzpicture}
  \begin{feynman}
    \vertex (a) {\(x,i\)};
    \vertex [right=1.2cm of a](b);
    \vertex [right=1.2cm of b] (c) {\(x',j\)};
    \vertex [below=0.5cm of b](d) {\(\quad\,\, G_1^{(0)} \delta_{ij}\)};
     
     \vertex at ($(b) + (0.35cm, 0.2cm)$) (u1); 
     \vertex [right=0.1cm of u1] (u2);
     \vertex at ($(u1) + (-0.3cm, -0.4cm)$) (d1); 
     \vertex [right=0.1cm of d1] (d2);
     
    \diagram* {
      (a) -- [very thick,solid] (b) -- [very thick,solid] (c),
      (d1) -- [solid] (u1),
      (d2) -- [solid] (u2),
    };
  \end{feynman}
\end{tikzpicture}
\hspace{1cm}
\begin{tikzpicture}
  \begin{feynman}
    \vertex (a) {\(x\)};
    \vertex [right=1.2cm of a](b);
    \vertex [right=1.2cm of b] (c) {\(x'\)};
    \vertex [below=0.5cm of b](d) {\(\quad G_2^{(0)}\)};
        
     \vertex at ($(b) + (0.25cm, 0.2cm)$) (u1);
     \vertex [right=0.1cm of u1] (u2);
     \vertex at ($(u1) + (-0.3cm, -0.4cm)$) (d1); 
     \vertex [right=0.1cm of d1] (d2);
        
    \diagram* {
      (a) -- [very thick,dash dot] (c),
       (d1) -- [solid] (u1),
      (d2) -- [solid] (u2),
    };
  \end{feynman}
\end{tikzpicture}
\caption{Top: Diagrammatic representation of the standard  massive propagators: $G_1$ with $m_1^2= \frac{\lambda}{2N} |\vec{\phi}_0|^2\equiv u$, and $G_2$ with $m_2^2=\frac{3\lambda}{2N} |\vec{\phi}_0|^2=3u$. Bottom: Diagrammatic representation of the euclidean zero-mode part of the massive propagators: $G^{(0)}_a=(V_d m_a^2)^{-1} \,(a=1,2)$.  }
\label{fig:App4}
\end{figure} 
Let us start with the diagram in Fig.~\ref{fig:Diana1}$a'$. Notice that the bubble itself, $\hat{\mathcal{B}}(z,z')$, can be split into a  connected bubble, ${\mathcal{B}}(z,z')$  (represented as a black square), plus an homogeneous disconnected part, ${{\mathcal{B}}_d}$ (drawn as a cut bubble):
\begin{eqnarray}
\hat{\mathcal{B}}(z,z')\quad=\quad
\begin{tikzpicture}[anchor=base, baseline=-0.1cm] 
  \begin{feynman}
    \vertex (a);
    \vertex at ($(a)$) (b) [blob,very thick,fill=black,minimum size=0.5cm] {};
    
    \vertex at ($(b) + (-0.4cm, -0.2cm)$) (l1) {\(_{z}\)}; 
    \vertex at ($(b) + (+0.45cm, -0.18cm)$) (l2) {\(_{z'}\)}; 
  \end{feynman}
\end{tikzpicture}
\quad=\quad
\begin{tikzpicture}[anchor=base, baseline=-0.1cm]
  \begin{feynman}
    \vertex (a);
    \vertex at ($(a)$) (b) [rectangle,very thick,fill=black,minimum size=0.5cm] {};
    
    \vertex at ($(b) + (-0.4cm, -0.2cm)$) (l1) {\(_{z}\)}; 
    \vertex at ($(b) + (+0.45cm, -0.18cm)$) (l2) {\(_{z'}\)}; 
  \end{feynman}
\end{tikzpicture}
+\,\,\,\,
\begin{tikzpicture}[anchor=base, baseline=-0.1cm]
  \begin{feynman}
    \vertex (a);
    \vertex at ($(a)$) (b) [blob,very thick,fill=black,minimum size=0.5cm] {};
          
    \vertex at ($(b) + (0.25cm, 0.4cm)$) (u1); 
     \vertex [right=0.1cm of u1] (u2);
     \vertex at ($(u1) + (-0.6cm, -0.8cm)$) (d1); 
     \vertex [right=0.1cm of d1] (d2);
    
    \diagram* {
      (d1) -- [solid] (u1),
      (d2) -- [solid] (u2),
    };
  \end{feynman}
\end{tikzpicture}
\quad \equiv \quad \mathcal{B}(z,z') \,\,\,\,+\,\,\,\, \mathcal{B}_d \label{App5}
\end{eqnarray}
Then, the diagram   in Fig.~\ref{fig:Diana1}$a'$  can be split as 
 %\begin{figure}[h!]
 %\centering
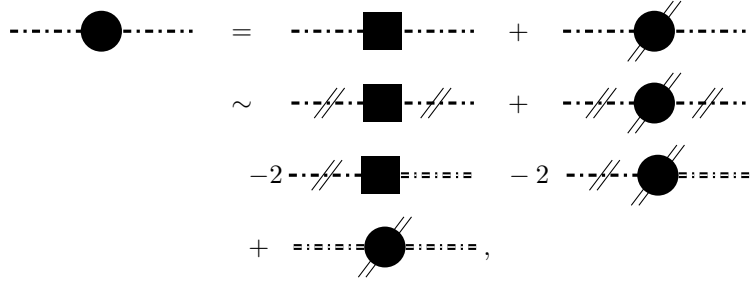
\begin{subequations} 
\begin{eqnarray*}
 \begin{tikzpicture}[anchor=base, baseline=-0.1cm]
  \begin{feynman}
    \vertex (a);
    \vertex [right=1.2cm of a] (b);
    \vertex [right=1.2 of b] (c); 
        
    \diagram* {
      (a) -- [very thick,dash dot] (b) -- [very thick,dash dot] (c),
    };

    \vertex at ($(b) + (0cm, 0cm)$) (n2) [blob,very thick,fill=black,minimum size=0.5cm] {};
  \end{feynman}
\end{tikzpicture}
\quad&=&\quad
\begin{tikzpicture}[anchor=base, baseline=-0.1cm]
  \begin{feynman}
    \vertex (a);
    \vertex [right=1.2cm of a] (b);
    \vertex [right=1.2 of b] (c); 
        
    \diagram* {
      (a) -- [very thick,dash dot] (b) -- [very thick,dash dot] (c),
    };

    \vertex at ($(b) + (0cm, 0cm)$) (n2) [rectangle,very thick,fill=black,minimum size=0.5cm] {};
  \end{feynman}
\end{tikzpicture}
\quad+\quad
 \begin{tikzpicture}[anchor=base, baseline=-0.1cm]
  \begin{feynman}
    \vertex (a);
    \vertex [right=1.2cm of a] (b);
    \vertex [right=1.2 of b] (c); 
        
      \vertex at ($(b) + (0.25cm, 0.4cm)$) (u1); 
     \vertex [right=0.1cm of u1] (u2);
     \vertex at ($(u1) + (-0.6cm, -0.8cm)$) (d1); 
     \vertex [right=0.1cm of d1] (d2);
        
    \diagram* {
      (a) -- [very thick,dash dot] (b) -- [very thick,dash dot] (c),
       (d1) -- [solid] (u1),
      (d2) -- [solid] (u2),
    };
    
    \vertex at ($(b) + (0cm, 0cm)$) (n2) [blob,very thick,fill=black,minimum size=0.5cm] {};
  \end{feynman}
\end{tikzpicture}
\\
&\sim&\quad
\begin{tikzpicture}[anchor=base, baseline=-0.1cm]
  \begin{feynman}
    \vertex (a);
    \vertex [right=1.2cm of a] (b);
    \vertex [right=1.2 of b] (c); 
        
     \vertex at ($(a) + (0.6cm, 0.2cm)$) (u1); 
     \vertex [right=0.1cm of u1] (u2);
     \vertex at ($(u1) + (-0.3cm, -0.4cm)$) (d1); 
     \vertex [right=0.1cm of d1] (d2);    
     
      \vertex at ($(c) + (-0.4cm, 0.2cm)$) (u3); 
     \vertex [right=0.1cm of u3] (u4);
     \vertex at ($(u3) + (-0.3cm, -0.4cm)$) (d3); 
     \vertex [right=0.1cm of d3] (d4);    
        
    \diagram* {
      (a) -- [very thick,dash dot] (b) -- [very thick,dash dot] (c),
      (d1) -- [solid] (u1),
      (d2) -- [solid] (u2),
      (d3) -- [solid] (u3),
      (d4) -- [solid] (u4),
    };
    
    \vertex at ($(b) + (0cm, 0cm)$) (n2) [rectangle,very thick,fill=black,minimum size=0.5cm] {};
  \end{feynman}
\end{tikzpicture}
\quad+\quad
 \begin{tikzpicture}[anchor=base, baseline=-0.1cm]
  \begin{feynman}
    \vertex (a);
    \vertex [right=1.2cm of a] (b);
    \vertex [right=1.2 of b] (c); 
        
      \vertex at ($(b) + (0.25cm, 0.4cm)$) (u5); 
     \vertex [right=0.1cm of u5] (u6);
     \vertex at ($(u5) + (-0.6cm, -0.8cm)$) (d5); 
     \vertex [right=0.1cm of d5] (d6);   
             
     \vertex at ($(a) + (0.6cm, 0.2cm)$) (u1); 
     \vertex [right=0.1cm of u1] (u2);
     \vertex at ($(u1) + (-0.3cm, -0.4cm)$) (d1); 
     \vertex [right=0.1cm of d1] (d2);    
     
      \vertex at ($(c) + (-0.4cm, 0.2cm)$) (u3); 
     \vertex [right=0.1cm of u3] (u4);
     \vertex at ($(u3) + (-0.3cm, -0.4cm)$) (d3); 
     \vertex [right=0.1cm of d3] (d4);       
        
    \diagram* {
      (a) -- [very thick,dash dot] (b) -- [very thick,dash dot] (c),
      (d1) -- [solid] (u1),
      (d2) -- [solid] (u2),
      (d3) -- [solid] (u3),
      (d4) -- [solid] (u4),
       (d5) -- [solid] (u5),
      (d6) -- [solid] (u6),
    };

    \vertex at ($(b) + (0cm, 0cm)$) (n2) [blob,very thick,fill=black,minimum size=0.5cm] {};
  \end{feynman}
\end{tikzpicture}
\\
&&\hspace{-0.19cm}-2\,
\begin{tikzpicture}[anchor=base, baseline=-0.1cm]
  \begin{feynman}
    \vertex (a);
    \vertex [right=1.2cm of a] (b);
    \vertex [right=1.2 of b] (c); 
        
     \vertex at ($(a) + (0.6cm, 0.2cm)$) (u1); 
     \vertex [right=0.1cm of u1] (u2);
     \vertex at ($(u1) + (-0.3cm, -0.4cm)$) (d1); 
     \vertex [right=0.1cm of d1] (d2);    

    \diagram* {
      (a) -- [very thick,dash dot] (b) -- [thick,dash dot,double] (c),
      (d1) -- [solid] (u1),
      (d2) -- [solid] (u2),
    };
    
    \vertex at ($(b) + (0cm, 0cm)$) (n2) [rectangle,very thick,fill=black,minimum size=0.5cm] {};
  \end{feynman}
\end{tikzpicture}
\hspace{0.38cm}-2\hspace{0.21cm}
 \begin{tikzpicture}[anchor=base, baseline=-0.1cm]
  \begin{feynman}
    \vertex (a);
    \vertex [right=1.2cm of a] (b);
    \vertex [right=1.2 of b] (c); 
        
      \vertex at ($(b) + (0.25cm, 0.4cm)$) (u5); 
     \vertex [right=0.1cm of u5] (u6);
     \vertex at ($(u5) + (-0.6cm, -0.8cm)$) (d5); 
     \vertex [right=0.1cm of d5] (d6);   
             
     \vertex at ($(a) + (0.6cm, 0.2cm)$) (u1); 
     \vertex [right=0.1cm of u1] (u2);
     \vertex at ($(u1) + (-0.3cm, -0.4cm)$) (d1); 
     \vertex [right=0.1cm of d1] (d2);    
        
    \diagram* {
      (a) -- [very thick,dash dot] (b) -- [thick,dash dot,double] (c),
      (d1) -- [solid] (u1),
      (d2) -- [solid] (u2),
       (d5) -- [solid] (u5),
      (d6) -- [solid] (u6),
    };
  
    \vertex at ($(b) + (0cm, 0cm)$) (n2) [blob,very thick,fill=black,minimum size=0.5cm] {};
  \end{feynman}
\end{tikzpicture}
\\
&&\hspace{-0.19cm}+\hspace{0.28cm}
\begin{tikzpicture}[anchor=base, baseline=-0.1cm]
  \begin{feynman}
    \vertex (a);
    \vertex [right=1.2cm of a] (b);
    \vertex [right=1.2 of b] (c); 
        
      \vertex at ($(b) + (0.25cm, 0.4cm)$) (u1); 
     \vertex [right=0.1cm of u1] (u2);
     \vertex at ($(u1) + (-0.6cm, -0.8cm)$) (d1); 
     \vertex [right=0.1cm of d1] (d2);
        
    \diagram* {
      (a) -- [thick,dash dot,double] (b) -- [thick,dash dot,double] (c),
       (d1) -- [solid] (u1),
      (d2) -- [solid] (u2),
    };
   
    \vertex at ($(b) + (0cm, 0cm)$) (n2) [blob,very thick,fill=black,minimum size=0.5cm] {};
  \end{feynman}
\end{tikzpicture}\,,
\end{eqnarray*}
\end{subequations}%\caption{...}
%\label{fig:App6}
%\end{figure}
where in the last line we discarded  the  connected contributions, which vanish at long distances. In formulas:
\begin{eqnarray}
&&\int_{z,z'} \hat{G}_2(x,z) \hat{\mathcal{B}}(z,z') \hat{G}_2(z',x') =\int_{z,z'}   \left(G_2(x,z)-G^{(0)}_2\right)  (\mathcal{B}(z,z')+ \mathcal{B}_d) \left(G_2 (x,z)-G^{(0)}_2\right) \\
&&\sim {G^{(0)}_2}^2 \int_{z,z'}    (\mathcal{B}(z,z')+\mathcal{B}_d) -2 G^{(0)}_2 \int_{z,z'}      (\mathcal{B}(z,z')+ \mathcal{B}_d  )   G_{2}(z',x') +  {\mathcal{B}_d}  \int_{z,z'}     G_2 (x,z) G_{2}(z',x'), \nonumber
\end{eqnarray} where the integrals go over all dS spacetime and  in the last line we  have only kept the disconnected contributions.

Now, the disconnected diagrams can be easily computed. Recalling  that $\int_{z'} G_{1}(z',x')=\int_{z} G_{1}(x,z)=G^{(0)}_1 V_d$ and taking into account  that $\int_{z}   \hat{\mathcal{B}}(z,z')$ cannot give an IR enhancement, we obtain that at leading order only the last term contributes 
 \begin{eqnarray}
&&\int_{z,z'} \hat{G}_2(x,z) \hat{\mathcal{B}}(z,z') \hat{G}_2(z',x') \sim  V_d^2   {G^{(0)}_2}^2  {\mathcal{B}_d},
\end{eqnarray}   with the   diagrammatic representation of this result  given by  

\begin{eqnarray} 
 \begin{tikzpicture}[anchor=base, baseline=-0.1cm]
  \begin{feynman}
    \vertex (a);
    \vertex [right=1.2cm of a] (b);
    \vertex [right=1.2 of b] (c); 
        
    \diagram* {
      (a) -- [very thick,dash dot] (b) -- [very thick,dash dot] (c),
    };

    \vertex at ($(b) + (0cm, 0cm)$) (n2) [blob,very thick,fill=black,minimum size=0.5cm] {};
  \end{feynman}
\end{tikzpicture}
\quad&\sim&\quad 
\begin{tikzpicture}[anchor=base, baseline=-0.1cm]
  \begin{feynman}
    \vertex (a);
    \vertex [right=1.2cm of a] (b);
    \vertex [right=1.2 of b] (c); 
        
     \vertex at ($(b) + (0.25cm, 0.4cm)$) (u5); 
     \vertex [right=0.1cm of u5] (u6);
     \vertex at ($(u5) + (-0.6cm, -0.8cm)$) (d5); 
     \vertex [right=0.1cm of d5] (d6);   
             
     \vertex at ($(a) + (0.6cm, 0.2cm)$) (u1); 
     \vertex [right=0.1cm of u1] (u2);
     \vertex at ($(u1) + (-0.3cm, -0.4cm)$) (d1); 
     \vertex [right=0.1cm of d1] (d2);    
     
     \vertex at ($(c) + (-0.4cm, 0.2cm)$) (u3); 
     \vertex [right=0.1cm of u3] (u4);
     \vertex at ($(u3) + (-0.3cm, -0.4cm)$) (d3); 
     \vertex [right=0.1cm of d3] (d4);       
        
    \diagram* {
      (a) -- [very thick,dash dot] (b) -- [very thick,dash dot] (c),
      (d1) -- [solid] (u1),
      (d2) -- [solid] (u2),
      (d3) -- [solid] (u3),
      (d4) -- [solid] (u4),
       (d5) -- [solid] (u5),
      (d6) -- [solid] (u6),
    };

    \vertex at ($(b) + (0cm, 0cm)$) (n2) [blob,very thick,fill=black,minimum size=0.5cm] {};
  \end{feynman}
\end{tikzpicture}\,.
\end{eqnarray}
 The remaining diagrams can be split similarly, and the corresponding results are: 
 % \begin{figure}[h!]
 %\centering
 \begin{eqnarray} 
   \begin{tikzpicture}[anchor=base, baseline=-0.1cm]
  \begin{feynman}
    \vertex (a);
    \vertex [right=1.2cm of a] (b);
    \vertex [right=1.2 of b] (c); 
    
    \diagram* {
      (a) -- [very thick,solid] (b) -- [very thick,solid] (c),
    };
    
    \vertex at ($(b) + (0cm, 0.25cm)$) (n1) [blob,very thick,fill=white,minimum size=0.7cm] {};
    \vertex at ($(b) + (0cm, 0cm)$) (n2) [blob,very thick,fill=black,minimum size=0.5cm] {};
  \end{feynman}
\end{tikzpicture}
\quad&\sim& -\hspace{0.08cm}
 \begin{tikzpicture}[anchor=base, baseline=-0.1cm]
  \begin{feynman}
    \vertex (a);
    \vertex [right=1.2cm of a] (b);
    \vertex [right=1.2 of b] (c); 
             
     \vertex at ($(a) + (0.6cm, 0.2cm)$) (u1); 
     \vertex [right=0.1cm of u1] (u2);
     \vertex at ($(u1) + (-0.3cm, -0.4cm)$) (d1); 
     \vertex [right=0.1cm of d1] (d2);    
     
     \vertex at ($(c) + (-0.4cm, 0.2cm)$) (u3); 
     \vertex [right=0.1cm of u3] (u4);
     \vertex at ($(u3) + (-0.3cm, -0.4cm)$) (d3); 
     \vertex [right=0.1cm of d3] (d4);       
    
    \vertex at ($(b) + (0.25cm, 0.4cm)$) (u5); 
     \vertex [right=0.1cm of u5] (u6);
     \vertex at ($(u5) + (-0.6cm, -0.8cm)$) (d5); 
     \vertex [right=0.1cm of d5] (d6);
     
     \vertex at ($(b) + (0.1cm, 0.8cm)$) (u7);
     \vertex [right=0.1cm of u7] (u8);
     \vertex at ($(u7) + (-0.3cm, -0.4cm)$) (d7); 
     \vertex [right=0.1cm of d7] (d8);    
    
     \vertex at ($(b) + (0cm, 0.25cm)$) (n1) [blob,very thick,fill=white,minimum size=0.7cm] {};
    \vertex at ($(b) + (0cm, 0cm)$) (n2) [blob,very thick,fill=black,minimum size=0.5cm] {};
    
    \diagram* {
      (a) -- [very thick,solid] (b) -- [very thick,solid] (c),
      (d1) -- [solid] (u1),
      (d2) -- [solid] (u2),
      (d3) -- [solid] (u3),
      (d4) -- [solid] (u4),
      (d5) -- [solid] (u5),
      (d6) -- [solid] (u6),
      (d7) -- [solid] (u7),
      (d8) -- [solid] (u8),
    };
  \end{feynman}
\end{tikzpicture}\,,
\\
\begin{tikzpicture}[anchor=base, baseline=-0.1cm]
  \begin{feynman}
    \vertex (a);
    \vertex [right=0.6cm of a] (n1);
    \vertex [right=1.2cm of n1] (n2);
    \vertex [right=0.6cm of n2] (b);
        
    \diagram* {
      (a) -- [very thick,solid] (n1) -- [very thick,solid] (n2) -- [very thick,solid] (b),
       (n1) -- [very thick,dash dot,half left] (n2),
    };
    
    \vertex at ($(n1) + (0.1cm, 0.26cm)$) (c) [blob,very thick,fill=black,minimum size=0.5cm] {};
  \end{feynman}
\end{tikzpicture}
+
\begin{tikzpicture}[anchor=base, baseline=-0.1cm]
  \begin{feynman}
    \vertex (a);
    \vertex [right=0.6cm of a] (n1);
    \vertex [right=1.2cm of n1] (n2);
    \vertex [right=0.6cm of n2] (b);

    \diagram* {
      (a) -- [very thick,solid] (n1) -- [very thick,solid] (n2) -- [very thick,solid] (b),
       (n1) -- [very thick,dash dot,half left] (n2),
    };
    
    \vertex at ($(n2) + (-0.1cm, 0.26cm)$) (c) [blob,very thick,fill=black,minimum size=0.5cm] {};
  \end{feynman}
\end{tikzpicture}
\quad&\sim& \hspace{-0.12cm}-2
\begin{tikzpicture}[anchor=base, baseline=-0.1cm]
  \begin{feynman}
    \vertex (a);
    \vertex [right=0.6cm of a] (n1);
    \vertex [right=0.6cm of n1] (m);
    \vertex [right=1.2cm of n1] (n2);
    \vertex [right=0.6cm of n2] (b);
        
      \vertex at ($(a) + (0.4cm, 0.2cm)$) (u1); 
     \vertex [right=0.1cm of u1] (u2);
     \vertex at ($(u1) + (-0.3cm, -0.4cm)$) (d1); 
     \vertex [right=0.1cm of d1] (d2);    
     
     \vertex at ($(b) + (-0.2cm, 0.2cm)$) (u3); 
     \vertex [right=0.1cm of u3] (u4);
     \vertex at ($(u3) + (-0.3cm, -0.4cm)$) (d3); 
     \vertex [right=0.1cm of d3] (d4);       
    
    \vertex at ($(n2) + (0.15cm, 0.65cm)$) (u5); 
     \vertex [right=0.1cm of u5] (u6);
     \vertex at ($(u5) + (-0.6cm, -0.8cm)$) (d5); 
     \vertex [right=0.1cm of d5] (d6);
     
     \vertex at ($(m) + (0.1cm, 0.2cm)$) (u7);
     \vertex [right=0.1cm of u7] (u8);
     \vertex at ($(u7) + (-0.3cm, -0.4cm)$) (d7); 
     \vertex [right=0.1cm of d7] (d8);        
        
       \vertex at ($(m) + (0.1cm, 0.75cm)$) (u9);
     \vertex [right=0.1cm of u9] (u10);
     \vertex at ($(u9) + (-0.3cm, -0.4cm)$) (d9); 
     \vertex [right=0.1cm of d9] (d10);       
        
    \diagram* {
      (a) -- [very thick,solid] (n1) -- [very thick,solid] (n2) -- [very thick,solid] (b),
       (n1) -- [very thick,dash dot,half left] (n2),
        (d1) -- [solid] (u1),
      (d2) -- [solid] (u2),
      (d3) -- [solid] (u3),
      (d4) -- [solid] (u4),
      (d5) -- [solid] (u5),
      (d6) -- [solid] (u6),
      (d7) -- [solid] (u7),
      (d8) -- [solid] (u8),
      (d9) -- [solid] (u9),
      (d10) -- [solid] (u10),
    };
    
     \vertex at ($(n2) + (-0.1cm, 0.26cm)$) (c) [blob,very thick,fill=black,minimum size=0.5cm] {};
  \end{feynman}
\end{tikzpicture}\,,
\\
\begin{tikzpicture}[anchor=base, baseline=-0.1cm]
\begin{feynman}
    \vertex (a);
    \vertex [right=0.6cm of a] (n1);
    \vertex [right=1.2cm of n1] (n2);
    \vertex [right=0.6cm of n2] (b);
        
    \diagram* {
      (a) -- [very thick,solid] (n1) -- [very thick,solid] (n2) -- [very thick,solid] (b),
       (n1) -- [very thick,dash dot,half left] (n2),
    };
  \end{feynman}
\end{tikzpicture}
+
\begin{tikzpicture}[anchor=base, baseline=-0.1cm]
\begin{feynman}
    \vertex (a);
    \vertex [right=0.6cm of a] (n1);
    \vertex [right=1.2cm of n1] (n2);
    \vertex [right=0.6cm of n2] (b);
        
    \diagram* {
      (a) -- [very thick,solid] (n1) -- [very thick,solid] (n2) -- [very thick,solid] (b),
       (n1) -- [very thick,dash dot,half left] (n2),
    };
    
    \vertex at ($(n1) + (0.6cm, 0.5cm)$) (c) [blob,very thick,fill=black,minimum size=0.5cm] {};
  \end{feynman}
\end{tikzpicture}
\quad&\sim&\quad
\begin{tikzpicture}[anchor=base, baseline=-0.1cm]
  \begin{feynman}
    \vertex (a);
    \vertex [right=0.6cm of a] (n1);
    \vertex [right=0.6cm of n1] (m);
    \vertex [right=1.2cm of n1] (n2);
    \vertex [right=0.6cm of n2] (b);
        
      \vertex at ($(a) + (0.4cm, 0.2cm)$) (u1); 
     \vertex [right=0.1cm of u1] (u2);
     \vertex at ($(u1) + (-0.3cm, -0.4cm)$) (d1); 
     \vertex [right=0.1cm of d1] (d2);    
     
     \vertex at ($(b) + (-0.2cm, 0.2cm)$) (u3); 
     \vertex [right=0.1cm of u3] (u4);
     \vertex at ($(u3) + (-0.3cm, -0.4cm)$) (d3); 
     \vertex [right=0.1cm of d3] (d4);       
     
     \vertex at ($(m) + (0.1cm, 0.2cm)$) (u7);
     \vertex [right=0.1cm of u7] (u8);
     \vertex at ($(u7) + (-0.3cm, -0.4cm)$) (d7); 
     \vertex [right=0.1cm of d7] (d8);        
        
       \vertex at ($(m) + (0.1cm, 0.75cm)$) (u9);
     \vertex [right=0.1cm of u9] (u10);
     \vertex at ($(u9) + (-0.3cm, -0.4cm)$) (d9); 
     \vertex [right=0.1cm of d9] (d10);       
        
    \diagram* {
      (a) -- [very thick,solid] (n1) -- [very thick,solid] (n2) -- [very thick,solid] (b),
       (n1) -- [very thick,dash dot,half left] (n2),
        (d1) -- [solid] (u1),
      (d2) -- [solid] (u2),
      (d3) -- [solid] (u3),
      (d4) -- [solid] (u4),
      (d7) -- [solid] (u7),
      (d8) -- [solid] (u8),
      (d9) -- [solid] (u9),
      (d10) -- [solid] (u10),
    };
  \end{feynman}
\end{tikzpicture}
 - 
\begin{tikzpicture}[anchor=base, baseline=-0.1cm]
  \begin{feynman}
    \vertex (a);
    \vertex [right=0.6cm of a] (n1);
    \vertex [right=0.6cm of n1] (m);
    \vertex [right=1.2cm of n1] (n2);
    \vertex [right=0.6cm of n2] (b);
        
      \vertex at ($(a) + (0.4cm, 0.2cm)$) (u1); 
     \vertex [right=0.1cm of u1] (u2);
     \vertex at ($(u1) + (-0.3cm, -0.4cm)$) (d1); 
     \vertex [right=0.1cm of d1] (d2);    
     
     \vertex at ($(b) + (-0.2cm, 0.2cm)$) (u3); 
     \vertex [right=0.1cm of u3] (u4);
     \vertex at ($(u3) + (-0.3cm, -0.4cm)$) (d3); 
     \vertex [right=0.1cm of d3] (d4);       
    
    \vertex at ($(m) + (0.25cm, 0.9cm)$) (u5); 
     \vertex [right=0.1cm of u5] (u6);
     \vertex at ($(u5) + (-0.6cm, -0.8cm)$) (d5); 
     \vertex [right=0.1cm of d5] (d6);
     
     \vertex at ($(m) + (0.1cm, 0.2cm)$) (u7);
     \vertex [right=0.1cm of u7] (u8);
     \vertex at ($(u7) + (-0.3cm, -0.4cm)$) (d7); 
     \vertex [right=0.1cm of d7] (d8);        
        
       \vertex at ($(n1) + (-0.1cm, 0.5cm)$) (u9);
     \vertex [right=0.1cm of u9] (u10);
     \vertex at ($(u9) + (0.3cm, -0.4cm)$) (d9); 
     \vertex [right=0.1cm of d9] (d10);
     
     \vertex at ($(n2) + (0.05cm, 0.5cm)$) (u11);
     \vertex [right=0.1cm of u11] (u12);
     \vertex at ($(u11) + (-0.3cm, -0.4cm)$) (d11); 
     \vertex [right=0.1cm of d11] (d12);       
        
    \diagram* {
      (a) -- [very thick,solid] (n1) -- [very thick,solid] (n2) -- [very thick,solid] (b),
       (n1) -- [very thick,dash dot,half left] (n2),
        (d1) -- [solid] (u1),
      (d2) -- [solid] (u2),
      (d3) -- [solid] (u3),
      (d4) -- [solid] (u4),
      (d5) -- [solid] (u5),
      (d6) -- [solid] (u6),
      (d7) -- [solid] (u7),
      (d8) -- [solid] (u8),
      (d9) -- [solid] (u9),
      (d10) -- [solid] (u10),
      (d11) -- [solid] (u11),
      (d12) -- [solid] (u12),
    };
    
    \vertex at ($(n1) + (0.6cm, 0.5cm)$) (c) [blob,very thick,fill=black,minimum size=0.5cm] {};
  \end{feynman}
\end{tikzpicture}\,.
\end{eqnarray}
%\caption{...}
%\end{figure}
Therefore, to evaluate this kind of diagrams we need to compute the cut bubbles and the proper symmetry factors for each diagram. The latter can be done as usual, while we   provide some details to achieve  the  former computation next. 

 Let us first focus on the partially-resummed bubble in Eq.~\eqref{App2-a}. We start by computing the leading order contribution of  the  fundamental bubble,  and of the diagram with two  and three fundamental bubbles, as shown in Eqs.~\eqref{App8a},\eqref{App8b} and \eqref{App8c},
%  \begin{figure}[h!]
%  \centering
\begin{subequations}\label{App8} 
 \begin{eqnarray} 
  \begin{tikzpicture}[anchor=base, baseline=-0.1cm]
  \begin{feynman}
    \vertex (a);        
    \vertex at ($(a) + (0cm, 0cm)$) (n) [blob,very thick,fill=white,minimum size=0.5cm] {};
  \end{feynman}
\end{tikzpicture}
\quad&=&\quad 2 \hat{G}_1^2(z,z') \sim  2 G^{(0)2}_1, \label{App8a} \\ 
 \begin{tikzpicture}[anchor=base, baseline=-0.1cm]
  \begin{feynman}
    \vertex (a);
    \vertex at ($(a) + (0cm, 0cm)$) (n1) [blob,very thick,fill=white,minimum size=0.5cm] {};
    \vertex [right=0.5cm of n1] (n2) [blob,very thick,fill=white,minimum size=0.5cm] {};
  \end{feynman}
\end{tikzpicture}
\quad&=&\quad -2\frac{\lambda}{2}\int_{y_1} \hat{G}_1^2(z,y_1)\hat{G}_1^2(y_1,z')\sim 2 \frac{\lambda V_d}{2}{G^{(0)}_1}^2 {G^{(0)}_1}^2, \label{App8b} \\
\begin{tikzpicture}[anchor=base, baseline=-0.1cm]
  \begin{feynman}
    \vertex (a);
    \vertex at ($(a) + (0cm, 0cm)$) (n1) [blob,very thick,fill=white,minimum size=0.5cm] {};
    \vertex [right=0.5cm of n1] (n2) [blob,very thick,fill=white,minimum size=0.5cm] {};
    \vertex [right=0.5cm of n2] (n3) [blob,very thick,fill=white,minimum size=0.5cm] {};
  \end{feynman}
\end{tikzpicture}
\quad&=&\quad 2\left(\frac{\lambda}{2}\right)^2\int_{y_1, y_2}  \hat{G}_1^2(z,y_1)\hat{G}_1^2(y_1,y_2)\hat{G}_1^2(y_2,z') \nonumber \label{App8c} \\
&\sim& \quad 2\left(\frac{\lambda V_d}{2}{G^{(0)}_1}^2\right)^2   {G^{(0)}_1}^2,\\
\begin{tikzpicture}[anchor=base, baseline=-0.1cm]
  \begin{feynman}
    \vertex (a);
    \vertex at ($(a) + (0cm, 0cm)$) (n1) [blob,very thick,fill=white,minimum size=0.5cm] {};
    \vertex [right=0.5cm of n1] (n2) [blob,very thick,fill=white,minimum size=0.5cm] {};
    \vertex [right=0.8cm of n2] (n3) {\(\dots\)};
    \vertex [right=0.75cm of n3] (n4) [blob,very thick,fill=white,minimum size=0.5cm] {};
    
    \draw [decoration={brace}, decorate] ($(n4) + (0cm, -0.35cm)$) -- ($(n1) + (0cm, -0.35cm)$) node[pos=0.5, below] {  \(n\)};    
  \end{feynman}
\end{tikzpicture}
\quad&=&\quad 2 \left(\frac{\lambda}{2}\right)^n\int_{y_1, y_2,\cdots,y_n}  \hat{G}_1^2(z,y_1)\hat{G}_1^2(y_1,y_2)\cdots\hat{G}_1^2(y_n,z') \nonumber \\
&\sim& \quad 2\left(\frac{\lambda V_d}{2}{G^{(0)}_1}^2\right)^n   {G^{(0)}_1}^2, \label{App8d} \\ 
\begin{tikzpicture}[anchor=base, baseline=-0.1cm]
  \begin{feynman}
    \vertex (a);        
    \vertex at ($(a) + (0cm, 0cm)$) (n) [blob,very thick,fill=black!40!white,minimum size=0.5cm] {};
  \end{feynman}
\end{tikzpicture}
\quad &\sim& \quad B_d= 2\sum_{n=0}^{+\infty} \left(\frac{\lambda V_d}{2}{G^{(0)}_1}^2\right)^n   {G^{(0)}_1}^2=  \frac{2{G^{(0)}_1}^2}{1-  \frac{\lambda V_d}{2}{G^{(0)}_1}^2 },  \label{App8e}
 \end{eqnarray}
 \end{subequations}
% \caption{...}
% \label{fig:App8}
% \end{figure}
% \begin{eqnarray}
% &&2 \hat{G}_1^2(z,z') \sim  2 G^{(0)2}_1 , \\
% &&-2\frac{\lambda}{2}\int_{y_1} \hat{G}_1^2(z,y_1)\hat{G}_1^2(y_1',z')\sim 2 \frac{\lambda V_d}{2}{G^{(0)}_1}^2 {G^{(0)}_1}^2, \\
% &&2\left(\frac{\lambda}{2}\right)^2\int_{y_1, y_2}  \hat{G}_1^2(z,y_1)\hat{G}_1^2(y_1,y_2)\hat{G}_1^2(y_2,z')\sim 2\left(\frac{\lambda V_d}{2}{G^{(0)}_1}^2\right)^2   {G^{(0)}_1}^2, \\
%   &&2 \left(\frac{\lambda}{2}\right)^n\int_{y_1, y_2,\cdots,y_n}  \hat{G}_1^2(z,y_1)\hat{G}_1^2(y_1,y_2)\cdots\hat{G}_1^2(y_n,z')\sim 2\left(\frac{\lambda V_d}{2}{G^{(0)}_1}^2\right)^n   {G^{(0)}_1}^2\\
% &&\sim B_d= 2\sum_{n=0}^{+\infty} \left(\frac{\lambda V_d}{2}{G^{(0)}_1}^2\right)^n   {G^{(0)}_1}^2=  \frac{2{G^{(0)}_1}^2}{1-  \frac{\lambda V_d}{2}{G^{(0)}_1}^2 },
% \end{eqnarray}
where we have taken into account the symmetry factors of the diagrams.
 
There are two main steps we have followed in the derivation of the result in Eq.~\eqref{App8e}: the first one is to note that given a diagram with $n$ fundamental bubbles,    once it is disconnected the value at leading order is the same as the fully disconnected diagram (i.e., the one where all the propagators are cut), up to a sign, with a minus ($-$)   if the number of cut lines is odd and a plus ($+$) if it is even. The next step consist in counting the amount of disconnected diagrams with a given sign and sum all of them taking into account the relative sign. These procedure, which might seem complicated at first, it  is not so much, since one can  show that   the  series becomes an alternated one so that once combined with the  additional  minus ($-$) sign accompanying each vertex,  the contribution of each diagram goes  always with a plus ($+$) sign (see Eqs.~\eqref{App8}). 

In a similar way, we can sum the partially-resummed bubbles to obtain the solid black one, as  
% Fig.~\ref{fig:App9},  
% \begin{figure}[h!]
%  \centering
\begin{subequations}\label{App9}
 \begin{eqnarray}
\begin{tikzpicture}[anchor=base, baseline=-0.1cm]
  \begin{feynman}
    \vertex (a);
    \vertex at ($(a) + (0cm, 0cm)$) (n1) [blob,very thick,fill=black!40!white,minimum size=0.5cm] {};
    \vertex [right=1cm of n1] (n2) [blob,very thick,fill=black!40!white,minimum size=0.5cm] {};
    \vertex [right=1cm of n2] (n3) [blob,very thick,fill=black!40!white,minimum size=0.5cm] {};
    
    \diagram* {
      (n1) -- [very thick,dash dot] (n2) -- [very thick,dash dot] (n3),
    };    
  \end{feynman}
\end{tikzpicture} 
\quad&\sim&\quad 2\left(\frac{\lambda}{2}\right)^2 2 \left(\frac{B_dV_d }{2}\right)^2\left(-\frac{  G^{(0)2}_2 |\vec{\phi}_0|^2}{N}\right), \label{App9a} \\
\begin{tikzpicture}[anchor=base, baseline=-0.1cm]
  \begin{feynman}
    \vertex (a);
    \vertex at ($(a) + (0cm, 0cm)$) (n1) [blob,very thick,fill=black!40!white,minimum size=0.5cm] {};
    \vertex [right=1cm of n1] (n2) [blob,very thick,fill=black!40!white,minimum size=0.5cm] {};
    \vertex [right=1cm of n2] (n3) [blob,very thick,fill=black!40!white,minimum size=0.5cm] {};
    \vertex [right=0.8cm of n3] (n4) {\(\dots\)};
    \vertex [right=0.75cm of n4] (n5) [blob,very thick,fill=black!40!white,minimum size=0.5cm] {};
    
    \diagram* {
      (n1) -- [very thick,dash dot] (n2) -- [very thick,dash dot] (n3),
    }; 
    
    \draw [decoration={brace}, decorate] ($(n5) + (0cm, -0.35cm)$) -- ($(n1) + (0cm, -0.35cm)$) node[pos=0.5, below] {  \(n+1\)}; 
    \end{feynman}
\end{tikzpicture}
\quad&=&\quad 2\left(\frac{\lambda}{2}\right)^{2 n} 2^n \left(\frac{B_d }{2}\right)^{n+1}\left(-\frac{  G^{(0)2}_2 V_d^2|\vec{\phi}_0|^2}{N}\right)^n, \label{App9b}\\
  \begin{tikzpicture}[anchor=base, baseline=-0.1cm]
  \begin{feynman}
    \vertex (a);        
    \vertex at ($(a) + (0cm, 0cm)$) (n) [blob,very thick,fill=black,minimum size=0.5cm] {};
  \end{feynman}
\end{tikzpicture}
\quad &\sim& \quad {\mathcal{B}_d}=  B_d\sum_{n=0}^{+\infty}   \left(-\frac{  \lambda V_d^2 B_d u G^{(0) }_2}{2}\right)^n=  \frac{B_d}{ \left(1+\frac{\lambda V_d B_d}{6}\right)}, \label{App9c}
 \end{eqnarray}
% \caption{...}
% \label{fig:App9}
% \end{figure}
\end{subequations}
% \begin{eqnarray}
% &\sim&2\left(\frac{\lambda}{2}\right)^2 2 \left(\frac{B_dV_d }{2}\right)^2\left(-\frac{  G^{(0)2}_2 |\vec{\phi}_0|^2}{N}\right)  \, , \\
% &\sim&2\left(\frac{\lambda}{2}\right)^{2 n} 2^n \left(\frac{B_d }{2}\right)^{n+1}\left(-\frac{  G^{(0)2}_2 V_d^2|\vec{\phi}_0|^2}{N}\right)^n  \,\\
% &\sim&{\mathcal{B}_d}=  B_d\sum_{n=0}^{+\infty}   \left(-\frac{  \lambda V_d^2 B_d u G^{(0) }_2}{2}\right)^n=  \frac{B_d}{ \left(1+\frac{\lambda V_d B_d}{6}\right)},
% \end{eqnarray}
where in the last line we have used that $|\vec{\phi}_0|^2/N= 2 u/\lambda$ and  $G^{(0)}_2=1/(3 V_d u)$.

%  \begin{figure}[h!]
%   
%  \includegraphics[width=15cm]{Diagramas1Ap}
% \end{figure}  
%  \begin{figure}[h!]
%   \includegraphics[width=15cm]{Diagramas2Ap}\end{figure}  
% 
%  \begin{figure}[h!]
% 
% \includegraphics[width=15cm]{Diagramas3Ap}
% 
%  % props-hat.png: 552x85 px, 300dpi, 4.67x0.72 cm, bb=0 0 132 20
% % \caption{Fig1: Th $\hat{G}_1$ and $\hat{G}_2$.}
% % \label{fig:props-hat}
% \end{figure}  

Having evaluated the bubbles, we can now express all the diagrams we need as a function of $u$, and the result is given in by:
\begin{subequations}\label{Diana10} 
\begin{eqnarray}
\begin{tikzpicture}[anchor=base, baseline=-0.1cm]
  \begin{feynman}
    \vertex (a);
    \vertex [right=1.2cm of a] (b);
    \vertex [right=1.2 of b] (c); 
        
    \diagram* {
      (a) -- [very thick,dash dot] (b) -- [very thick,dash dot] (c),
    };

    \vertex at ($(b) + (0cm, 0cm)$) (n2) [blob,very thick,fill=black,minimum size=0.5cm] {};
  \end{feynman}
\end{tikzpicture}
\quad&\sim&\quad \left(\frac{\lambda}{2}\right)^2 \mathcal{B}_d {G_2^{(0)}}^2 V_d^2 \left(\frac{2 u}{\lambda}\right), \label{App10a} \\
  \begin{tikzpicture}[anchor=base, baseline=-0.1cm]
  \begin{feynman}
    \vertex (a);
    \vertex [right=1.2cm of a] (b);
    \vertex [right=1.2 of b] (c); 
    
    \diagram* {
      (a) -- [very thick,solid] (b) -- [very thick,solid] (c),
    };
    
    \vertex at ($(b) + (0cm, 0.25cm)$) (n1) [blob,very thick,fill=white,minimum size=0.7cm] {};
    \vertex at ($(b) + (0cm, 0cm)$) (n2) [blob,very thick,fill=black,minimum size=0.5cm] {};
  \end{feynman}
\end{tikzpicture}
\quad&\sim&\quad -\frac{1}{N}\left(\frac{\lambda}{2}\right)^2 \mathcal{B}_d {G_1^{(0)}}^3 V_d^2, \label{App10b} \\
 \begin{tikzpicture}[anchor=base, baseline=-0.1cm]
  \begin{feynman}
    \vertex (a);
    \vertex [right=0.6cm of a] (n1);
    \vertex [right=1.2cm of n1] (n2);
    \vertex [right=0.6cm of n2] (b);

    \diagram* {
      (a) -- [very thick,solid] (n1) -- [very thick,solid] (n2) -- [very thick,solid] (b),
       (n1) -- [very thick,dash dot,half left] (n2),
    };
    
    \vertex at ($(n1) + (0.1cm, 0.26cm)$) (c) [blob,very thick,fill=black,minimum size=0.5cm] {};
  \end{feynman}
\end{tikzpicture}
+
\begin{tikzpicture}[anchor=base, baseline=-0.1cm]
  \begin{feynman}
    \vertex (a);
    \vertex [right=0.6cm of a] (n1);
    \vertex [right=1.2cm of n1] (n2);
    \vertex [right=0.6cm of n2] (b);

    \diagram* {
      (a) -- [very thick,solid] (n1) -- [very thick,solid] (n2) -- [very thick,solid] (b),
       (n1) -- [very thick,dash dot,half left] (n2),
    };
    
    \vertex at ($(n2) + (-0.1cm, 0.26cm)$) (c) [blob,very thick,fill=black,minimum size=0.5cm] {};
  \end{feynman}
\end{tikzpicture}
\quad&\sim&\quad \frac{4}{N}\left(\frac{\lambda}{2}\right)^3 \mathcal{B}_d {G_1^{(0)}}^3 G_2^{(0)}\left(\frac{2 u}{\lambda}\right) V_d^3, \label{App10c} \\
 \begin{tikzpicture}[anchor=base, baseline=-0.1cm]
\begin{feynman}
    \vertex (a);
    \vertex [right=0.6cm of a] (n1);
    \vertex [right=1.2cm of n1] (n2);
    \vertex [right=0.6cm of n2] (b);
        
    \diagram* {
      (a) -- [very thick,solid] (n1) -- [very thick,solid] (n2) -- [very thick,solid] (b),
       (n1) -- [very thick,dash dot,half left] (n2),
    };
  \end{feynman}
\end{tikzpicture}
+
\begin{tikzpicture}[anchor=base, baseline=-0.1cm]
\begin{feynman}
    \vertex (a);
    \vertex [right=0.6cm of a] (n1);
    \vertex [right=1.2cm of n1] (n2);
    \vertex [right=0.6cm of n2] (b);
        
    \diagram* {
      (a) -- [very thick,solid] (n1) -- [very thick,solid] (n2) -- [very thick,solid] (b),
       (n1) -- [very thick,dash dot,half left] (n2),
    };
    
    \vertex at ($(n1) + (0.6cm, 0.5cm)$) (c) [blob,very thick,fill=black,minimum size=0.5cm] {};
  \end{feynman}
\end{tikzpicture}
\quad&\sim&\quad 0. \label{App10d} 
\end{eqnarray}
\end{subequations}

In order to obtain the   final result we need to perform the integration in $u$, which in the $1/N$ expansion can be done systematically  using the Laplace method, as described in Appendix \ref{app-laplace}. The result cancels the constant in Eq.~\eqref{constant-resummed} in the main text, and makes the two-point functions to vanish in the long 
distance limit.

We finally mention here that the inhomogeneous 0-connected contribution \eqref{2pt-hatphi-NLO-disc-explicit}, which we dropped in the calculation on the sphere, will also be enhanced in dS. This enhancement is stronger the more derivatives with respect to $m^2$ act on the propagators $\hat{G}_1$ and $\hat{G}_2$, and therefore it will be more important for the 0-connected part than for the part we kept \eqref{resummed-2-pt-func}. This is why we kept those terms with the most derivatives acting on the $r$-dependent functions in \eqref{2pt-hatphi-NLO-disc-explicit}. However, it is straightforward to check that even with the stronger enhancement, the 0-connected part is still at least suppressed by one factor of $\sqrt{\lambda}$ with respect to the other parts. Moreover, this is true beyond the specific diagrams computed in previous section, and in particular it remains true for the infinite set of diagrams that we need to resum here in order to deal with the limiting value for $r \to \infty$. This follows directly from counting powers of $\lambda$ in the expansion \eqref{1/N-expanded-generic-product-text} with coefficients \eqref{double-exp-massless}, and taking into account that only one of the two functions $g(u)$ and $k(u)$ depends on $r$ and therefore receives an enhancement in dS of a factor $\lambda^{-1/2}$ for each derivative acting on it. Therefore, we can still ignore the 0-connected parts in dS in the current discussion, as it does not affect the limiting value of the two-point functions for $r \to \infty$, at least up to NNLO in $\sqrt{\lambda}$.  

% \section{0-connected contributions to the two-functions}

\end{document}